\documentclass[aps,pra,twocolumn,superscriptaddress,notitlepage,nofootinbib,longbibliography,colorlinks=true, allcolors=teal]{revtex4-2}
\usepackage{mathrsfs}
\usepackage{epsfig}
\usepackage{graphicx}
\usepackage{amsfonts}
\usepackage{amssymb}
\usepackage{amsmath}
\usepackage{dcolumn}
\usepackage{bm}
\usepackage{xcolor}
\usepackage{braket}
\usepackage{units}
\usepackage{xspace}

\DeclareMathOperator{\sech}{sech}
\usepackage{bbold}
\usepackage{orcidlink}

\usepackage[]{hyperref}

\newcommand{\ZJNU}{Department of Physics, Zhejiang Normal University, Jinhua 321004, China}
\newcommand{\SUSTECH}{Department of Physics, Southern University of Science and Technology, Shenzhen 518055, China}
\newcommand{\ECNU}{State Key Laboratory of Precision Spectroscopy, Quantum Institute for Light and Atoms, Department of Physics, East China Normal University, Shanghai 200062, China}
\newcommand{\BNL}{Beijing National Laboratory for Condensed Matter Physics, Institute of Physics, Chinese Academy of Sciences, Beijing 100190, China}
\newcommand{\UCAS}{School of Physical Sciences, University of Chinese Academy of Sciences, Beijing 100190, China}
\newcommand{\SLM}{Songshan Lake Materials Laboratory, Dongguan 523808, Guangdong, China}
\begin{document}
\title{Sub-Planck structures and sensitivity of the superposed photon-added or photon-subtracted squeezed-vacuum states}

\author{Naeem Akhtar\,\orcidlink{0000-0002-6756-2898}}
\email{naeem\_abbasi@zjnu.edu.cn}
\affiliation{\ZJNU}
\author{Jizhou Wu\,\orcidlink{0000-0003-4732-1437}}
\email{Corresponding author: wujz3@sustech.edu.cn}
\affiliation{\SUSTECH}
\author{Jia-Xin Peng\,\orcidlink{0000-0002-0119-0139}}
\affiliation{\ECNU}
\author{Wu-Ming Liu\,\orcidlink{0000-0002-5373-4417}}
\affiliation{\BNL}
\affiliation{\UCAS}
\affiliation{\SLM}
\author{Gao Xianlong\,\orcidlink{0000-0001-6914-3163}}
\email{gaoxl@zjnu.edu.cn}
\affiliation{\ZJNU}
\date{\today}

\begin{abstract}
The Wigner function of the compass state (a superposition of four coherent states) develops phase-space structures of dimension much less than the Planck scale~$\hbar$, which are crucial in determining the sensitivity of these states to phase-space displacements. In the present work we introduce compasslike states that may have connection to the contemporary experiments, which are obtained by either adding photons to or subtracting photons from the superposition of two squeezed-vacuum states.\;We show that when a significant quantity of photons is added (or subtracted), the Wigner functions of these states are shown to have phase-space structures of an area that is substantially smaller than the Planck scale. In addition, these states exhibit sensitivity to displacements that is much higher than the standard quantum limit. Finally, we show that both the size of the sub-Planck structures and the sensitivity of our states are strongly influenced by the average photon number, with the photon-addition case having a higher average photon number leading to the smaller sub-Planck structures and, consequently, being more sensitive to displacement than the photon-subtraction case.\;Our states offer unprecedented resolution to the external perturbations, making them suitable for quantum sensing applications.
\end{abstract}

\maketitle

\section{Introduction}\label{sec:intro}
Quantum mechanical states can be visualized in the phase space via the Wigner quasiprobability distribution~\cite{Wig32,Gerry05book,Sch01,Russel2021,leonhardt1997measuring}. The term ``Gaussian state" refers to a state having the Gaussian Wigner function~\cite{CarlosNB15,SCHUMAKER1986317}. The coherent state~\cite{Sch26} is an example of a Gaussian state. The Wigner function of the coherent state exhibits the Planck limit~\cite{Robertson1929,wheeler2014} in the phase space, 
which is also known as the standard quantum limit (SQL) or shot-noise limit. The Wigner function of certain non-Gaussian states may attain negative values~\cite{BUZEK19951,Walschaers2021,Freitas2021,Dodonov2002}, indicating that these states are nonclassical. The quantum superposition is the source of intriguing nonclassical properties of quantum states, such as quantum coherence~\cite{Streltsov2017,BUZEK19951}, squeezing~\cite{drummond2004quantum,Basit2023}, and entanglement~\cite{San92,San92E,San12}.
Nonclassical quantum states play a significant role in quantum-information processing~\cite{Mirrahimi2014}, tests of fundamental of physics~\cite{Bose1999,Marshall2003,Marshall2003b}, and applications in sensing and metrology~\cite{pan2014,mitchell2004}.

Nonclassical states are not always non-Gaussian: however, nonclassical states can be Gaussian in some cases. For example, a squeezed-vacuum state (SVS) is a common nonclassical state, but it possesses a Gaussian Wigner function~\cite{Walschaers2021,Freitas2021}.\;The Wigner function of the superposition of SVS is non-Gaussian and may have negative amplitudes~\cite{Happ2018,Barry1989}. Squeezed quantum states play an important role 
in performing enhanced quantum metrology~\cite{Maccone2004,drummond2004quantum}. Squeezed light has been utilized experimentally to carry out improved measurements~\cite{ligo2011gravitational,ligo2013gravitational}.

The superpositions of two coherent states with opposite phases (cat states) also possess non-Gaussian Wigner functions~\cite{Mil86,YS86,Schleich1991,knight1992}. Moreover, the superposition of four coherent states, which is known as the ``compass state''~\cite{Zurek2001}, exhibits nonclassical features in the Wigner function with dimensions far smaller than the SQL. The quantum states with sub-Planck structures are found to be very sensitive to environmental decoherence~\cite{Zurek2003} and have achieved prominent theoretical attentions in quantum metrology~\cite{Zurek2003,pan2013cat,Dalvit2006,Toscano06,ghosh2019b}.
The connection between sub-Planck structures and teleportation fidelity has been established~\cite{Eff6}. Sub-Fourier sensitivity is a classical analog of the sub-Planck structures~\cite{Eff5}.\;Compasslike states have been thoroughly investigated in several situations~\cite{Eff1,Eff2,Eff3,Eff7,Eff8,Eff9,Eff10,eff11,Eff12,Eff13,Eff14,Eff15,Naeem2022, Naeem2021,PhysRevA.99.063813,panigrahi2022compass}.\;Both theoretical~\cite{Prop1,Prop2,Prop3,prop4,Prop5,Prop6} and experimental studies~\cite{EXP1,EXP2,EXP4,EXP5} have been undertaken to achieve the controlled generation of such states.

In recent years, there has been a lot of focus on subtracting photons from or adding photons to the quantum states~\cite{Nha2004,takahashi2010,zhang1992,Hughes2014,QUESNE2001,SiKun2010,Yun2010,Dey2020,Meng12,Biswas2014,Ma2012,Vogel2018,TANG201586}.\;A non-Gaussian state can also be generated by adding or subtracting photons from a Gaussian state.\;For example, when photons are added to or subtracted from the Gaussian SVS, one may obtain two non-Gaussian squeezed states that have non-positive Wigner functions~\cite{SiKun2010,Yun2010,Dey2020,Meng12,Ma2012,TANG201586,Barnett2018}: photon-added squeezed-vacuum states\;(PASVS)~\cite{zhang1992} and photon-subtracted squeezed-vacuum states (PSSVS)~\cite{Meng12}. Both PASVS and PSSVS have attracted theoretical interest in quantum metrology~\cite{Lang2013,WANG2019102,Richard2014}.

The first theoretical investigation into the photon-addition operation is accomplished by adding photons to the coherent states~\cite{Agarwal1991}. Later, the photon-addition operation was successfully adopted in experiments by using a non-degenerate parametric amplifier with a weak coupling~\cite{Zavatta2004}. The PSSVS is currently the most successfully experimentally observed non-Gaussian SVS in quantum optics~\cite{Wenger2004,Ourjoumtsev2007}.

Schr\"{o}dinger cat states with higher amplitude can be used as qubits in quantum computing or as resources for quantum error-correcting coding~\cite{Cochrane1999,Ralph2003}.
\;Conventional methods are unable to produce Schr\"{o}dinger-cat-like states with the necessary amplitudes~\cite{Huang2015,sychev2017,ulanov2016loss}. Numerous theoretical~\cite{Gen2,Dakna1997,Takase2021,Podoshvedov2023} and experimental~\cite{Neergaard2006,Alexei2006,endo2023} research involving addition or subtractions of photons from the states have been carried out to achieve larger-amplitude, catlike states and even multicomponent cat states~\cite{Thekkadath2020}.

In the present work we introduce a few non-Gaussian SVSs that may also hold the properties of the compass state.\;In particular, we show that the Wigner function of the superpositions of two PASVSs (or PSSVSs) exhibit phase-space structures of an area, which varies inversely  with the number of photons added (or subtracted). When a large amount of the photons are added to or subtracted from our states the support area of these structures is substantially smaller than that found for coherent states. Similar sub-Planck structures are also found in the phase space of the mixed states related to the PASVSs and PSSVSs. We demonstrate that the average photon number in the states significantly influences the size of these sub-Planck structures, with the photon-addition case having higher average photon number, leading to smaller sub-Planck structures in the phase space than the photon-subtraction case.

To investigate the potential applications of these non-Gaussian states in quantum metrology, we analyze the overlap between these states and their slightly shifted analogs~\cite{Audenaert14}. The degree to which the state is sensitive against perturbations in the phase space can be determined from this overlap.
The sensitivity associated with coherent states cannot be improved by increasing the number of photons.\;Techniques using probes prepared in such states have the sensitivity at the SQL~\cite{Braun2018,Luis2010}.\;Here, we show that the sensitivity of our states is much higher than the SQL when the quantity of added (or subtracted) photons is relatively high.\;Furthermore, our superpositions exhibit this enhanced sensitivity in all phase-space directions, whereas the mixtures only do so for specific displacements. The varying average photon number in the states also 
contributed to the variation in the sensitivities between the photon-addition and -subtraction cases; it is shown that the photon-addition cases have higher sensitivity than the subtracted ones.

The structure of our paper is as follows. In Sec.~\ref{sec:intro_sub_Planck} we review the concept of the sub-Planck structures associated to the compass state. In Sec.~\ref{sec:Non_Gaussian} we review the Wigner functions of PASVS and PSSVS.
In Sec.~\ref{sec:sub_planck_nongaussian} we introduce our states and analyze their phase space by using the Wigner function, where we also discuss the sensitivity of our states against the phase-space perturbations. In Sec.~\ref{sec:conc} we provide our conclusion.

\section{Theory of Sub-Planck structures} \label{sec:intro_sub_Planck}
This section provides the background of the sub-Planck structures and is organized as follows.\;Section~\ref{subsec:priliminary} introduces the basic concepts that will be used in this article. In Sec.~\ref{subsec:compass} we review the sub-Planck structures that build in the phase space of the compass state.\;Section~\ref{subsec:sensitivity} explains the sensitivity to phase-space displacements associated to this compass state.

\subsection{Basic concepts}\label{subsec:priliminary}
The position operator $\hat{x}$ and the momentum operator $\hat{p}$ acts on an infinite-dimensional Hilbert space, forming the so-called Heisenberg-Weyl (HW) algebra~$\mathfrak{hw}(1)$~\cite{weyl1950theory,Per86,Gaz09} for a single degree of freedom. For convenience, we use the unit convention $\hbar =c = 1$ in this paper.\;The quantum uncertainty principle~\cite{Robertson1929,wheeler2014}
arising from commutator relations $\left[\hat{x},\hat{p}\right]=\text{i}$ limits the size of a phase-space structure~\cite{wheeler2014}, for example, represented by the Wigner function~\cite{Wig32}
for $\mathfrak{hw}(1)$ algebra and, more generally, by Moyal symbols~\cite{Moy49} for other symmetries~\cite{Per86}.
For convenience, we use the vector
\begin{equation}
    \bm{\zeta}:=(x,p)^{\top}
\end{equation}
to represent the position-momentum pair in the following.

A Schr\"{o}dinger coherent state is a nonspreading wave packet of the quantum harmonic oscillator~\cite{Sch26} and is an eigenstate of the annihilation operator: $\hat a \ket{\alpha}=\alpha\ket{\alpha}$ with $\alpha\in\mathbb{C}$. The coherent states are obtained by displacing the vacuum state $\ket{0}$, i.e.,
\begin{align}
	\ket{\alpha}=\hat{D}(\alpha)\ket{0},
	\end{align}
where
\begin{align}
\hat{D}(\alpha):=\exp(\alpha\hat{a}^\dagger-\alpha^*\hat{a})
\end{align}
is the displacement operator~\cite{Gaz09}.

The overlap between two coherent states $\ket{\alpha}$ and $\ket{\beta}$ is~\cite{gilmore1974}
\begin{align}
  \left|\Braket{\alpha|\beta}\right|^2=\text{e}^{-|\alpha|^2-|\beta|^2+2\beta^*\alpha}=\text{e}^{-|\alpha-\beta|^2},
\end{align}
which implies that two different coherent states are not orthogonal.

\begin{figure*}
    \centering
    (a)~\includegraphics[width=0.25\textwidth]{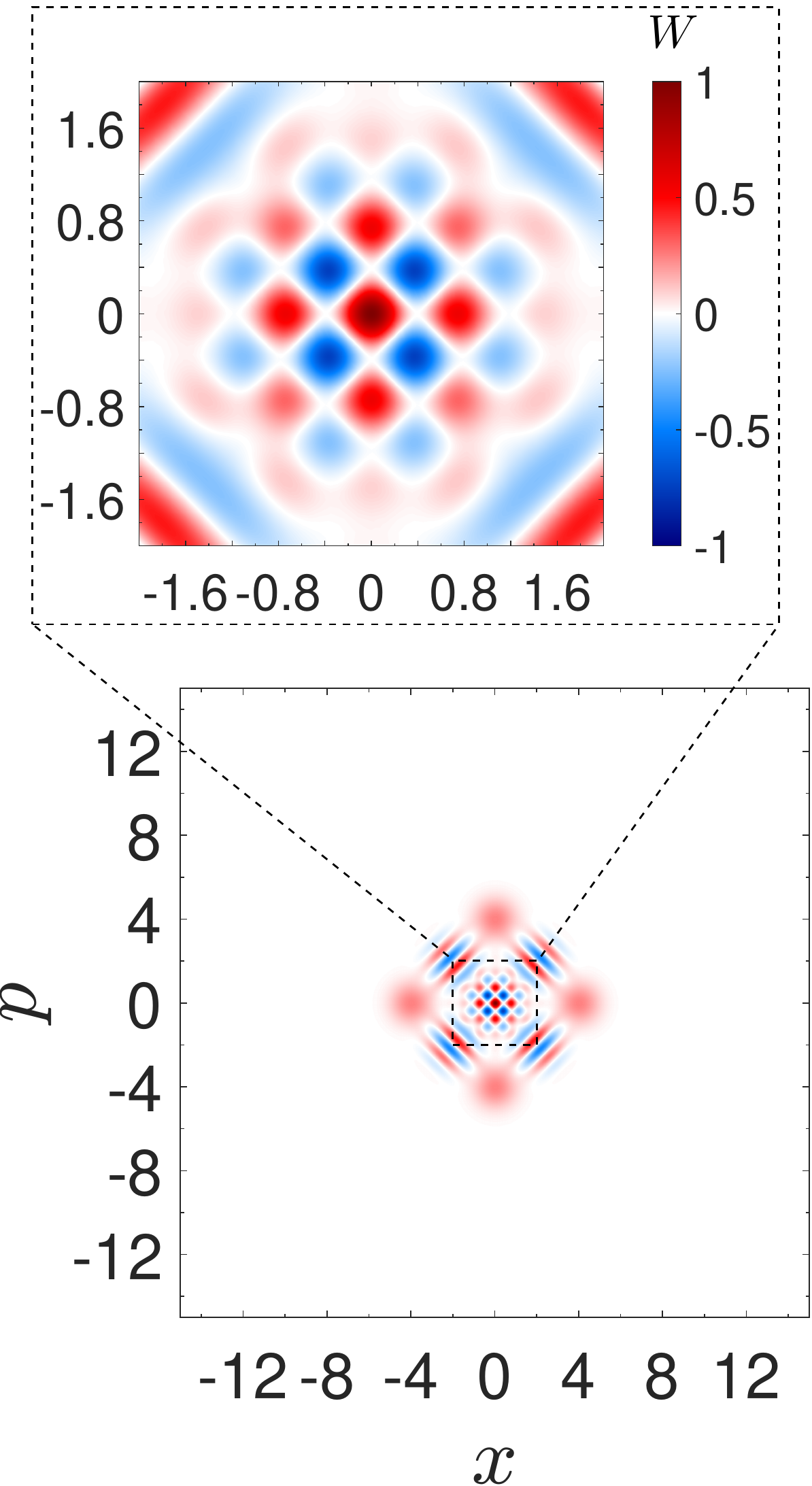}
    (b)~\includegraphics[width=0.25\textwidth]{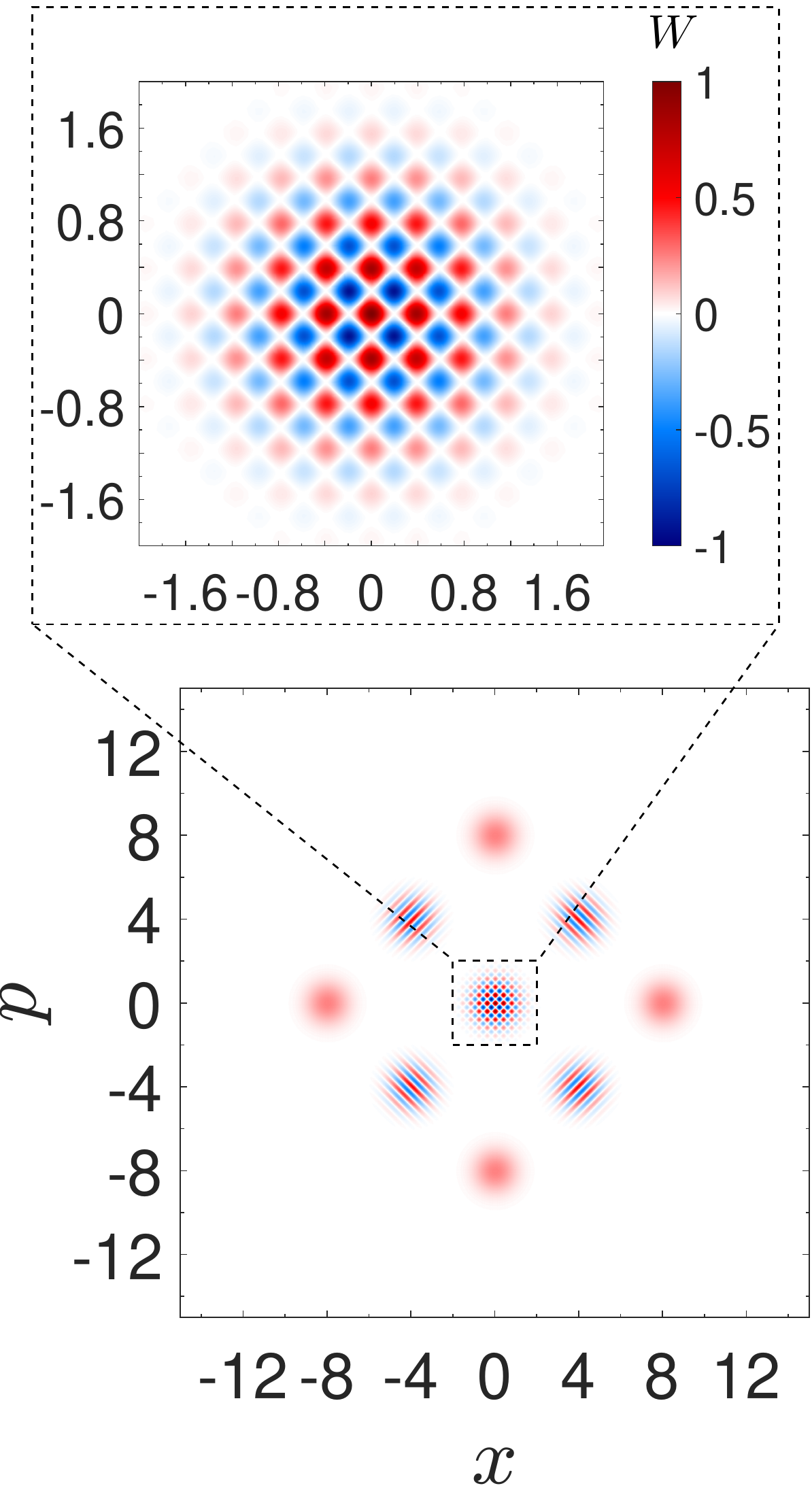}
    (c)~\includegraphics[width=0.25\textwidth]{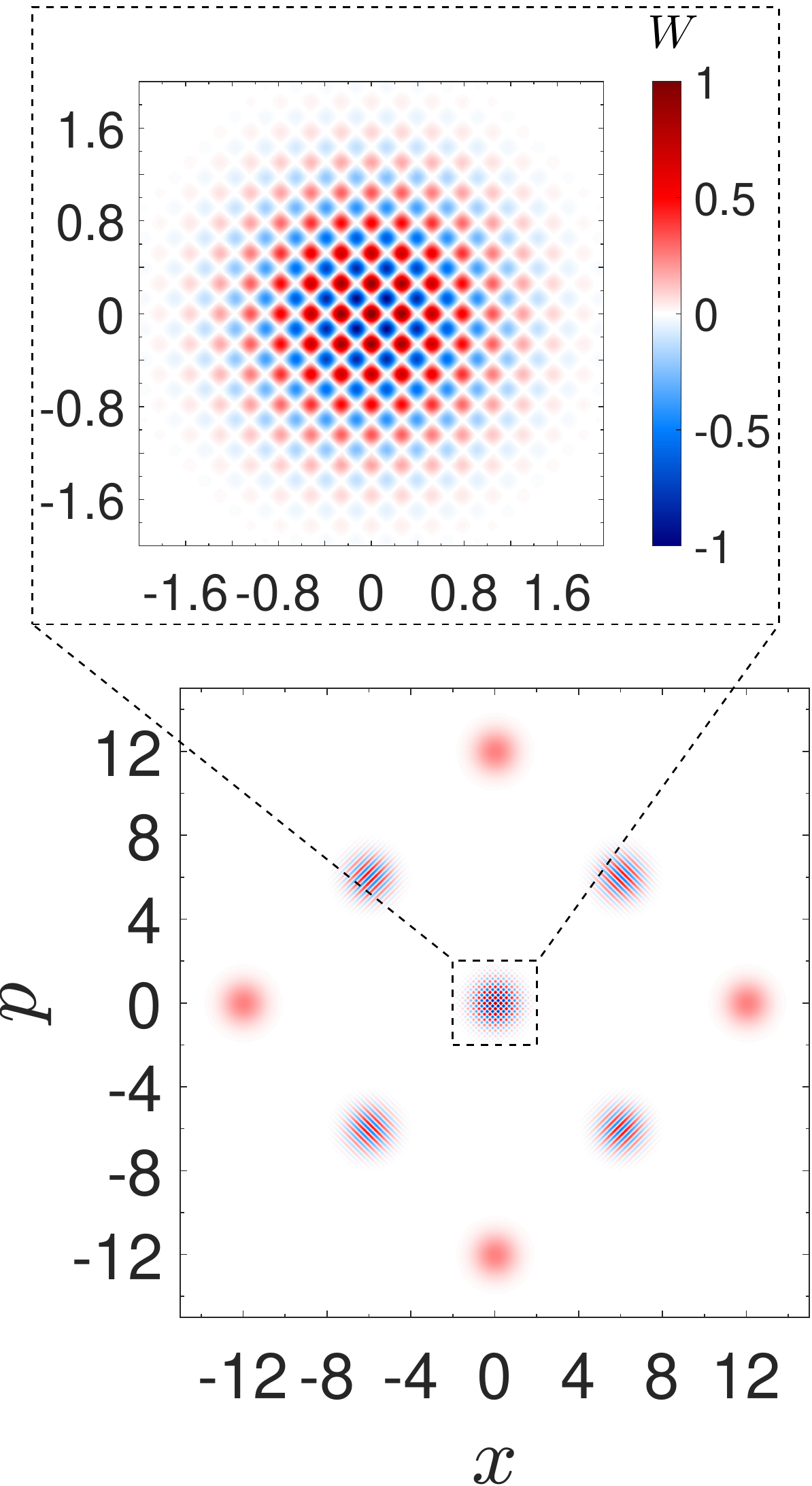}
    \caption{Wigner distribution of the compass state with (a)~$x_{0}=4$, (b)~$x_{0}=8$, and (c)~$x_{0}=12$. Insets represent the central interference pattern of each case.}
     \label{fig:compass_wigner}
\end{figure*}

The Wigner function for a generic quantum state
$\hat{\rho}$ is written as an expectation value of the parity kernel~\cite{CarlosNB15,Russel2021}
\begin{align}
    W_{\hat{\rho}}\left(\bm{\zeta}\right)
    :=\text{tr}\left[\hat{\rho}\hat{\Delta}(\alpha)\right],
\label{eq:wigner_general}
\end{align}
with
\begin{align}
 \hat{\Delta}(\alpha):=2\hat{D}(\alpha)\hat{\Pi}\hat{D}^\dagger(\alpha),\,
 \hat{\Pi}
:=\left(-1\right)^{\hat{a}^\dagger\hat{a}}
\end{align}
being the displaced parity operator.

The Wigner function for a coherent state is a strictly positive function
and appeared as a Gaussian of the form~\cite{leonhardt1997measuring} (we omit the normalization of states and their Wigner functions throughout the paper)
\begin{align}
  G(\bm{\zeta};\pm x_0,\pm p_0)=\text{e}^{-(x\mp x_0)^2-(p\mp p_0)^2},
  \label{eq:coh_wig}
\end{align}
where $(x_0,p_0)$ is the location of the coherent state in phase
space. The product of uncertainties of position and momentum for a coherent state has a lower
limit $\Delta x \Delta p=\nicefrac{1}{2}$~\cite{leonhardt1997measuring,Robertson1929,wheeler2014,gilmore1974}, which is also known as the \textit{Planck action} in the phase space.

It is a common belief that phase-space structures with areas smaller than the Planck scale either do not exist or have no observational consequences for physical quantum states. In fact, this is true for all Gaussian states (coherent, squeezed, thermal, etc.)~\cite{CarlosNB15,SCHUMAKER1986317} and even for other non-Gaussian states like cat states~\cite{Mil86,YS86} that exhibit rapid oscillations in one direction of phase space but an infinite Gaussian profile in the orthogonal direction~\cite{Naeem2021}.
However, this notion was refuted by Zurek~\cite{Zurek2001}, who demonstrated that the Wigner function of compass states develops phase-space structures with dimensions far smaller than the Planck scale, arguing that these structures play a vital role
in determining the sensitivity of these states against perturbations.

\subsection{Zurek compass state}\label{subsec:compass}
The Zurek compass state~\cite{Zurek2001} is obtained from the superposition of the following four coherent states
\begin{align}\label{eq:compass_state}
    \ket{\psi}:=\ket{\nicefrac{x_0}{\sqrt{2}}}+\ket{\nicefrac{-x_0}{\sqrt{2}}}+\ket{\nicefrac{\text{i}x_0}{\sqrt{2}}}+\ket{\nicefrac{-\text{i}x_0}{\sqrt{2}}},
\end{align}
with $x_0 \in \mathbb{R}$. Figure~\ref{fig:compass_wigner} depicts the Wigner function for this compass state for the cases of $x_{0}=4,8$, and 12.
Note that we normalize the Wigner functions throughout by using their maximum amplitudes, $|W_{\hat{\rho}}(0)|$. The Wigner function of the compass state (\ref{eq:compass_state}) can be represented as follows:
\begin{align}
 W_{\ket{\psi}}(\bm{\zeta})=W_{\circ}(\bm{\zeta})+W_{\Xi}(\bm{\zeta})+W_\boxplus(\bm{\zeta}),
 \label{eq:compass_wig}
\end{align}
where the first term,
\begin{align}
   W_{\circ}(\bm{\zeta}):=&\nonumber G(\bm{\zeta};x_0,0)+G(\bm{\zeta};-x_0,0)+G(\bm{\zeta};0,x_0)\\&+G(\bm{\zeta};0,-x_0),
\end{align}
represents the Wigner function of four coherent states that
appear in the phase space as Gaussian lobes. The second term in Eq.~\eqref{eq:compass_wig} is
\begin{equation}
W_{\Xi}(\bm{\zeta}):=\frac12\sum_{i_1,i_2=\pm 1}I(i_1 x,i_2 p)
\end{equation}
with
\begin{align}
     I(\bm{\zeta}):=G(\bm{\zeta};\nicefrac{x_0}{2},\nicefrac{x_0}{2})\cos\Big[x_0\left(x+p-\frac{x_0}{2}\right)\Big],
\end{align}
reflecting the Gaussian-modulated oscillations that appear far away from the phase-space origin.

\begin{figure}
\centering
\includegraphics[width=0.6\columnwidth]{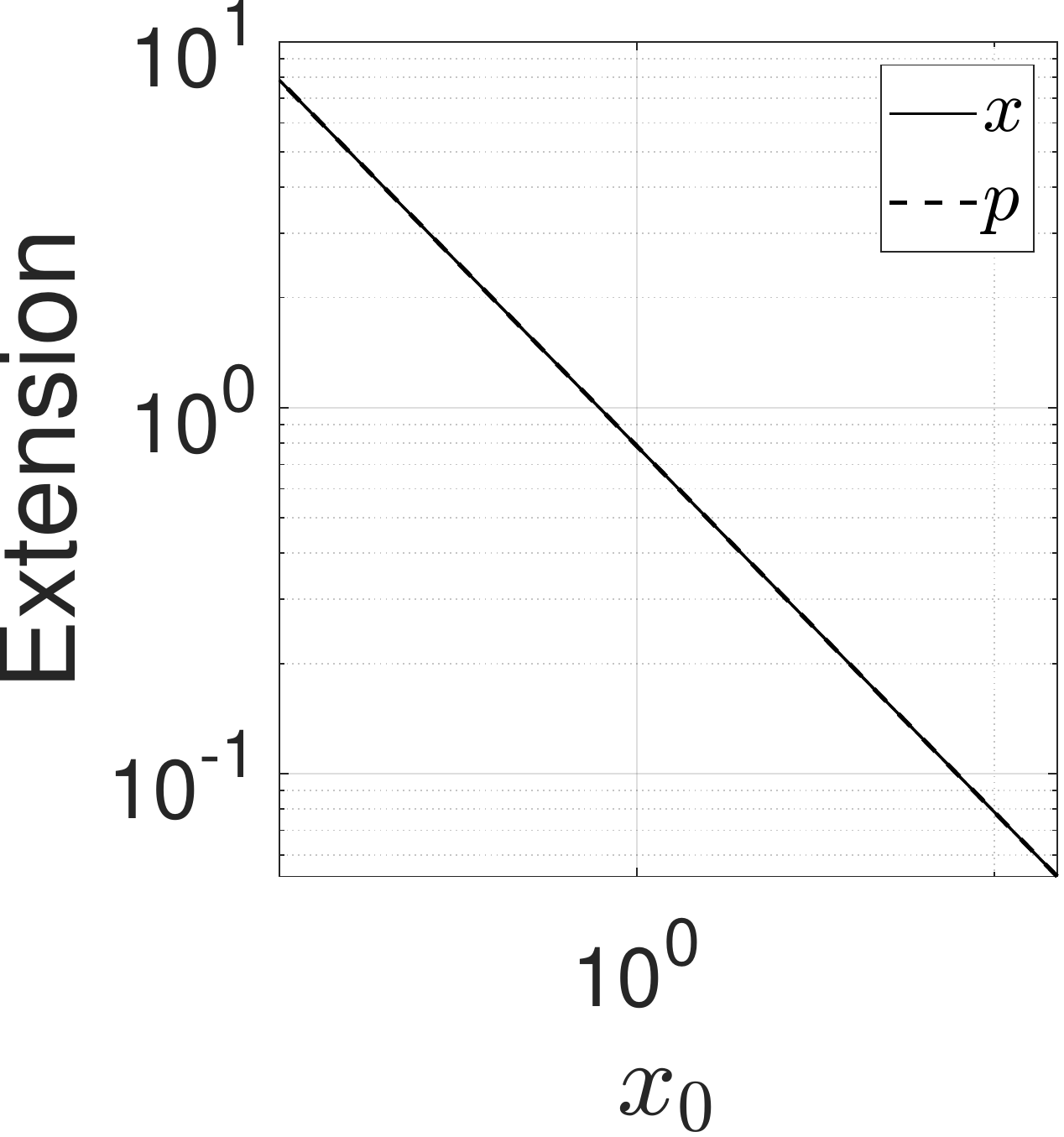}
\caption{Extension of the central phase-space structure versus $x_{0}$ of the compass state along the $x$ and $p$ directions.}
 \label{fig:compass_area_versus_n}
\end{figure}

The central pattern resembles a chessboard as shown in the insets of Fig.~\ref{fig:compass_wigner} and is generated by
\begin{align}\label{eq:chess0}
 W_\boxplus(\bm{\zeta}):=\frac12 G(\bm{\zeta};0,0)\Big[\cos(2x_0 x)+\cos(2x_0 p)\Big].
\end{align}
This pattern consists of tiles with alternate signs (a central chessboardlike pattern). The extension of each tile can be roughly estimated by calculating zeros of Eq.~\eqref{eq:chess0}, and it is found that it is proportional to $x^{-1}_0$ in all directions of phase space. Numerically, we use the half width at half maximum~(HWHM) of the central phase-space tile to represent the extension. In Fig.~\ref{fig:compass_area_versus_n} we show the log-log plot of the extension of the central tile along the $x$ and $p$ directions. It is demonstrated that for $x_0\gg1$, the extension of the central tile is constrained in all phase-space directions and can be simultaneously much smaller than the corresponding extensions of a coherent state. Note that the mixture of two cat states also contains the same sub-Planck structures that are found in the compass state~\cite{Naeem2021}.

The sub-Planck structures 
also emerge in the Wigner functions of non-Gaussian states with
the SU(1,1)~\cite{Naeem2022} and SU(2) symmetries~\cite{Naeem2021}. 
In particular, it has been found that the Wigner function of the superposition of four SU(1,1) (or SU(2)) coherent states 
also have sub-Planck structures similar to the compass state when represented on the Poincar\'e\ disk~\cite{Naeem2022} (or the sphere~\cite{Naeem2021}). 
The two-mode bosonic realization of the SU(1,1) implies that the sub-Planck structures in the phase space of the SU(1,1) compass state can be associated to the number of photons added to one of the modes
of the two-mode squeezed number states~\cite{Naeem2022},
and they arise at greater numbers of these added photons. The existence of the sub-Planck structures in the Wigner function of the SU(2) compass state can similarly be linked to the angular momentum; the
higher value of the angular momentum causes sub-Planck structures in the phase space~\cite{Naeem2021}.
In the subsequent sections, we will demonstrate how adding or subtracting photons from superpositions related to the one-mode Gaussian SVS can also cause the emergence of
sub-Planck structures
in the phase space of those states.

\subsection{Sensitivity of compass state}\label{subsec:sensitivity}
The sensitivity of the compass state to displacements is determined by calculating the overlap between it and its slightly displaced version~\cite{Audenaert14}. The overlap between a state $\hat{\rho}$ and its displaced version $\hat{D}(\delta\alpha)\hat{\rho}\hat{D}^\dagger(\delta\alpha)$ is
\begin{equation}
\label{eq:overlap_HW}
O_{\hat{\rho}}(\delta\alpha)
:=\text{tr}\{\hat{\rho}\hat{D}(\delta\alpha)\hat{\rho}\hat{D}^\dagger(\delta\alpha)\}	
=\left|\braket{\psi|\hat{D}(\delta\alpha)|\psi}\right|^2,
\end{equation}
where $\delta\alpha\in\mathbb{C}$ is an arbitrary displacement. Note that the last equality of above expression holds when the state is pure, $\hat{\rho}=\ket{\psi}\!\bra{\psi}$. The smaller the displacement $\delta \alpha$ needs to be in order to bring the overlap
to zero, the more sensitive the state is claimed to be against displacements~\cite{Toscano06}.
This overlap results in 
\begin{equation}
    O_{\ket{\alpha}}(\delta \alpha)=\text{e}^{-|\delta\alpha|^2},
    \label{eq:oalphadeltaalpha}
\end{equation}
for a coherent state $\ket{\alpha}$, indicating that the smallest noticeable displacement that vanishes
this overlap is above the Planck scale, $|\delta\alpha|>1$. It is interesting to note that the sensitivity to displacements in coherent states is independent of the quantity of quanta contained in the state, $\bar{n} =\braket{\hat{a}^\dagger\hat{a}}=|\alpha|^2$. Therefore increasing $\bar{n}$ will not improve the sensitivity and is solely limited by the \textit{shot noise} introduced by vacuum fluctuations~\cite{Braun2018,Luis2010}.

\begin{figure*}
    \centering
    (a)~\includegraphics[width=0.25\textwidth]{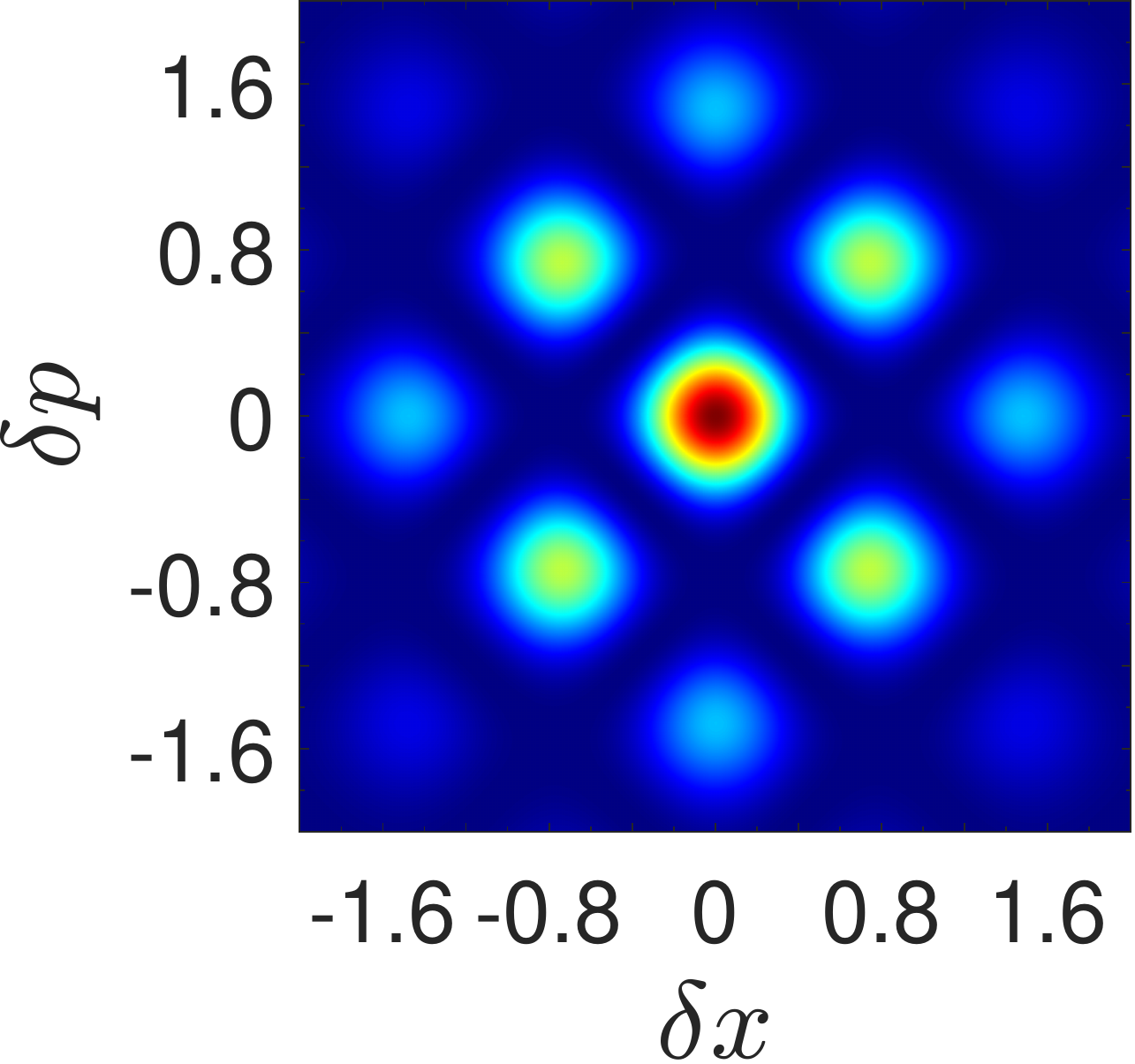}
    (b)~\includegraphics[width=0.25\textwidth]{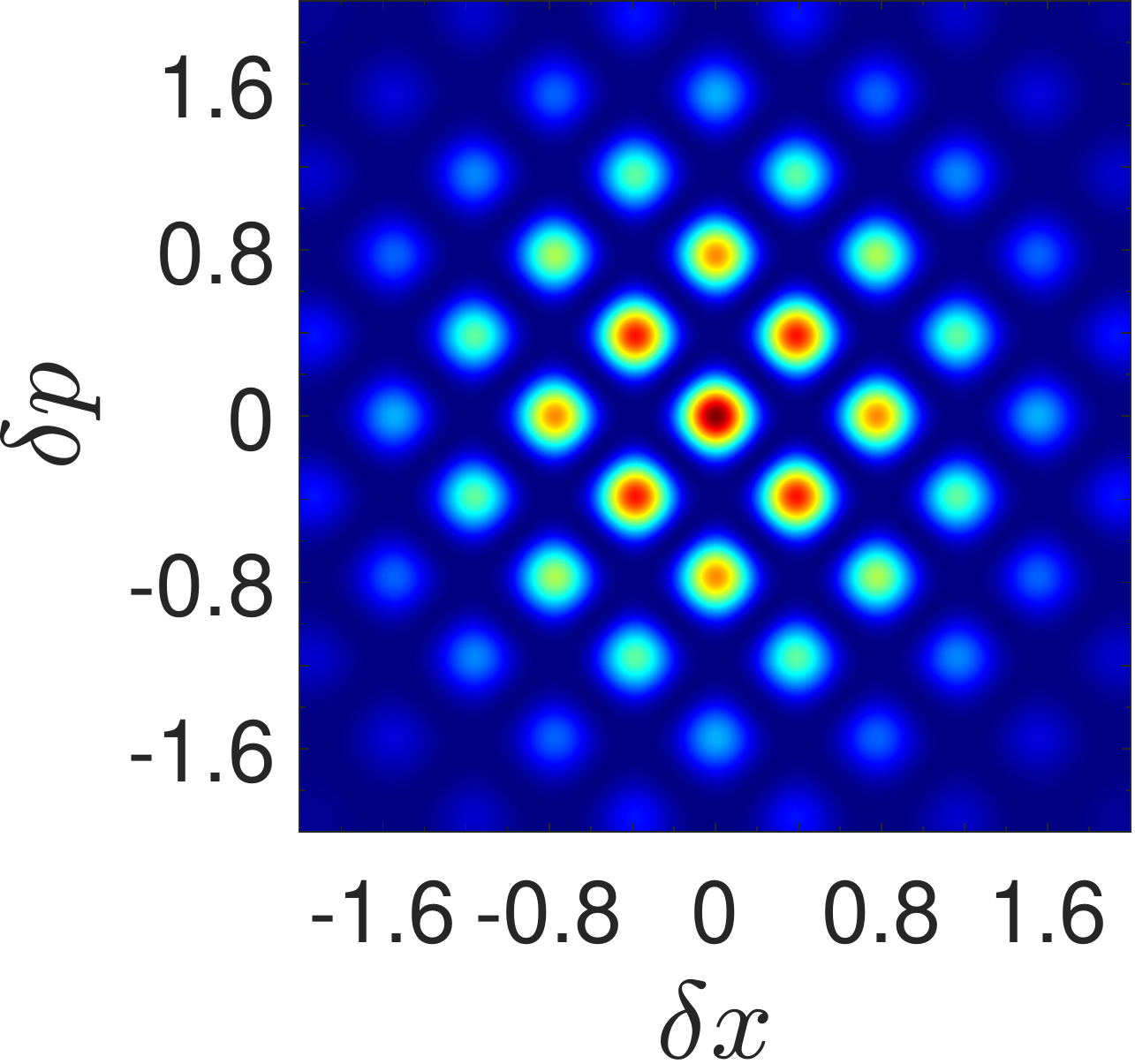}
    (c)~\includegraphics[width=0.305\textwidth]{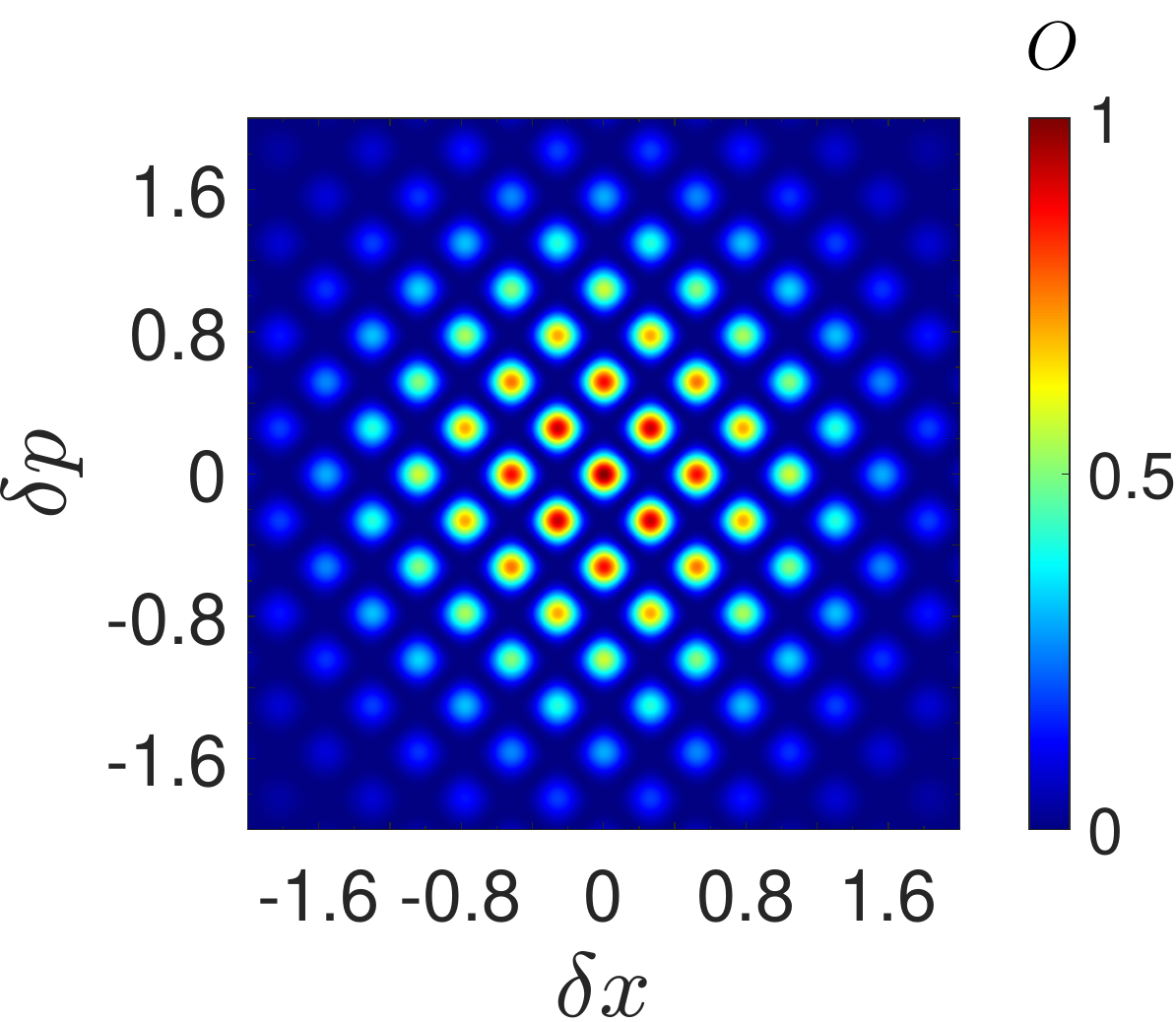}
    \caption{Overlap of the compass state with its $\delta \alpha$-displaced part with $\delta \alpha = (\delta x+\mathrm{i}\delta p)/\sqrt{2}$: (a)~$x_{0}=4$, (b)~$x_{0}=8$, and (c)~$x_{0}=12$.}
    \label{fig:compass_overlap}
\end{figure*}

We now discuss the sensitivity of the compass state (\ref{eq:compass_state}) to phase-space displacements. Assuming $x_0\gg 1$ and $|\delta\alpha|\ll1$ , the overlap (\ref{eq:overlap_HW}) for this compass state results in
\begin{align}
O_{\ket{\psi}}(\delta\alpha)=\frac14 \mathrm{e}^{-\frac12|\delta\alpha|^2}\big[\cos\left(x_0 \delta x\right)+\cos\left(x_0 \delta p\right)\big]^2,
\end{align}
with
\begin{equation}
\delta\alpha
=\delta x+\text{i}\delta p,\hspace{3mm} \delta x,\delta p\in\mathbb{R}.
\end{equation}
It can be concluded that $O_{\ket{\psi}}(\delta \alpha)$ becomes zero when either of the conditions is satisfied
\begin{equation}\label{eq:compassov}
	\delta x\pm\delta p=\frac{2m+1}{x_0}\pi,~m\in\mathbb{Z}.
\end{equation}
As illustrated in Fig.~\ref{fig:compass_overlap} for $O_{\ket{\psi}}(\delta \alpha)$ of cases when $x_{0}$ is increased from 4 to 12, the overlap vanishes for the displacements $|\delta \alpha|\sim x^{-1}_0$ and the arbitrary directions in the phase space.
As a result, it can be inferred that this sensitivity is proportional to $x^{-1}_0$ and that $\bar{n} =\nicefrac{x^2_0}{2}$ ties it to the number of excitations $\bar{n}$.
Therefore, in comparison to coherent states, a compass state with $\Bar{n}$ excitations has shown $\sqrt{\Bar{n}}$-enhanced sensitivity to displacements of any arbitrary directions in the phase space.\;Weak force measurements have been performed with Heisenberg-limited sensitivity using compass states~\cite{Dalvit2006,Toscano06}. In contrast, cat states have shown
the
sensitivity to displacements along the specific direction in the phase space~\cite{Naeem2021}.
It has been found that cat-state mixtures only exhibit this enhanced sensitivity for displacements along particular phase-space directions~\cite{Naeem2021,Naeem2022}.\;Hence, cat-state mixtures with sub-Planck structures in the Wigner function do not have the potential for metrology of compass states, for which the additional quantum coherence of the cat-state superposition provided by the second term in Eq.~\eqref{eq:compass_wig} plays a crucial role.

The SU(1,1) and SU(2) compass states have shown the same sensitivity to displacements as their HW counterparts~\cite{Naeem2021,Naeem2022}. The sensitivity of  the SU(1,1) compass state can be connected with the amount of the number of photons added to one of the modes of the
two-mode squeezed number state, and this sensitivity improves as the quantity of added photons increases~\cite{Naeem2022}. Similarly, the sensitivity of the SU(2) compass state improves as the angular momentum goes higher~\cite{Naeem2021}.
\;This addition of the photons increases the average photon number in the states, and it can be understood in a way similar to that for compass states of the harmonic oscillator, i.e., injecting more photons in the states improves its sensitivity. 

\section{Non-Gaussian SVS} \label{sec:Non_Gaussian}
The non-Gaussian Wigner functions of the PASVS and PSSVS illustrate the non-Gaussian nature of these states~\cite{SiKun2010,Yun2010,Dey2020,Meng12,Ma2012,TANG201586}. 
In this section we first provide a brief review of two non-Gaussian SVS, the PASVS and the PSSVS, in Sec.~\ref{subsec:PASVS} and~\ref{subsec:PSSVS}, respectively.\;These two states are heavily used in our construction of non-Gaussian states that manifest the sub-Planck structures in~Sec.~\ref{sec:sub_planck_nongaussian}. The Wigner functions of both PASVS and PSSVS are discussed in relation to the amount of photons added or subtracted in the following sections.

\subsection{PASVS}
\label{subsec:PASVS}

First, we review the Wigner function of the PASVS. The creation operator $\hat{a}^\dagger$ is repeatedly applied to SVS $\hat{S}(\pm r)\ket{0}$ to obtain a single-mode PASVS~\cite{zhang1992}:
\begin{align}\label{eq:pasvs}
\ket{\psi^{\pm}_{\text{PA}}}:=\hat{a}^{\dagger n}\hat{S}(\pm r)\ket{0}\;\text{with}\,n\in\mathbb{N}.
\end{align}
The subscript ``PA'' is the shorthand for ``PASVS'', 
and
we introduce``$\pm$'' in the squeezing operator $\hat{S}(\pm r)$ with the definition~\cite{Kok2010}
\begin{align}
\hat{S}(\pm r):=&\exp\bigg[\pm\frac{r}{2}\left(\hat{a}^{\dagger2}-\hat{a}^2\right)\bigg],
\end{align}
which allows us to preserve part of the expressions introduced in this section for later uses in Sec.~\ref{sec:sub_planck_nongaussian}.

\begin{figure*}
    \centering
    (a)~\includegraphics[width=0.25\textwidth]{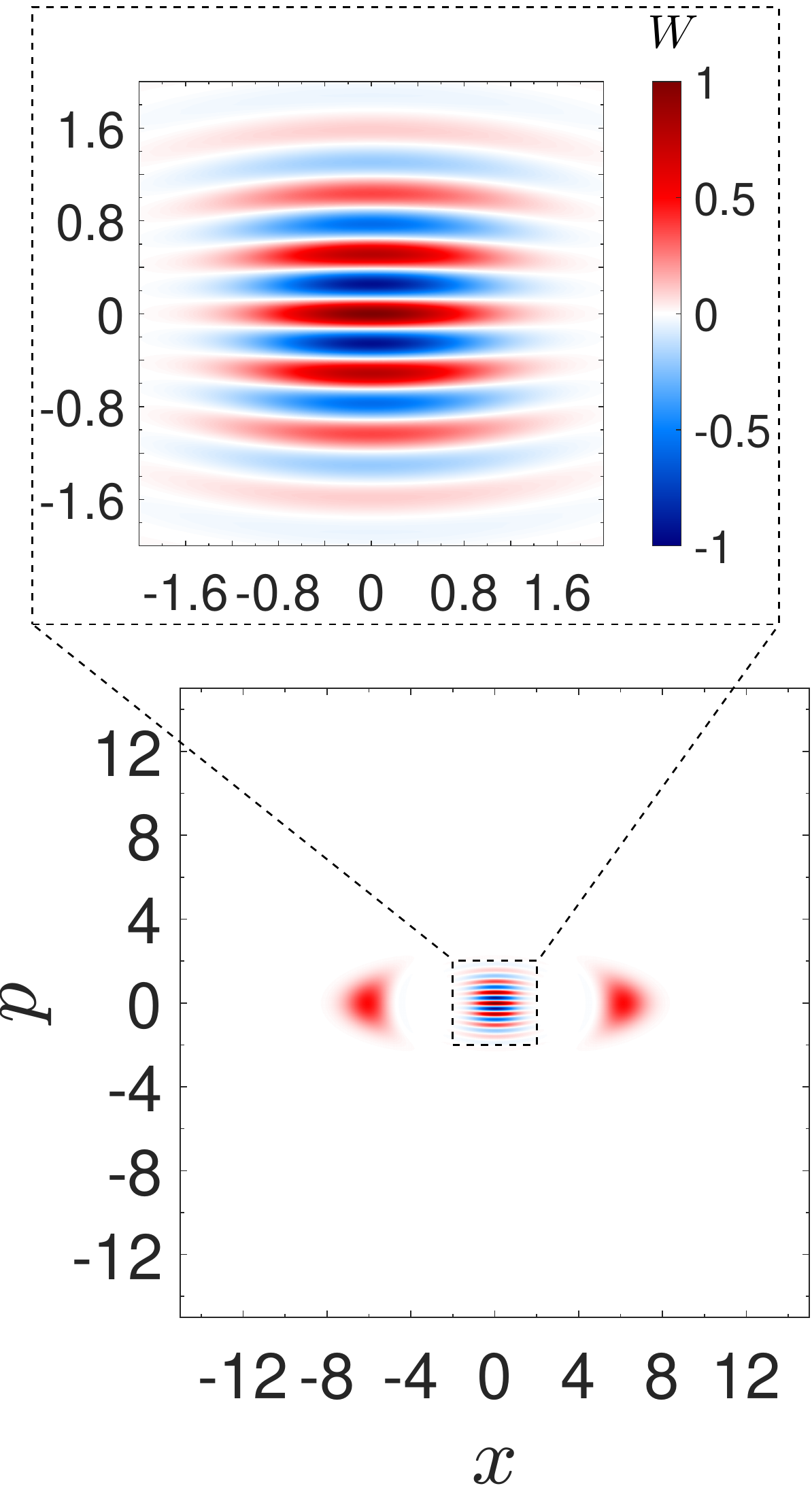}
    (b)~\includegraphics[width=0.25\textwidth]{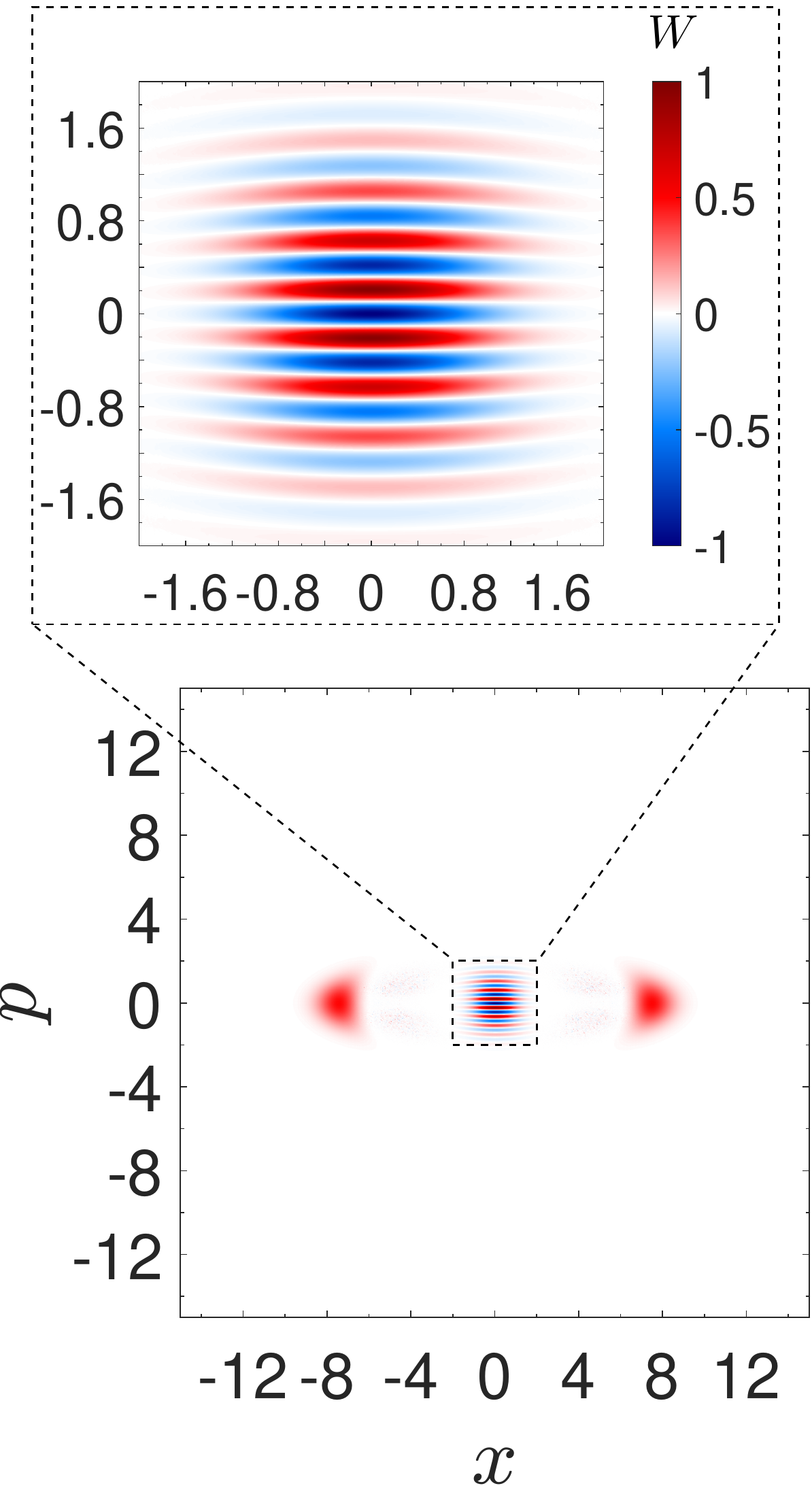}
    (c)~\includegraphics[width=0.25\textwidth]{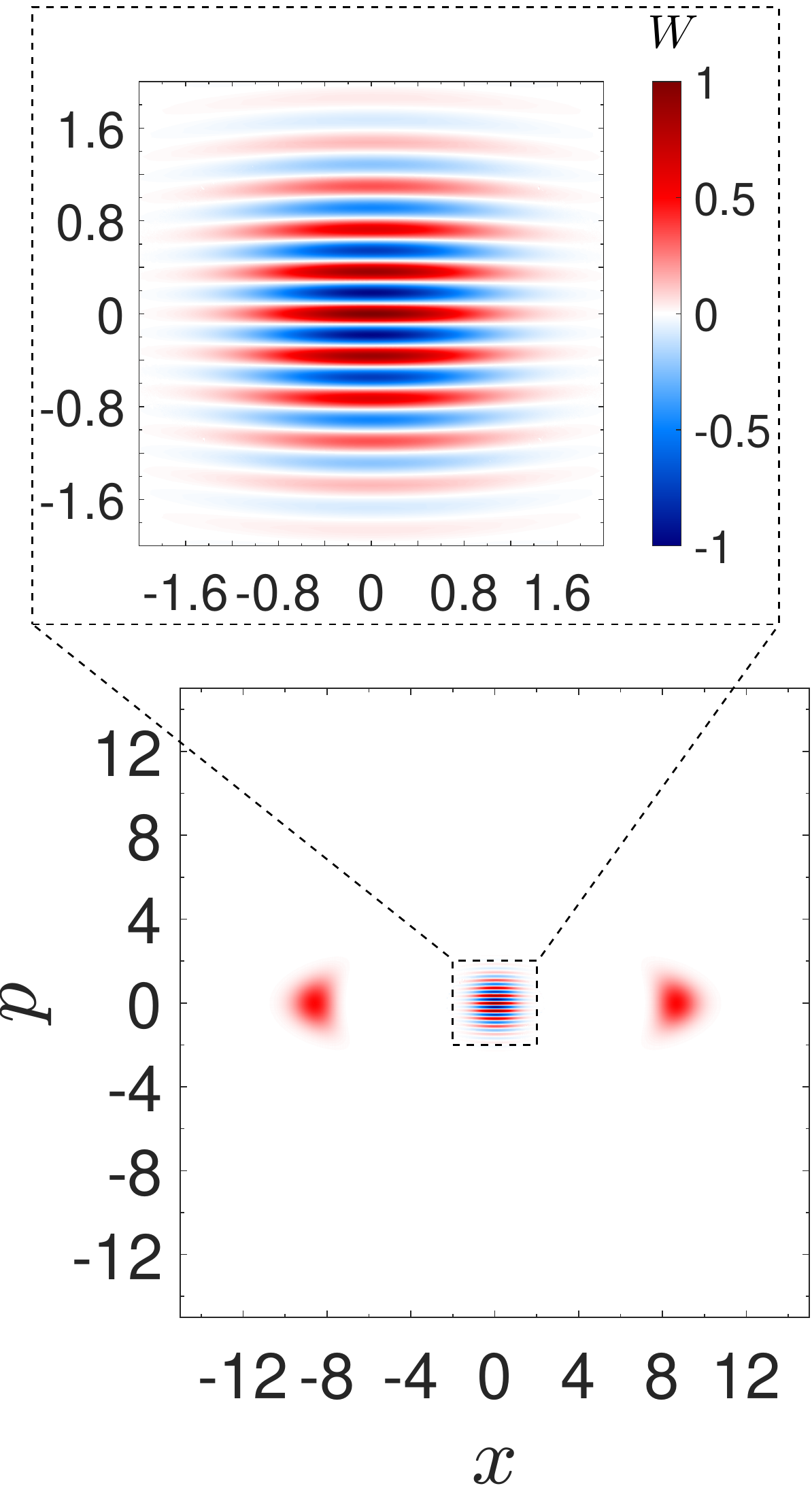}
    \caption{Wigner distribution of the PASVS with (a)~$n=10$, (b)~$n=15$. and (c)~$n=20$. In all cases $r=0.5$. Insets represent the central interference pattern of each case.}
    \label{fig:PASVS_wigner}
\end{figure*}

\begin{figure*}
    \centering
    (a)~\includegraphics[width=0.25\textwidth]{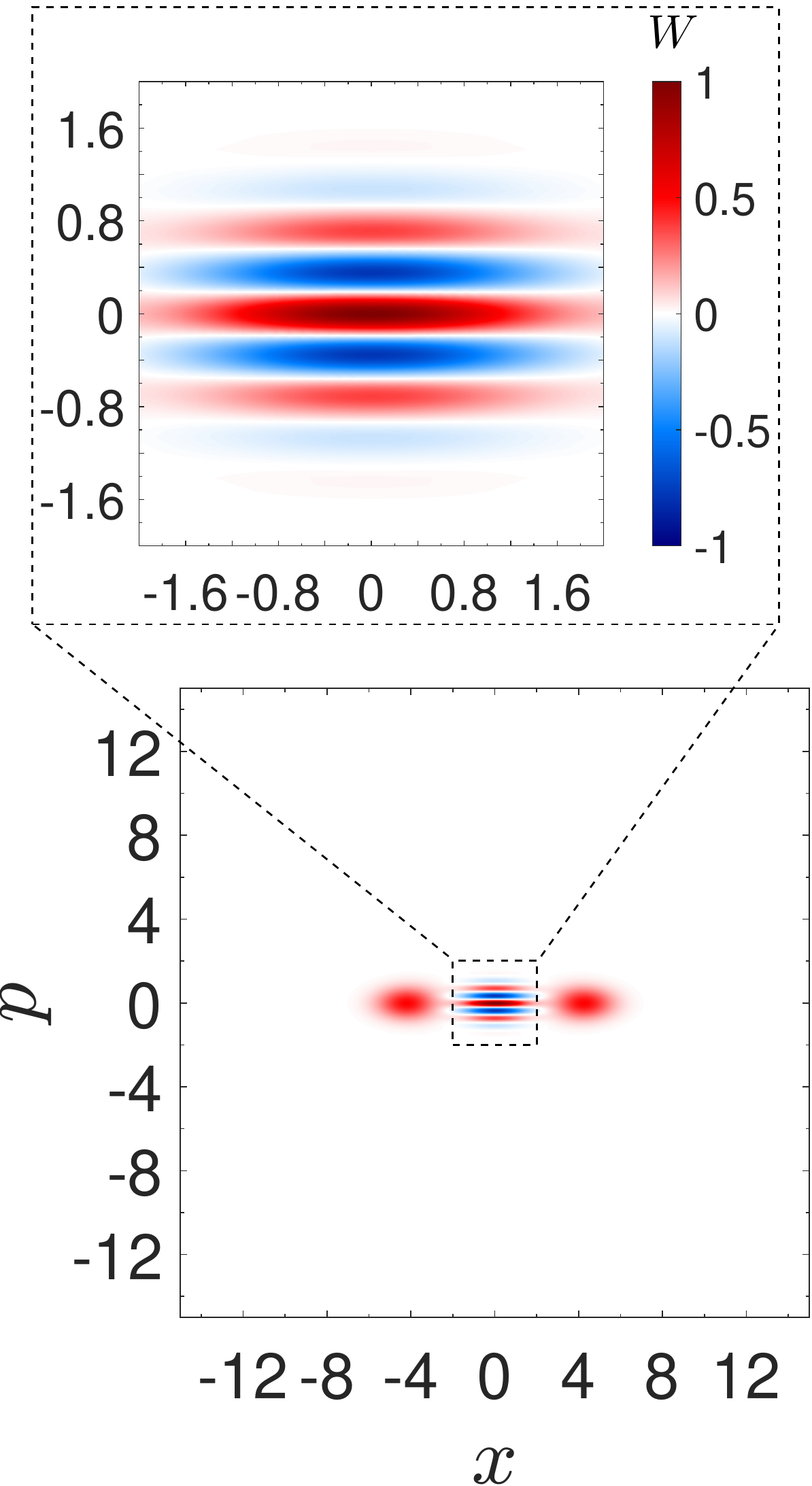}
    (b)~\includegraphics[width=0.25\textwidth]{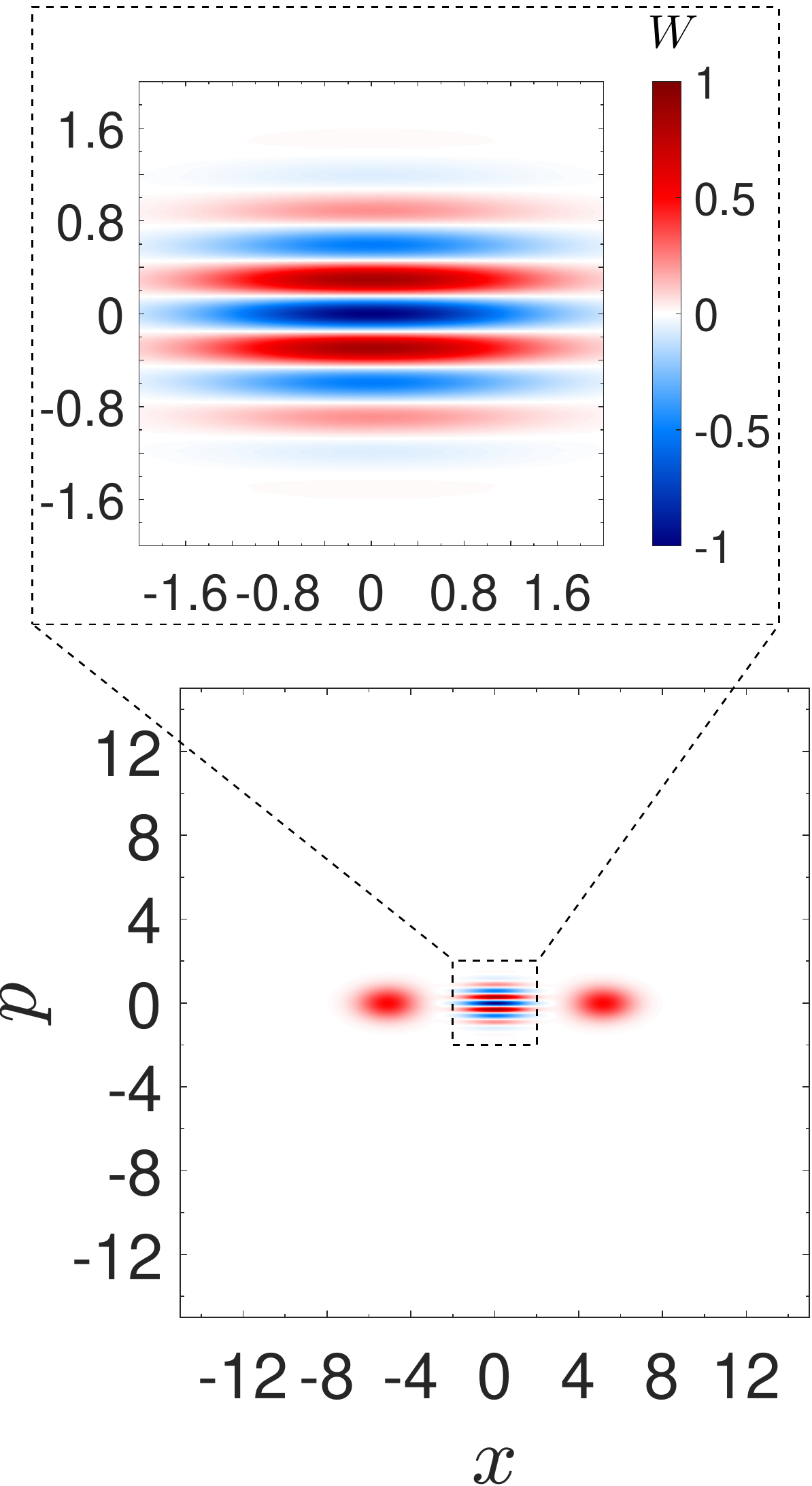}
    (c)~\includegraphics[width=0.25\textwidth]{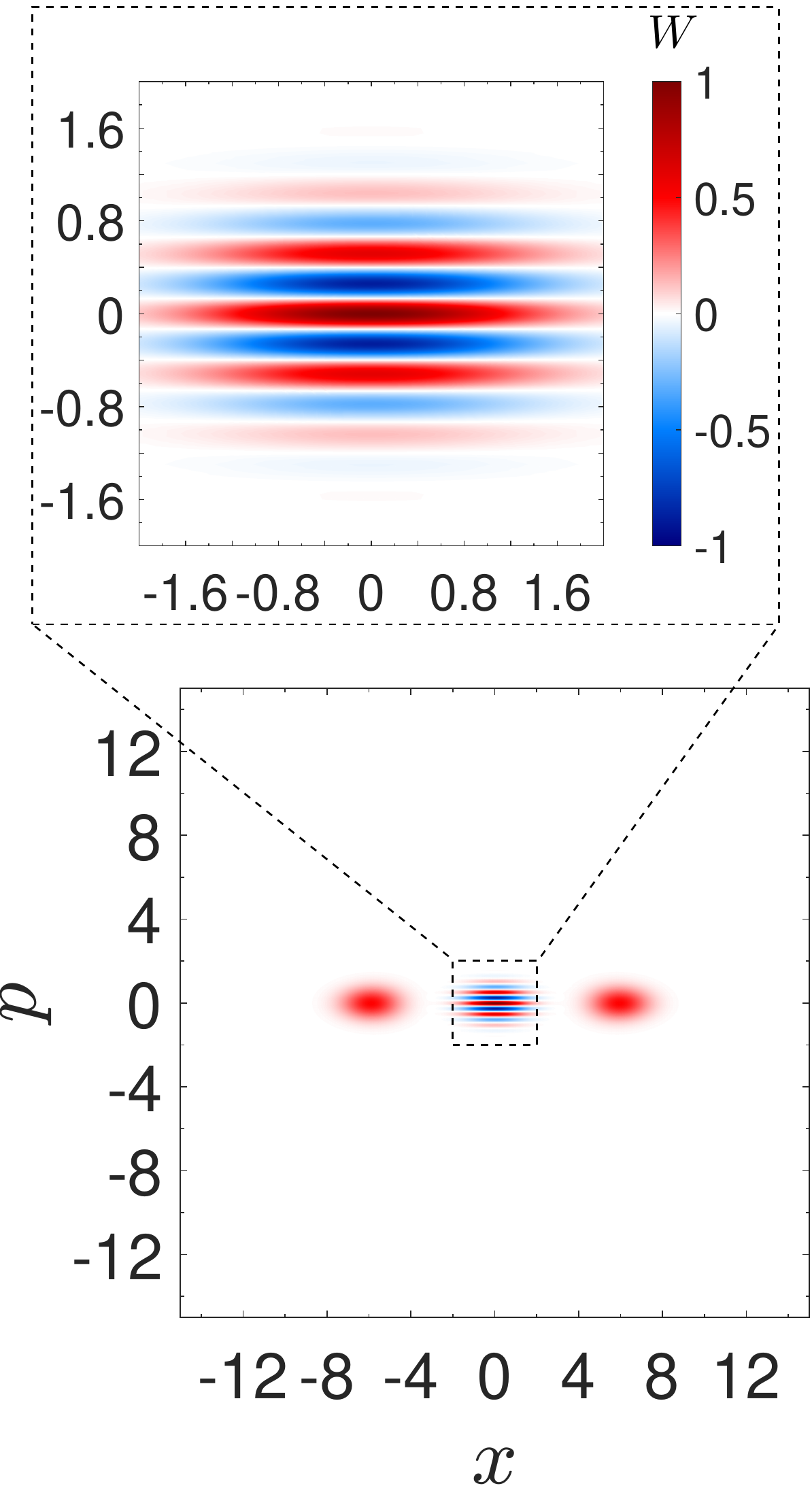}
    \caption{Wigner distribution of the PSSVS with (a)~$n=10$, (b)~$n=15$, and (c)~$n=20$. In all cases $r=0.5$. Insets represent the central interference pattern of each case.}
    \label{fig:PSSVS_wigner}
\end{figure*}

Using Eq.~\eqref{eq:wigner_general}, the Wigner function of PASVS is easily found to be~\cite{SiKun2010,Yun2010,Ma2012,TANG201586}
\begin{align}
    W_{\ket{\psi^{\pm}_{\text{PA}}}}(\bm{\zeta})
    =&\nonumber\frac{\exp\left(\chi_{\pm }\right)\left[\pm \sinh(2r)\right]^n}{\pi4^n}
    \sum_{l=0}^{n}\frac{\left(n!\right)^{2}\left[\mp 2 \coth (r)\right]^{l}}{l!\left[(n-l)!\right]^{2}}\\&\left|H_{n-l}\left[-\mathrm{i}\sqrt{\pm 2\coth(r)}\bar{\alpha}_{\pm}\right]\right|^2,
    \label{eq:expr_PASV1}
\end{align}
where $H_m$ represents the Hermite polynomial, and
\begin{align}
\chi_{\pm}:=\pm\sinh (2r)\left(\alpha^{*2}+\alpha^2\right)-2|\alpha|^2\cosh (2r),
\end{align}
with
\begin{align}
\bar{\alpha}_{\pm}:=\alpha \cosh (r)\mp \alpha^* \sinh(r).
\end{align}
The non-Gaussian shape of the Wigner function $W_{\ket{\psi^+_{\text{PA}}}}(\bm{\zeta})$, which is shown in Fig.~\ref{fig:PASVS_wigner} for the cases when $n$ is chosen as 10, 15, and 20, indicates that PASVS is a non-Gaussian state. We can clearly see the interference pattern that emerges
in the form of an oscillating pattern in the $p$ direction in the phase space. As the number of photons $n$ rises, this pattern gets more pronounced (the frequency of the oscillations is increased).  Moreover, the existence of these negative peaks in the Wigner function shows that the PASVS is a nonclassical state as well. Another indication of the nonclassicality of this state is the squeezing effect in one of the quadratures, which is visible in the plots. Note that the PASVS is the Gaussian SVS when $n=0$.

\subsection{PSSVS}
\label{subsec:PSSVS}

\begin{figure*}
\centering
(a)~\includegraphics[width=0.25\textwidth]{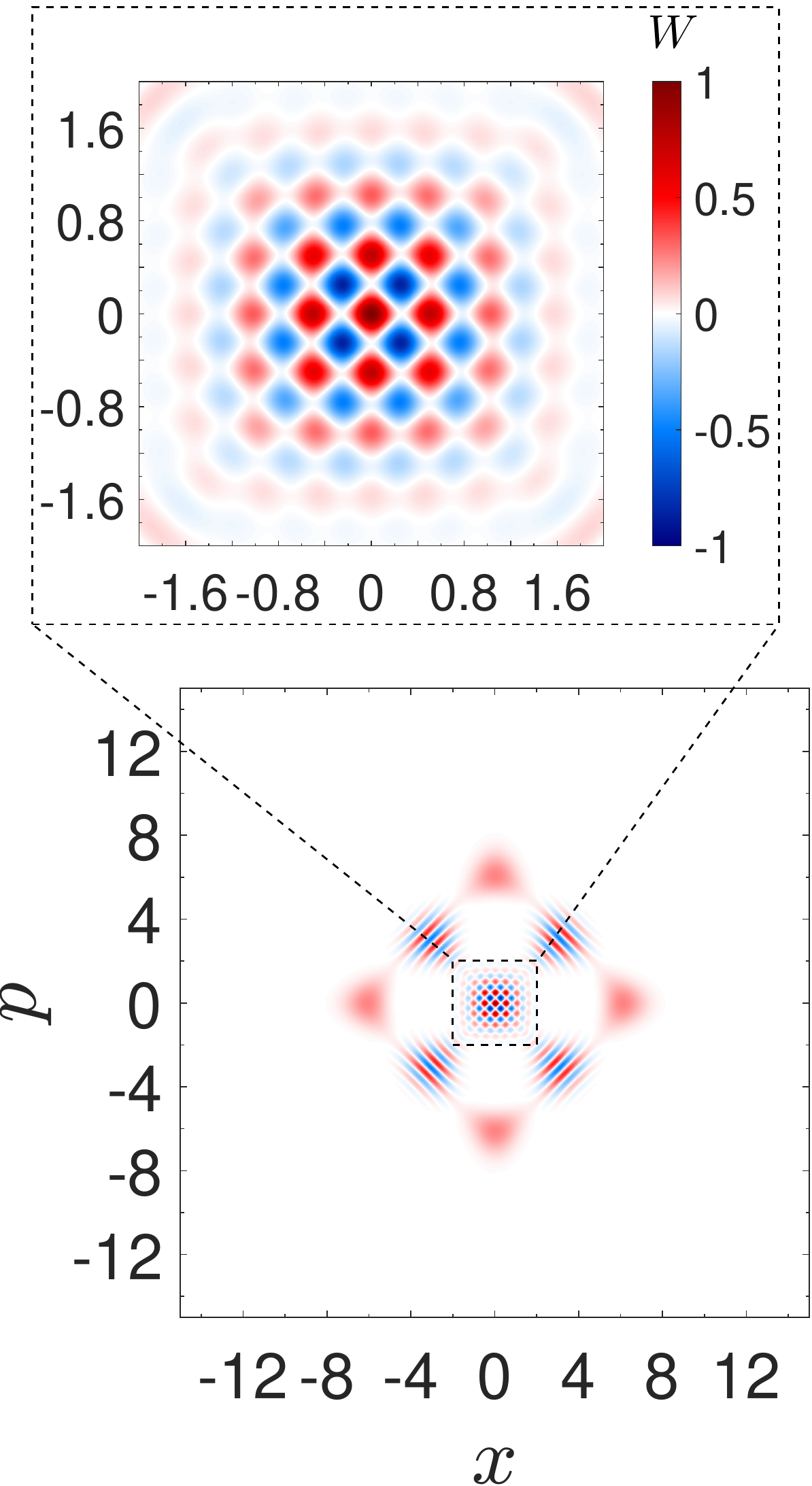}
(b)~\includegraphics[width=0.25\textwidth]{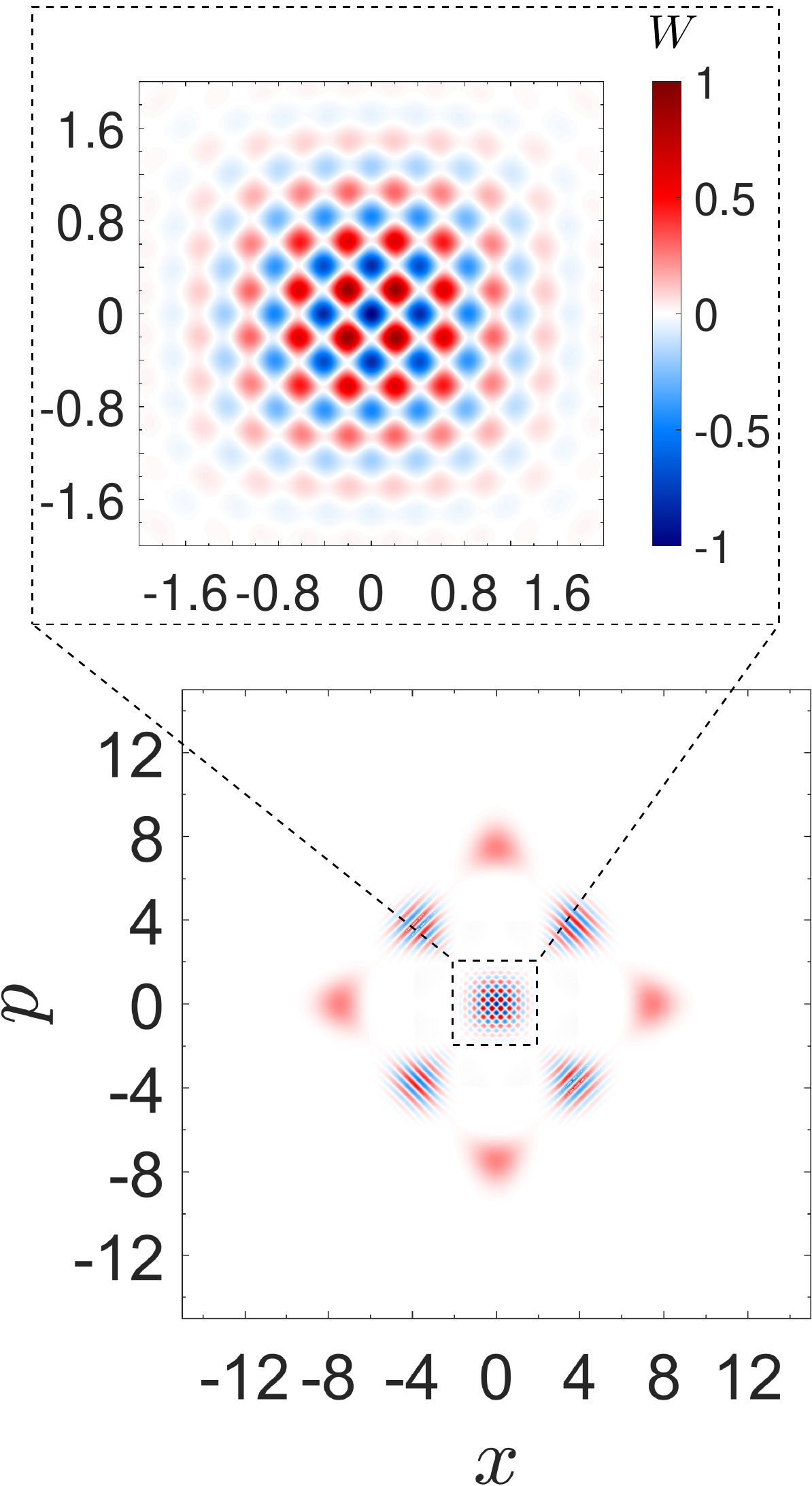}
(c)~\includegraphics[width=0.25\textwidth]{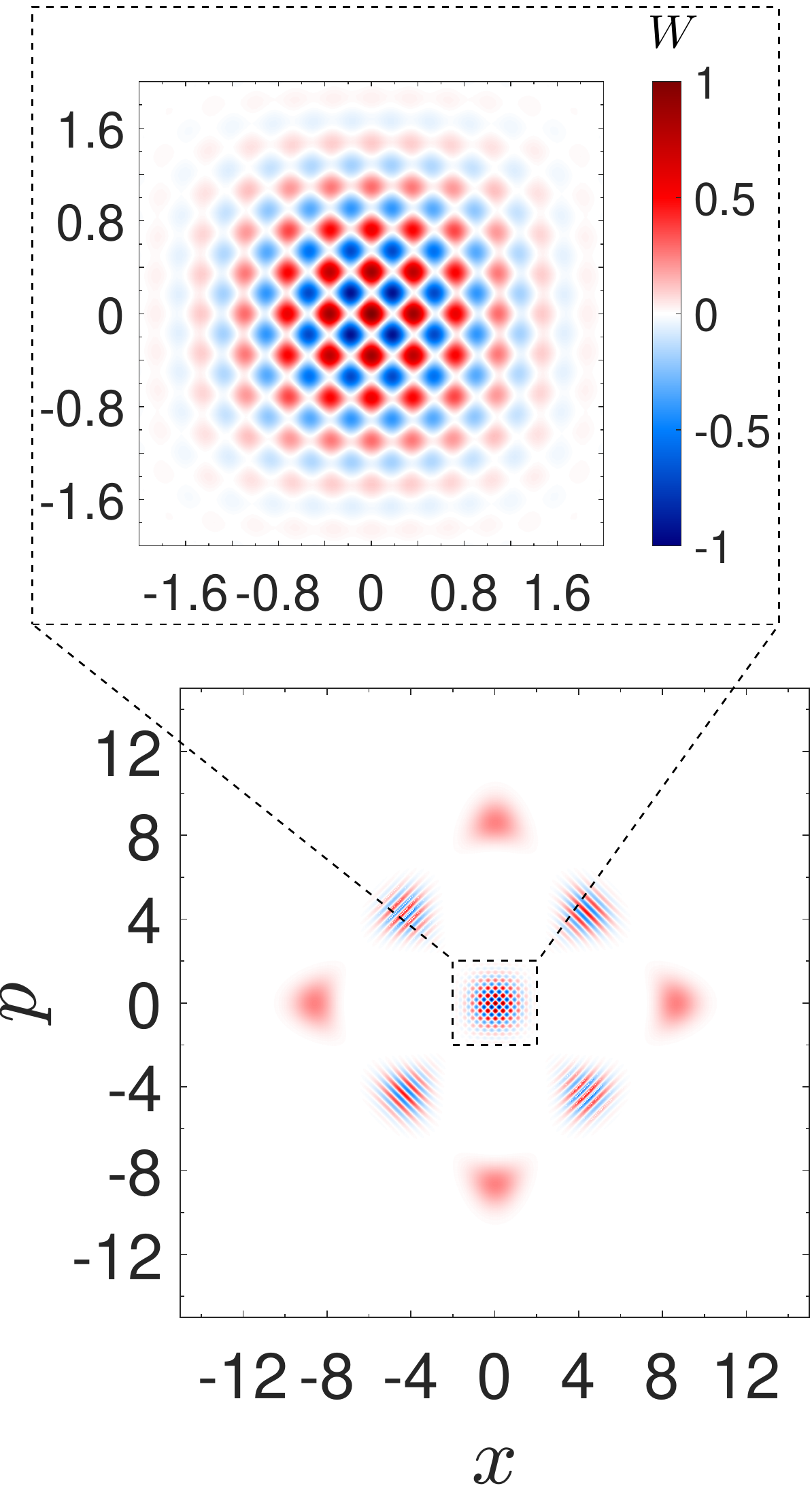}
\caption{Wigner distribution of the pure SPASVS with (a)~$n=10$, (b)~$n=15$, and (c)~$n=20$. In all cases $r=0.5$ and $c_1=\nicefrac{1}{\sqrt{2}}$. Insets represent the central interference pattern of each case.}
    \label{fig:SPASVS_pure_wigner}
\end{figure*}

We now review the Wigner function of the PSSVS. The PSSVS~\cite{Dey2020,Meng12} is obtained by repeatedly applying 
the annihilation operator $\hat{a}$ to the SVS, as
\begin{align}\label{eq:spssv}
\ket{\psi^{{\pm}}_{\text{PS}}}:=\hat{a}^{n}\hat{S}(\pm r)\ket{0},
\end{align}
where the subscript ``PS'' is the shorthand for ``PSSVS''.
The Wigner function of this state is also in a
non-Gaussian form and is written as
\begin{align}
    W_{\ket{\psi^{\pm}_{\text{PS}}}}(\bm{\zeta})
    =&\nonumber\frac{\exp\left(\chi_{\pm }\right)[\pm \sinh(2r)]^n}{\pi4^n}
    \sum_{l=0}^{n}\frac{\left(n!\right)^{2}[\mp 2 \tanh (r)]^{l}}{l!\left[(n-l)!\right]^{2}}\\&\left|H_{n-l}\left[-\mathrm{i}\sqrt{\pm 2\tanh(r)}\bar{\alpha}_{\pm}\right]\right|^2.
    \label{eq:expr_PSSV1}
\end{align}
We plot the Wigner function $W_{\ket{\psi^+_{\text{PS}}}}(\bm{\zeta})$ in Fig.~\ref{fig:PSSVS_wigner},
with $r$ being $0.5$ and
$n$ being 10, 15, and 20.
This Wigner function
exhibits the interference pattern around the origin of the phase space and oscillates along the $p$ direction in the phase space. As $n$ grows, the frequency of this oscillating pattern increases.
The simplest case $n=0$ 
of the PSSVS corresponds to the Gaussian SVS.\;Another indicator of the nonclassical nature of this state is the squeezing effect in one of the quadratures, which is indicative of the nonclassicality of state.

In summary, for nonzero values of $n$, the Wigner function of PASVS and PSSVS maintains non-Gaussianity. It is interesting to note that both PASVS and PSSVS exhibit similar phase-space features as catlike states. The addition or subtraction of photons from the Gaussian SVS has been employed both theoretically~\cite{Gen2,Dakna1997,Takase2021,Podoshvedov2023} and experimentally~\cite{Neergaard2006,Alexei2006} to produce catlike states.
Both PASVS and PSSVS have been found very useful for the quantum metrology~\cite{Lang2013,WANG2019102,Richard2014}. When photons are added to or subtracted from the Gaussian SVS, the average photon number of the resulting state grows~\cite{WANG2019102,Richard2014}.
It has been shown that for the same number of photons applied on the Gaussian SVS, the subsequent PASVS has a higher average photon number than the PSSVS~\cite{WANG2019102,Richard2014}.\;This means that the PASVS has a better potential for metrology than the PSSVS~\cite{WANG2019102}.

\section{Superposition of non-Gaussian SVS}\label{sec:sub_planck_nongaussian}

In this section we introduce the quantum states associated to non-Gaussian SVSs and present their phase-space analysis by using the Wigner function~\cite{Wig32,Gerry05book,Sch01,Russel2021,leonhardt1997measuring}. 

The superpositions involving the Gaussian SVSs have been analyzed previously~\cite{Barry1989,Happ2018}. Theoretically, a nonlinear harmonic oscillator can be used to create some specified superpositions of two SVSs. The superposition of two Gaussian SVSs with opposite phases is given by

\begin{align}\label{eq:ssv}
\ket{\psi_{\text{SSV}}}:=c_1 \hat{S}(r)\ket{0}+c_2 \hat{S}(-r)\ket{0},
\end{align}
with ``SSV'' in the subscript for the ``superposition of SVSs'' and the probability amplitudes $c_1$ and $c_2$ fulfilling
$|c_1|^2+|c_2|^2=1$.
The Wigner function corresponding to this state exhibits non-Gaussian and nonclassical properties~\cite{Barry1989,Happ2018}.

\begin{figure*}
(a)~\includegraphics[width=0.25\textwidth]{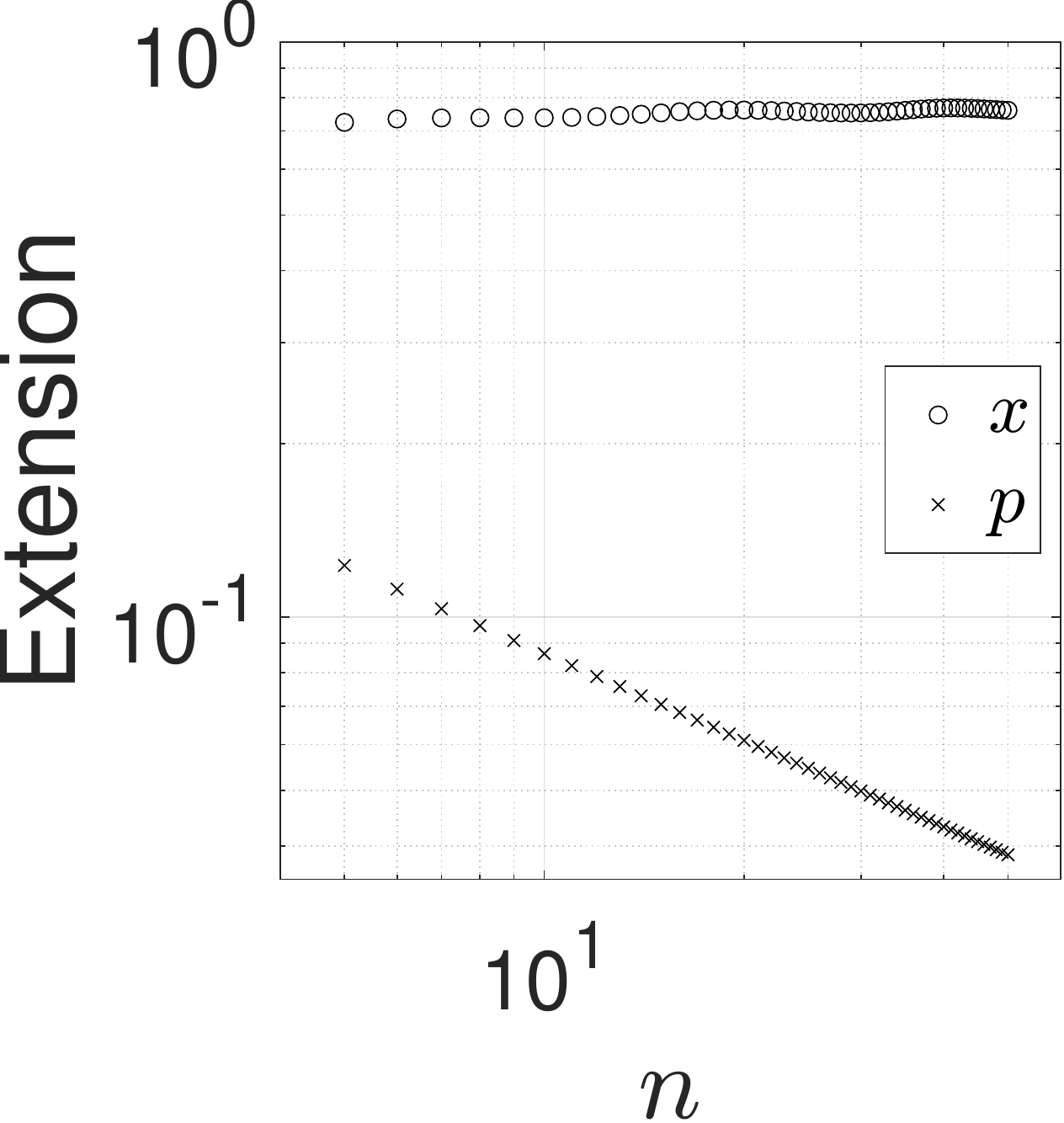}
(b)~\includegraphics[width=0.25\textwidth]{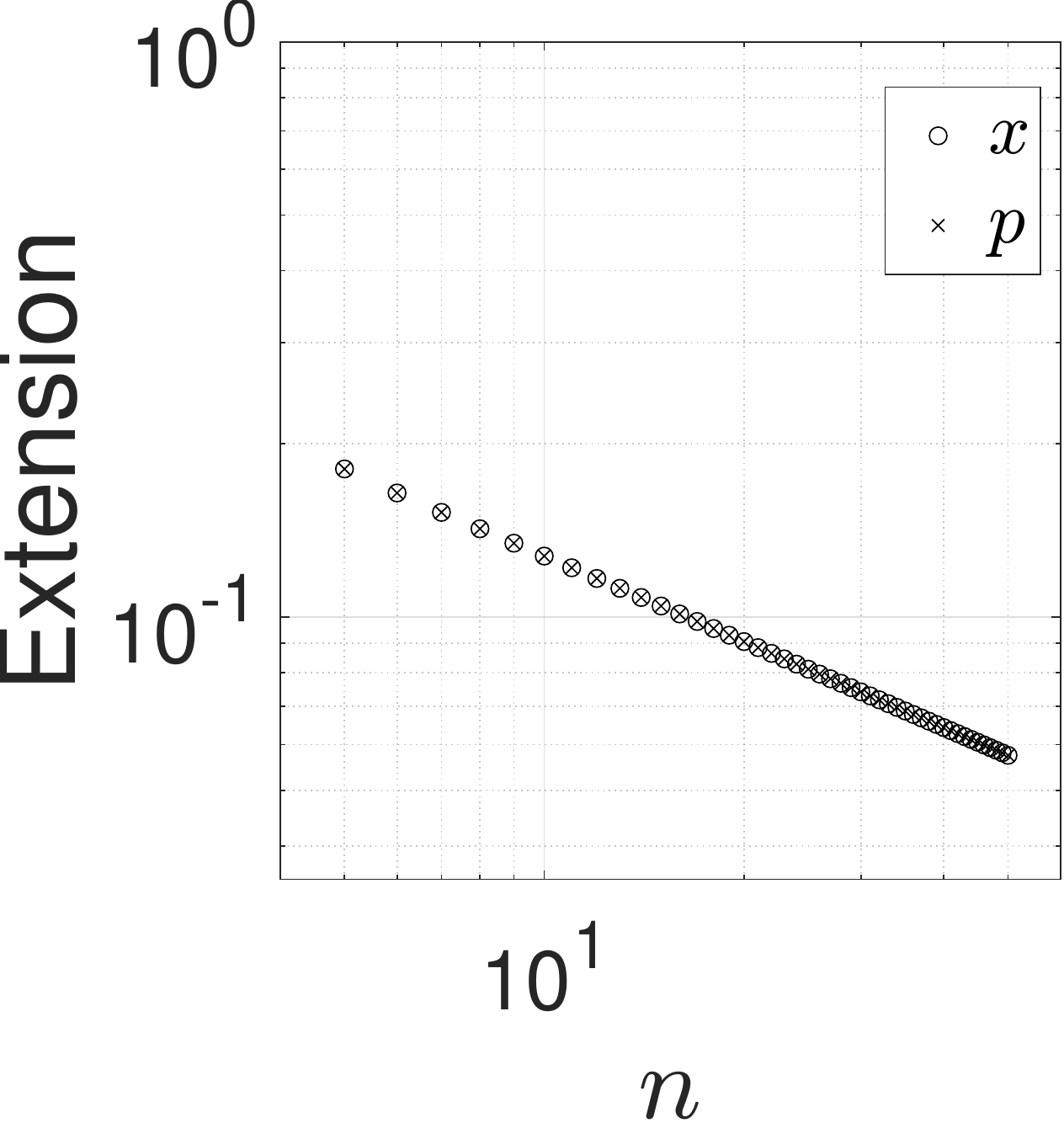}
(c)~\includegraphics[width=0.25\textwidth]{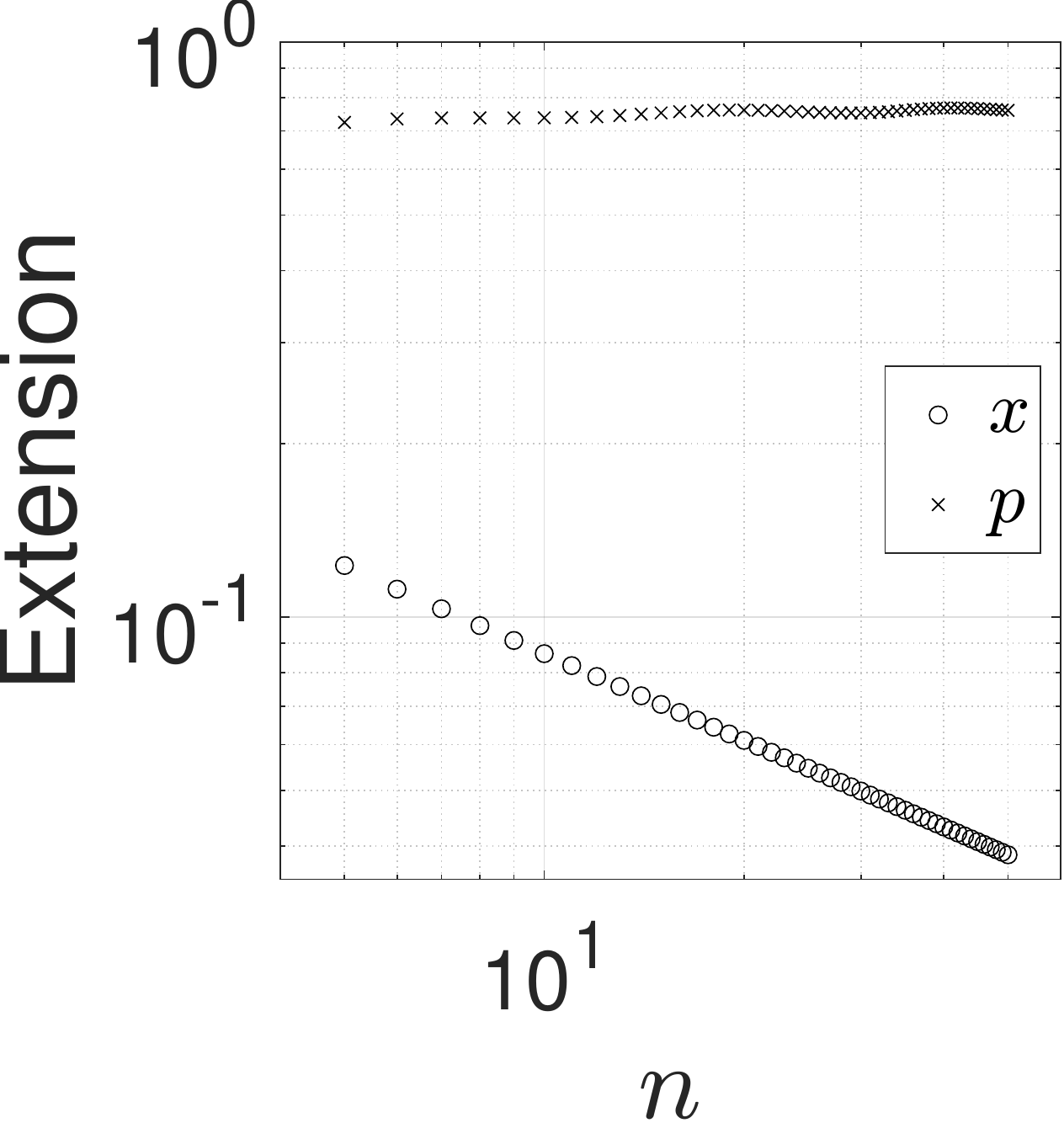}
\caption{Extension of the central phase-space structure vs the photon number $n$ of the SPASVS state with $n$ chosen from 5 to 50 for (a)~$c_{1}=\nicefrac{1}{10}$, (b)~$c_{1}=\nicefrac{1}{\sqrt{2}}$, and (c)~$c_{1}=\nicefrac{3\sqrt{11}}{10}$.}
\label{fig:SPASVS_area_versus_n}
\end{figure*}

\begin{figure*}
\centering
(a)~\includegraphics[width=0.25\textwidth]{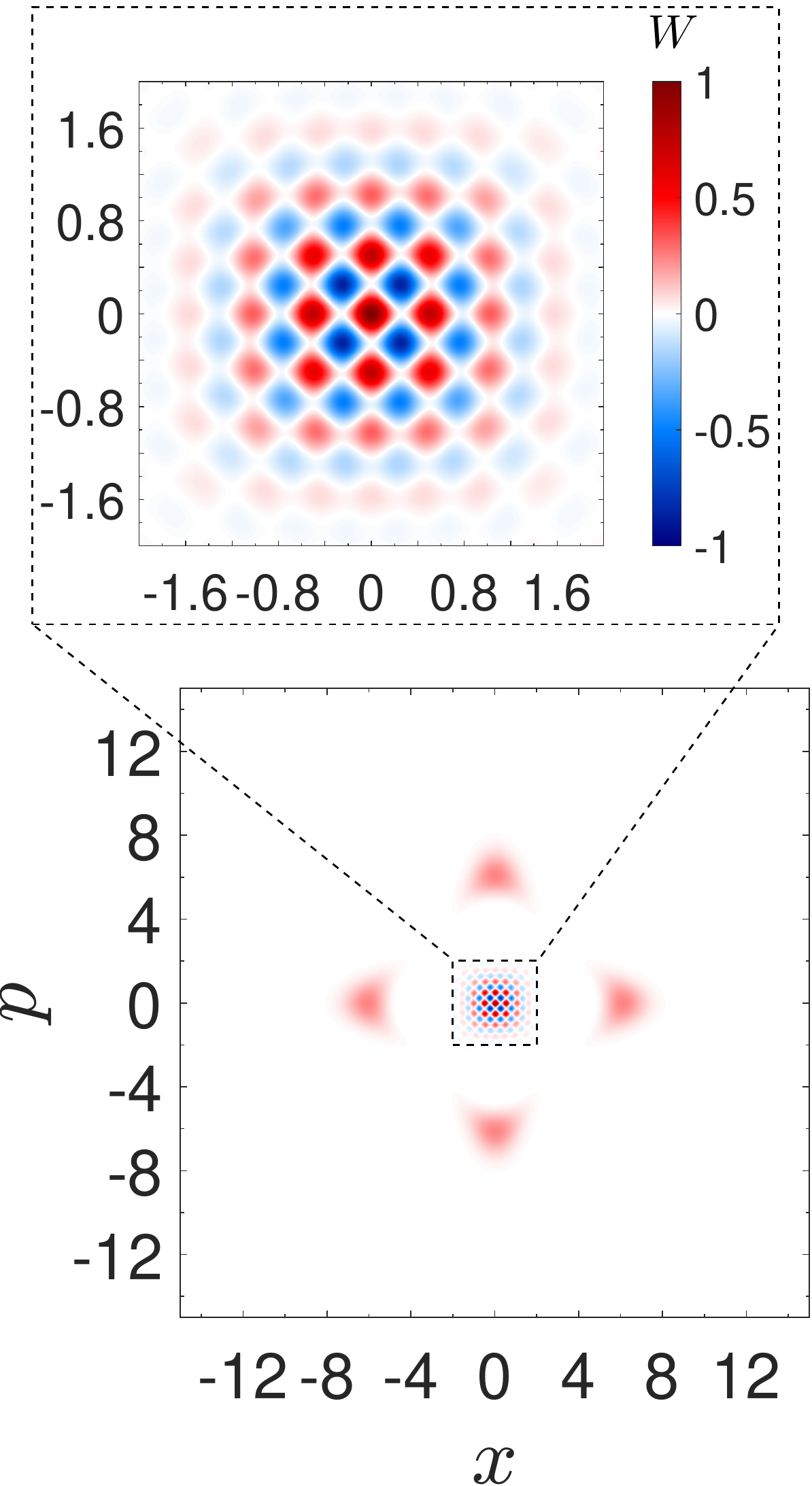}
(b)~\includegraphics[width=0.25\textwidth]{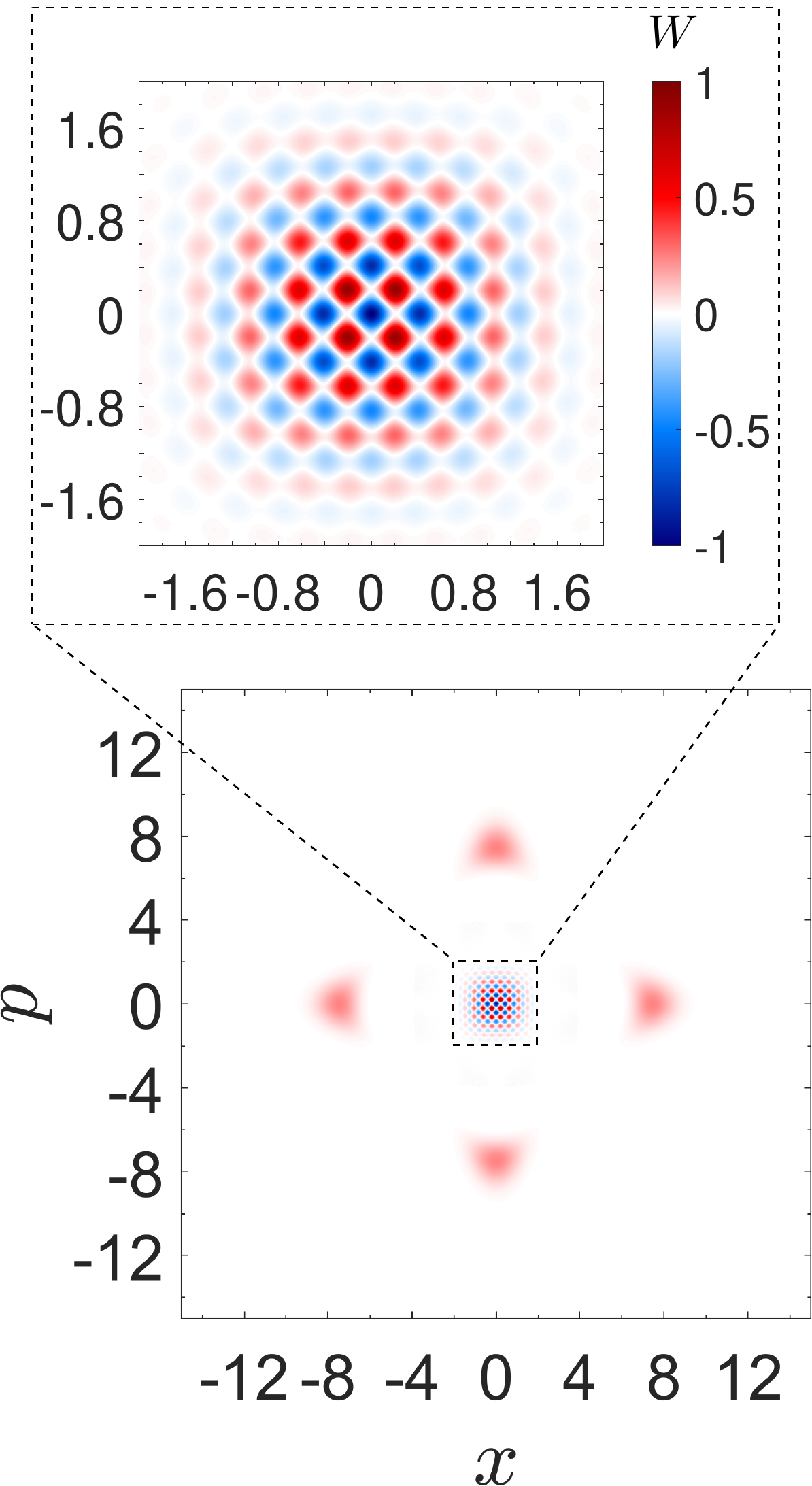}
(c)~\includegraphics[width=0.25\textwidth]{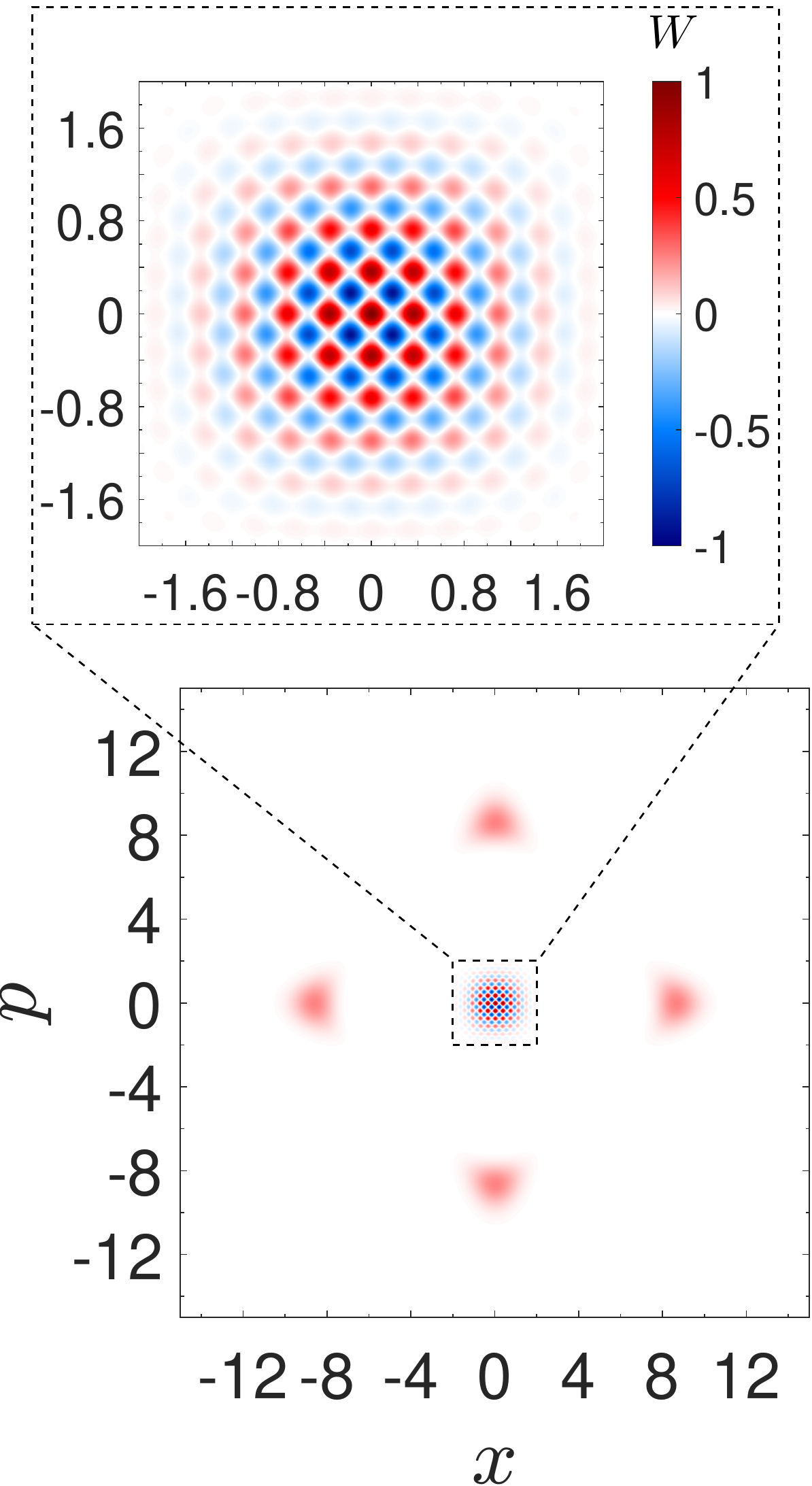}
\caption{Wigner distribution of the mixed-state SPASVS with (a)~$n=10$, (b)~$n=15$, and (c)~$n=20$. In all cases $r=0.5$ and $c_1=\nicefrac{1}{\sqrt{2}}$. Insets represent the central interference pattern of each case.}
\label{fig:SPASVS_mixed_wigner}
\end{figure*}

Now we introduce the superposition related to non-Gaussian SVSs. The addition of $n$ photons to the superposed state (\ref{eq:ssv}) leads to the superposition of two photon-added squeezed-vacuum states (SPASVS), that is,
\begin{align}\label{eq:spa}
\ket{\psi_{\text{SPA}}}:=\hat{a}^{\dagger n}\ket{\psi_{\text{SSV}}}=c_1\ket{\psi^{+}_{\text{PA}}}+c_2\ket{\psi^{-}_{\text{PA}}},
\end{align}
with the subscript ``SPA''
a shorthand
for ``SPASVS''. 
Similarly, the subtraction of $n$ photons from the superposition (\ref{eq:ssv}) results in the superposition of two photon-subtracted squeezed-vacuum states (SPSVS) of the following form:
\begin{align}\label{eq:sps}
\ket{\psi_{\text{SPS}}}:=\hat{a}^{ n}\ket{\psi_{\text{SSV}}}=c_1\ket{\psi^+_{\text{PS}}}+c_2\ket{\psi^-_{\text{PS}}},
\end{align}
where the subscript ``SPS'' is the short form of ``SPSSVS''. Here, we concentrate on the sub-Planck structures in the Wigner function of these states and their sensitivity to displacements.

\begin{figure*}
\centering
(a)~\includegraphics[width=0.25\textwidth]{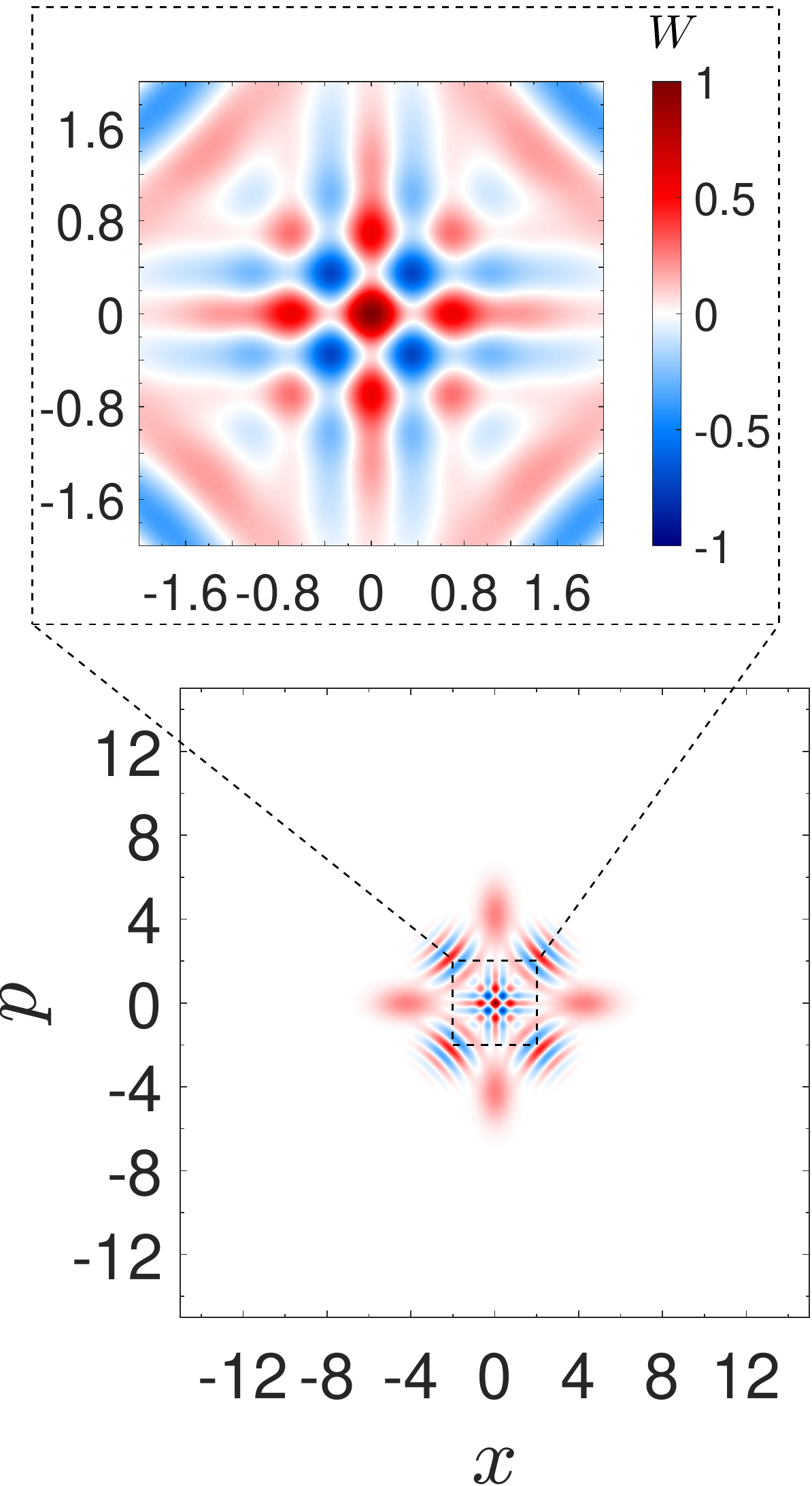}
(b)~\includegraphics[width=0.25\textwidth]{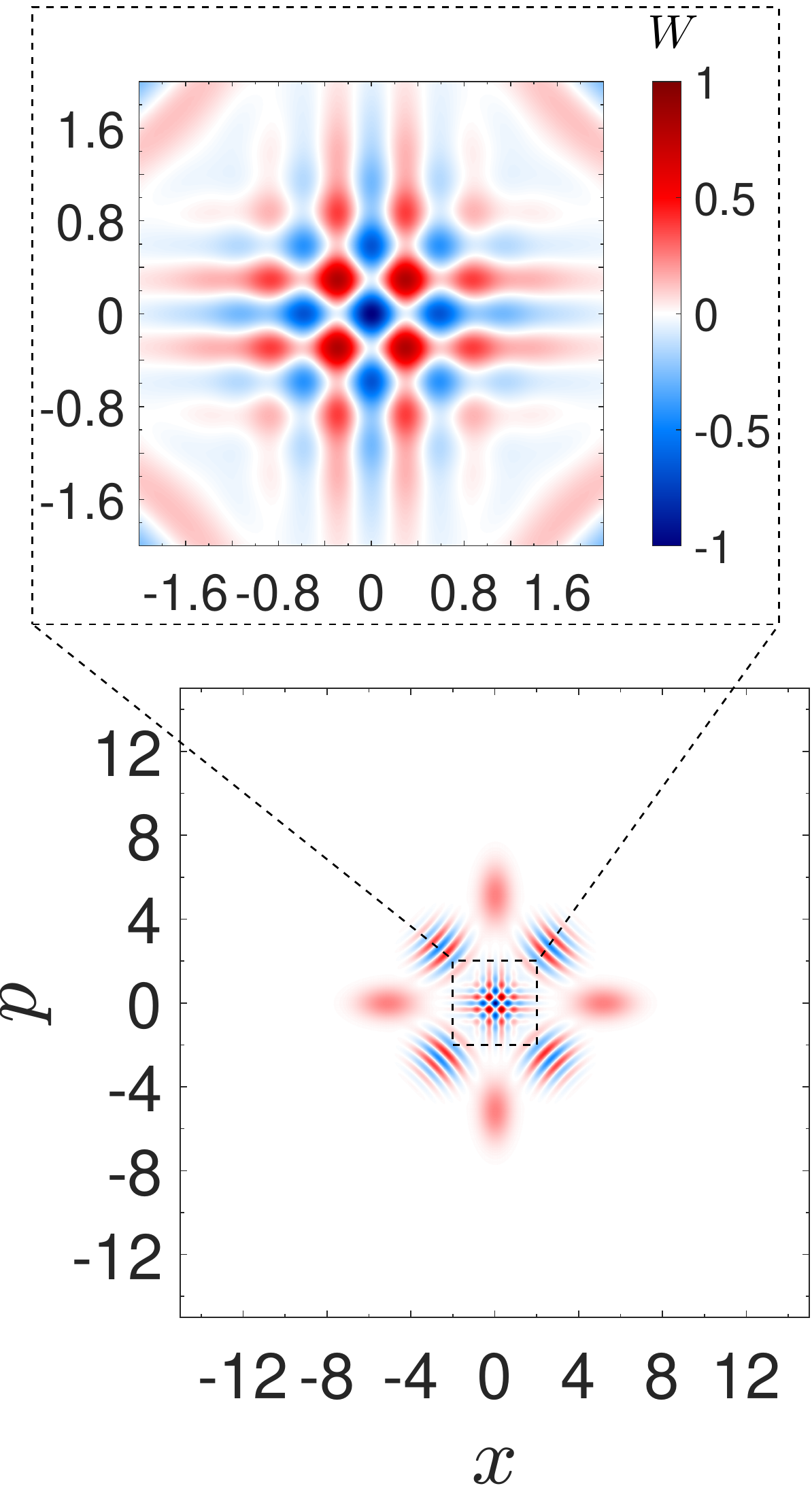}
(c)~\includegraphics[width=0.25\textwidth]{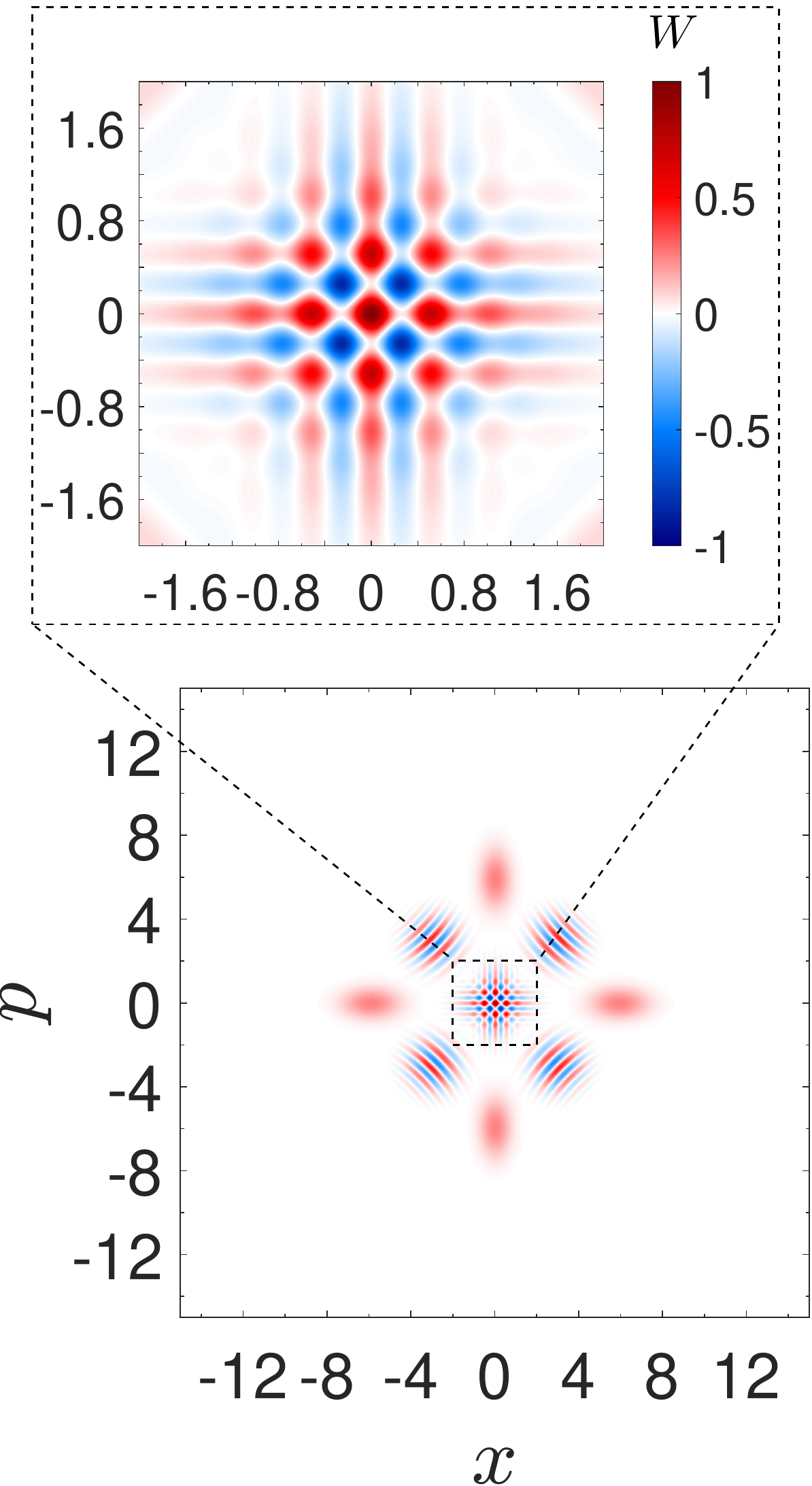}
\caption{Wigner distribution of the pure SPSSVS with (a)~$n=10$, (b)~$n=15$, and (c)~$n=20$. In all cases $r=0.5$ and $c_1=\nicefrac{1}{\sqrt{2}}$. Insets represent the central interference pattern of each case.}
\label{fig:SPSSVS_pure_wigner}
\end{figure*}

\begin{figure*}
(a)~\includegraphics[width=0.25\textwidth]{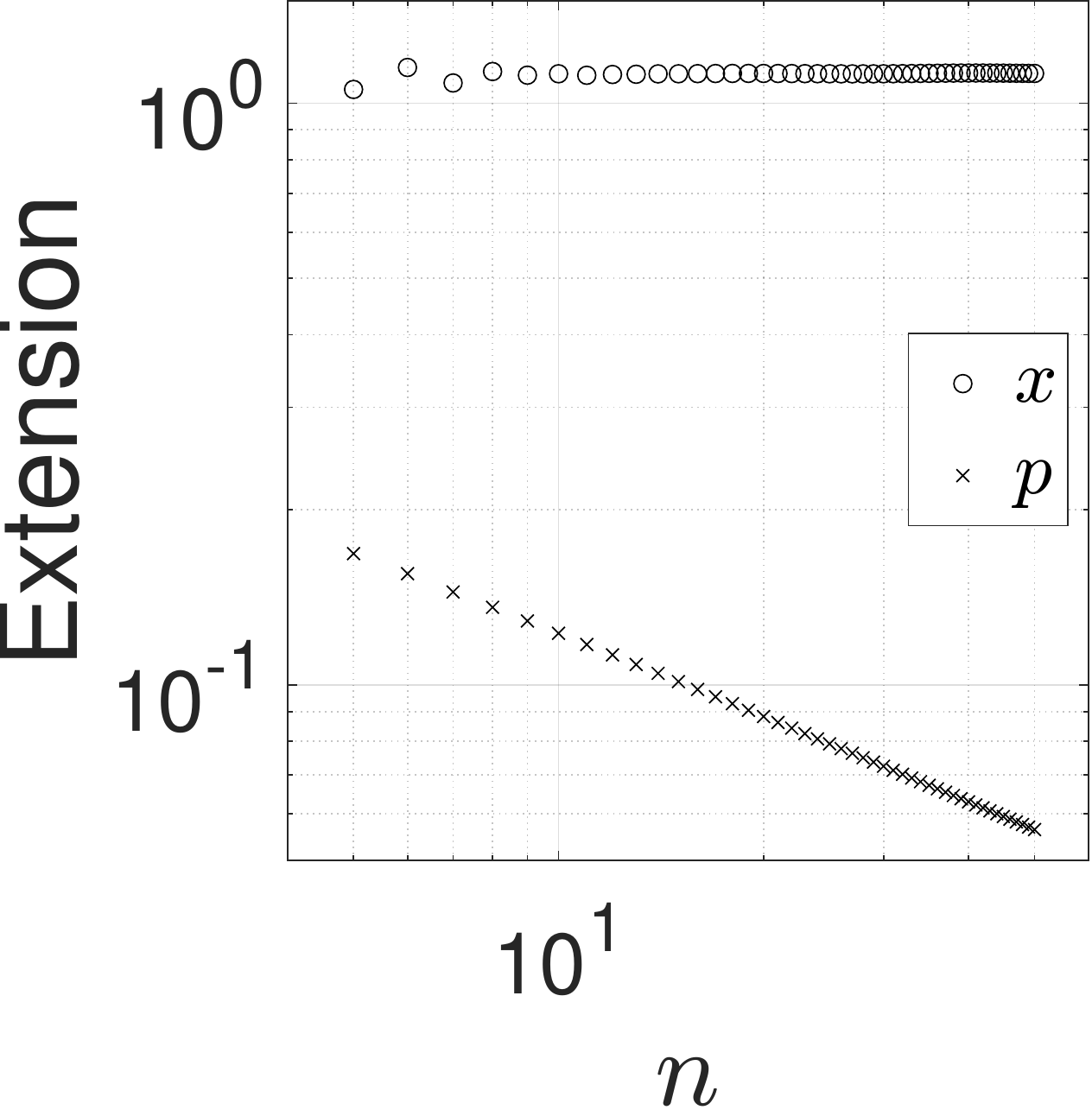}
(b)~\includegraphics[width=0.25\textwidth]{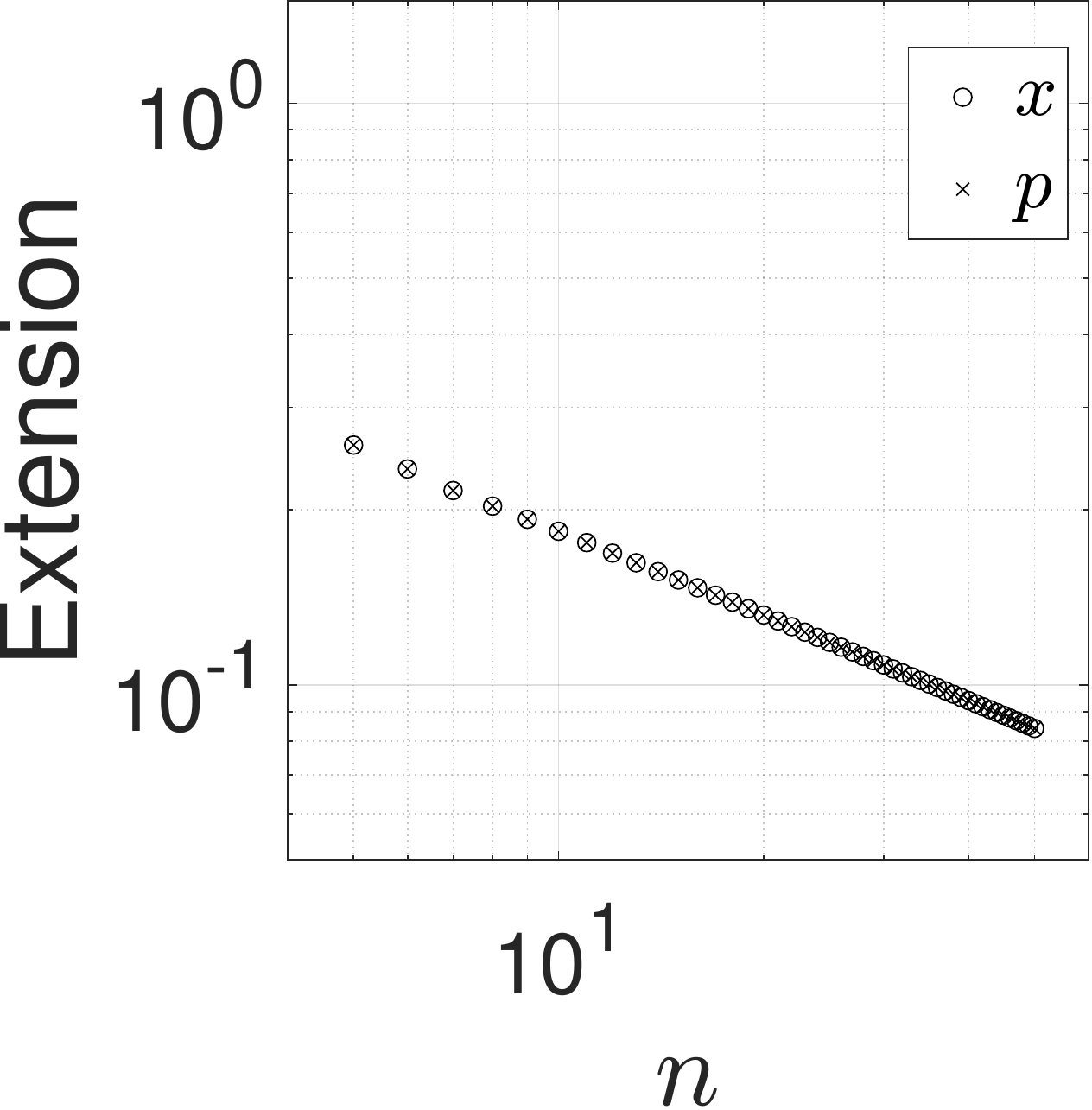}
(c)~\includegraphics[width=0.25\textwidth]{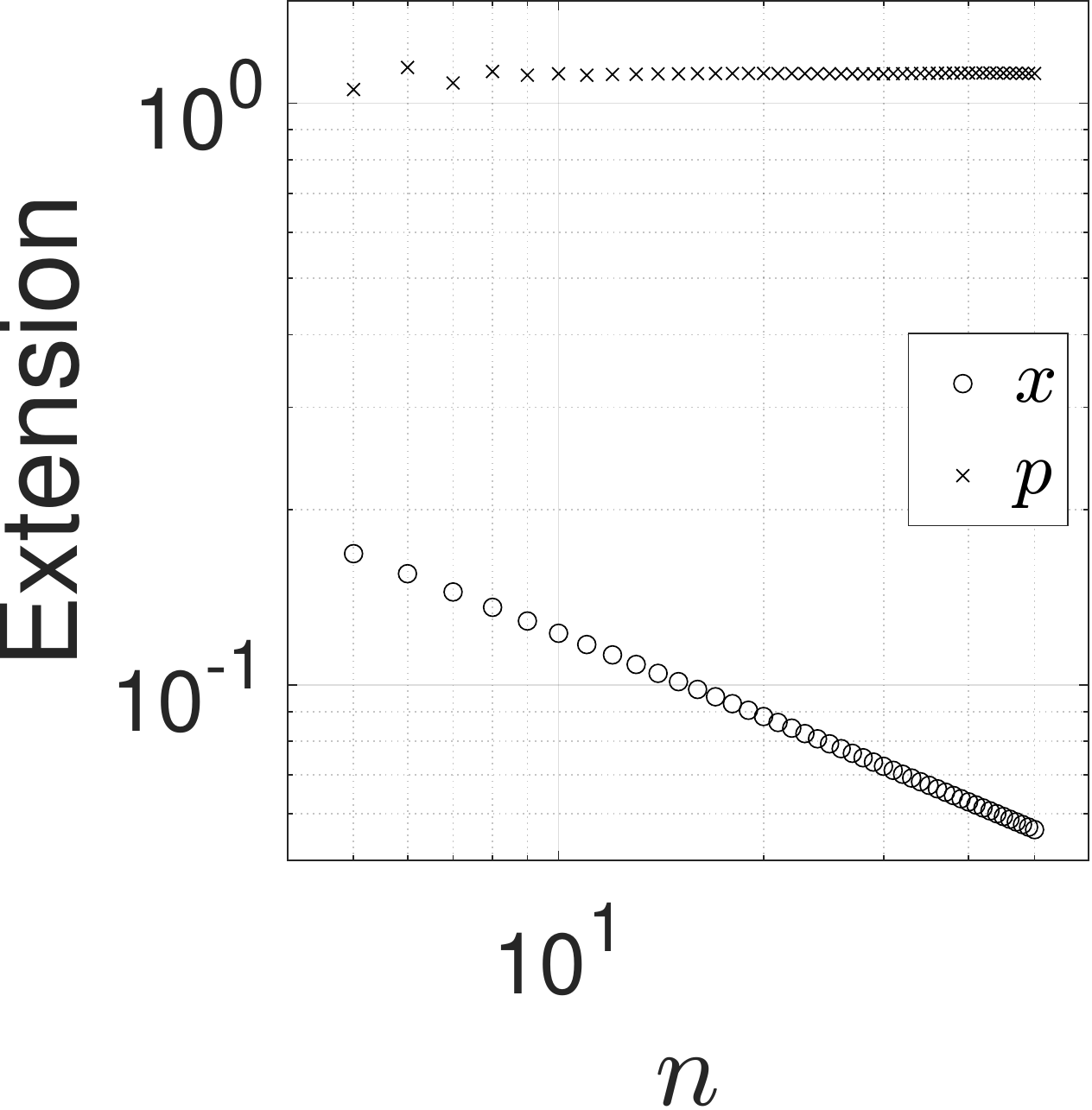}
\caption{Extension of the central phase-space structure versus the photon number $n$ of the SPSSVS state with $n$ chosen from 5 to 50 for (a)~$c_{1}=\nicefrac{1}{10}$, (b)~$c_{1}=\nicefrac{1}{\sqrt{2}}$, and
(c)~$c_{1}=\nicefrac{3\sqrt{11}}{10}$.}
\label{fig:SPSSVS_area_versus_n}
\end{figure*}

\begin{figure*}
\centering
(a)~\includegraphics[width=0.25\textwidth]{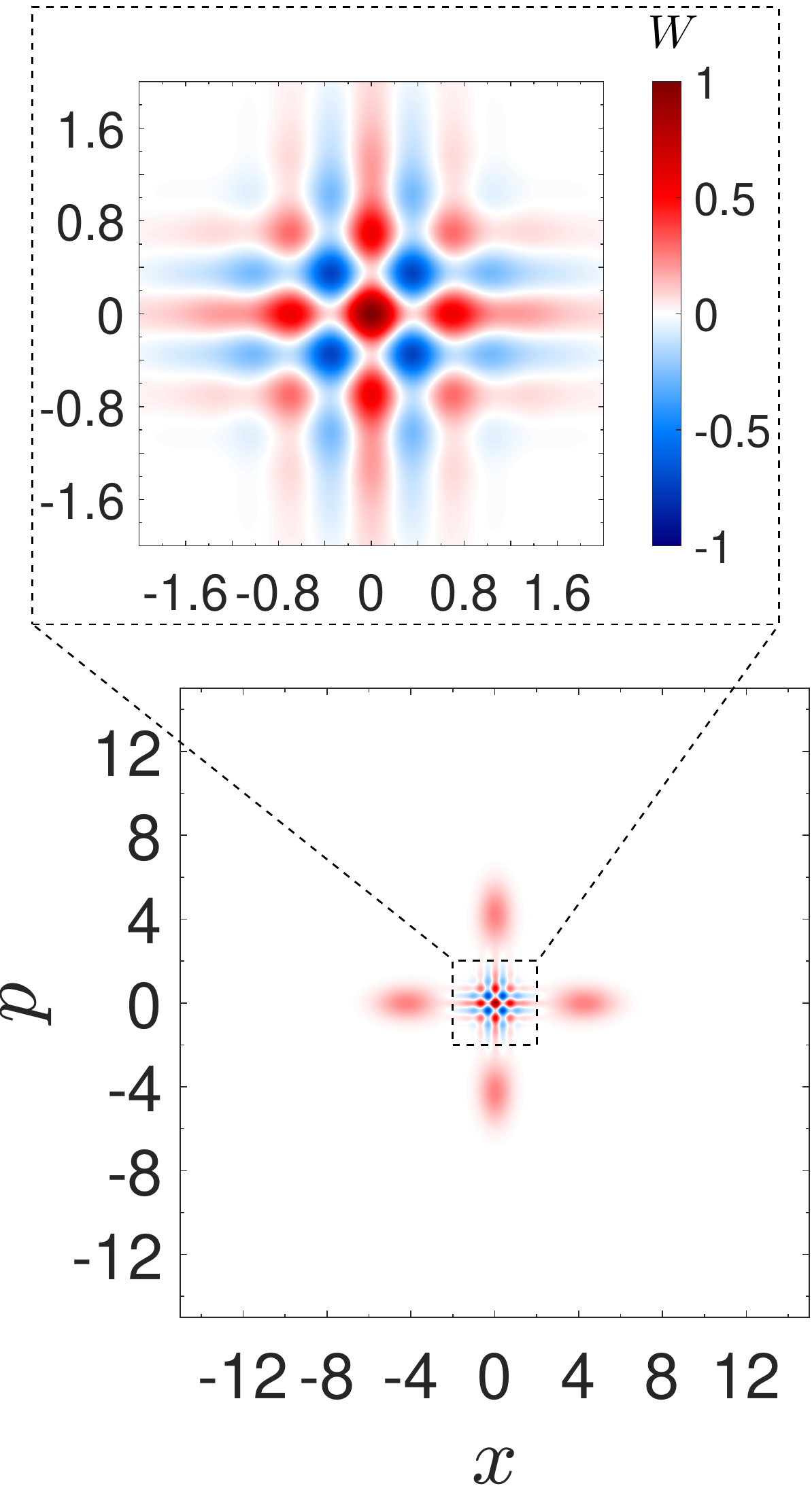}
(b)~\includegraphics[width=0.25\textwidth]{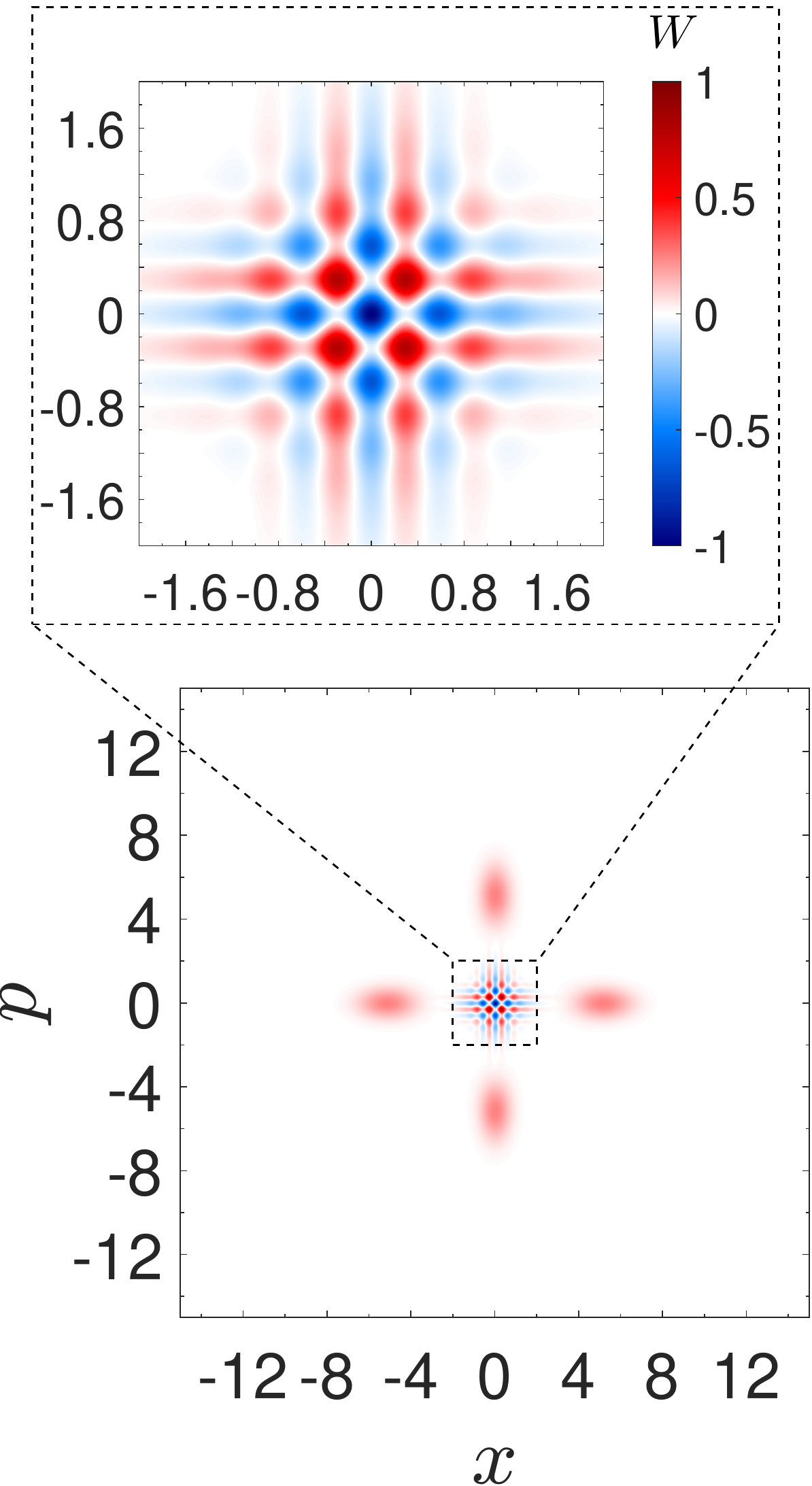}
(c)~\includegraphics[width=0.25\textwidth]{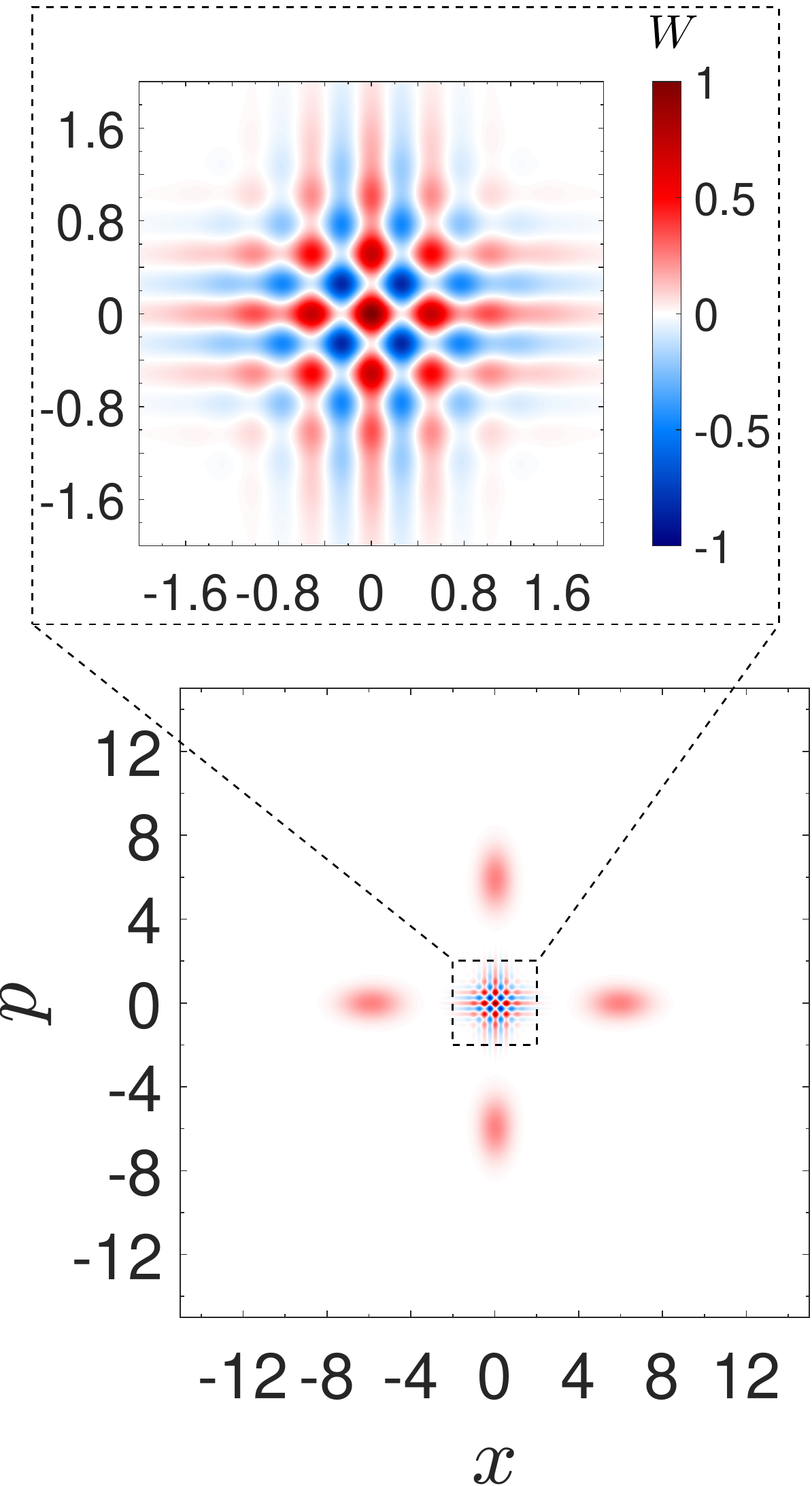}
\caption{Wigner distribution of the mixed-state SPSSVS with (a)~$n=10$, (b)~$n=15$, and (c)~$n=20$. In all cases $r=0.5$ and $c_1=\nicefrac{1}{\sqrt{2}}$. Insets represent the central interference pattern of each case.}
    \label{fig:SPSSVS_mixed_wigner}
\end{figure*}

This section is structured as follows. In Sec.~\ref{subsec:PA} we discuss Wigner functions corresponding to $\ket{\psi_{\text{SPA}}}$ and $\ket{\psi_{\text{SPS}}}$. Here, we describe how the addition and subtraction of photons lead to the sub-Planck structures in the phase space.
In Sec.~\ref{subsec:sens} we discuss the sensitivity to displacements associated with these two superpositions.

\subsection{Photons addition versus photons subtraction}\label{subsec:PA}
The Wigner function of the SPASVS~\eqref{eq:spa}
can be obtained by using Eq.~\eqref{eq:wigner_general} as (see Appendix~\ref{appendix:appendixA} for detailed derivations)
\begin{align}\label{eq:wig_spa}
W_{\ket{\text{SPA}}}(\bm{\zeta})=2 \operatorname{Re}\left[I_{\Xi}(\bm{\zeta})\right]+W_\boxplus(\bm{\zeta}),
\end{align}
and is shown in Fig.~\ref{fig:SPASVS_pure_wigner} for the cases when $n=10,15$, and 20. The first term in~\eqref{eq:wig_spa}
\begin{align}\label{eq:cross_1}
I_{\Xi}(\bm{\zeta}):=&\nonumber\frac{c_1^*c_2\exp\left(\xi\right)\left[-\text{i} \tanh(2 r)\right]^n}{\pi 4^n\cosh(r)\sqrt{1+\tanh^2(r)}}\sum^n_{l=0}\frac{(n!)^2\left[-2\text{i}\coth(r)\right]^l}{[(n-l)!]^2}\\&H_{n-l}\left[\text{i}\Omega\alpha_{-}\right]H_{n-l}\left[-\Omega\alpha^*_+\right],
\end{align}
provides the interference pattern that appears far away from the phase-space origin,
where
\begin{align}
\Omega:=\frac{\sqrt{\tanh(2r)}}{\sinh(r)},
\end{align}
and
\begin{align}
\xi:=-\tanh(2r)\left(\alpha^2-\alpha^{2 *}\right)-2|\alpha|^2\sech(2r),
\end{align}
with
\begin{align}
\alpha_{\pm}:=\alpha^*\sinh(r)\pm\alpha \cosh(r)
\end{align}
being the hyperbolic-rotated $\alpha$.

\begin{figure*}
\centering
(a)~\includegraphics[width=0.25\textwidth]{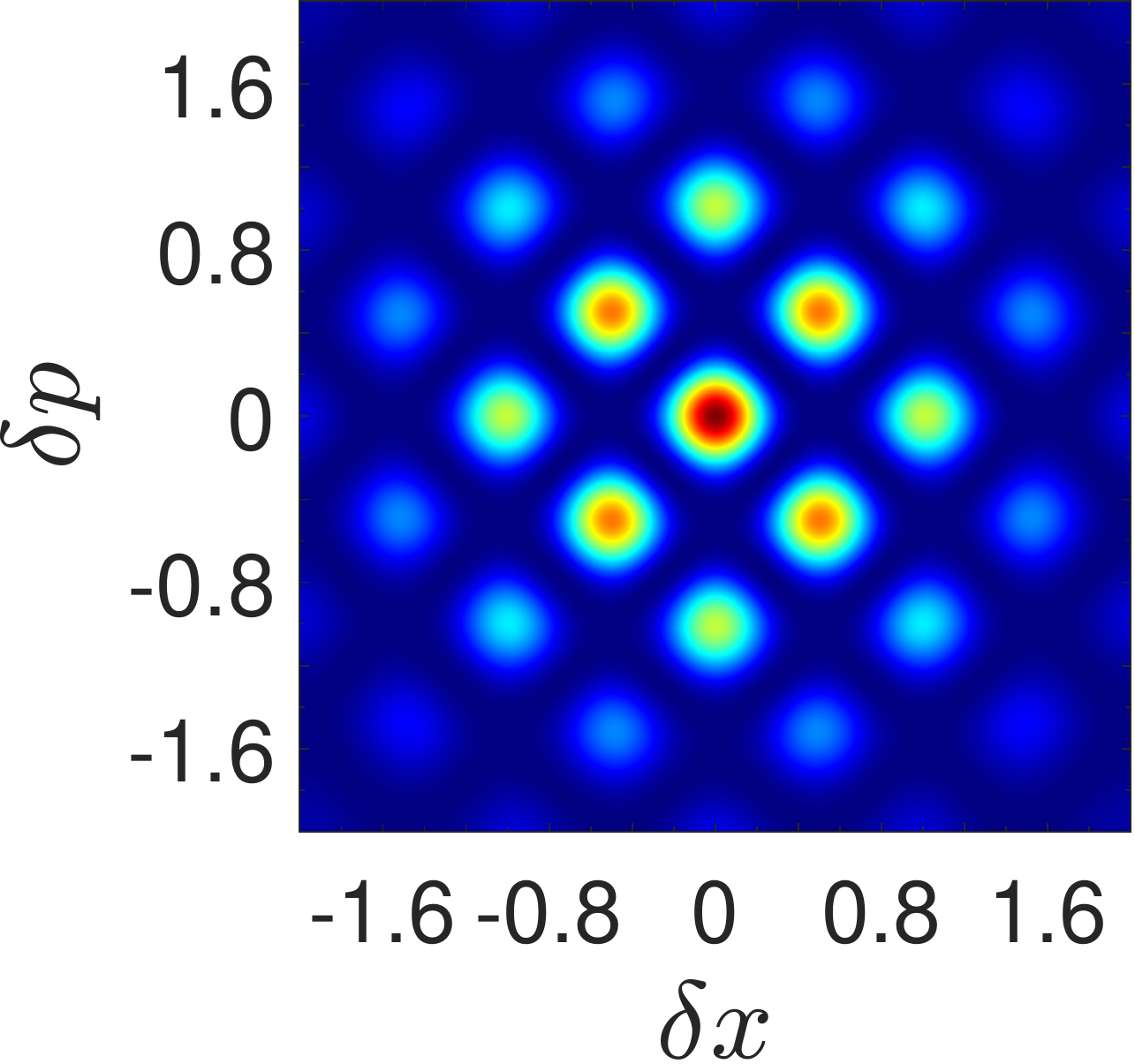}
(b)~\includegraphics[width=0.25\textwidth]{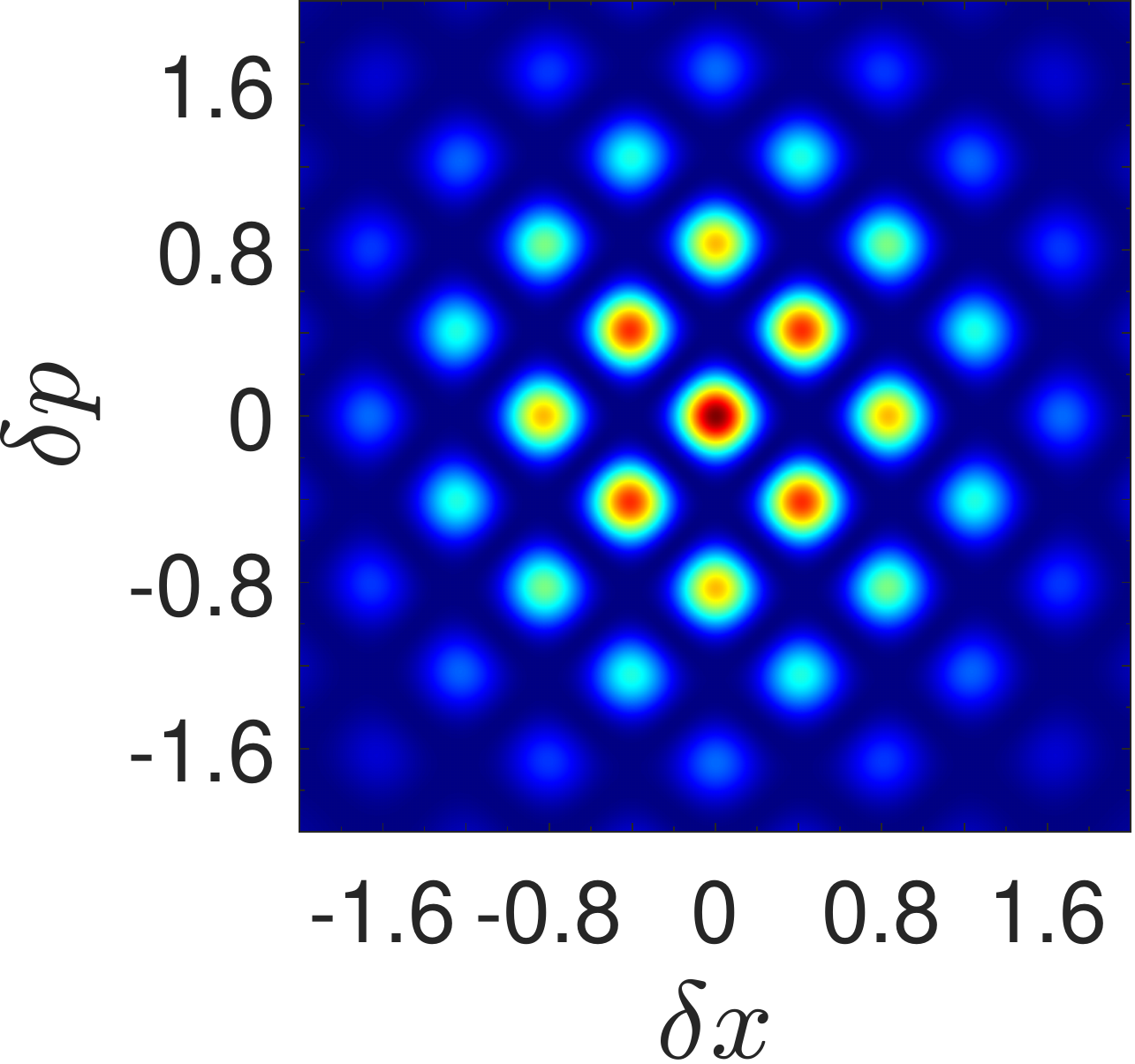}
(c)~\includegraphics[width=0.305\textwidth]{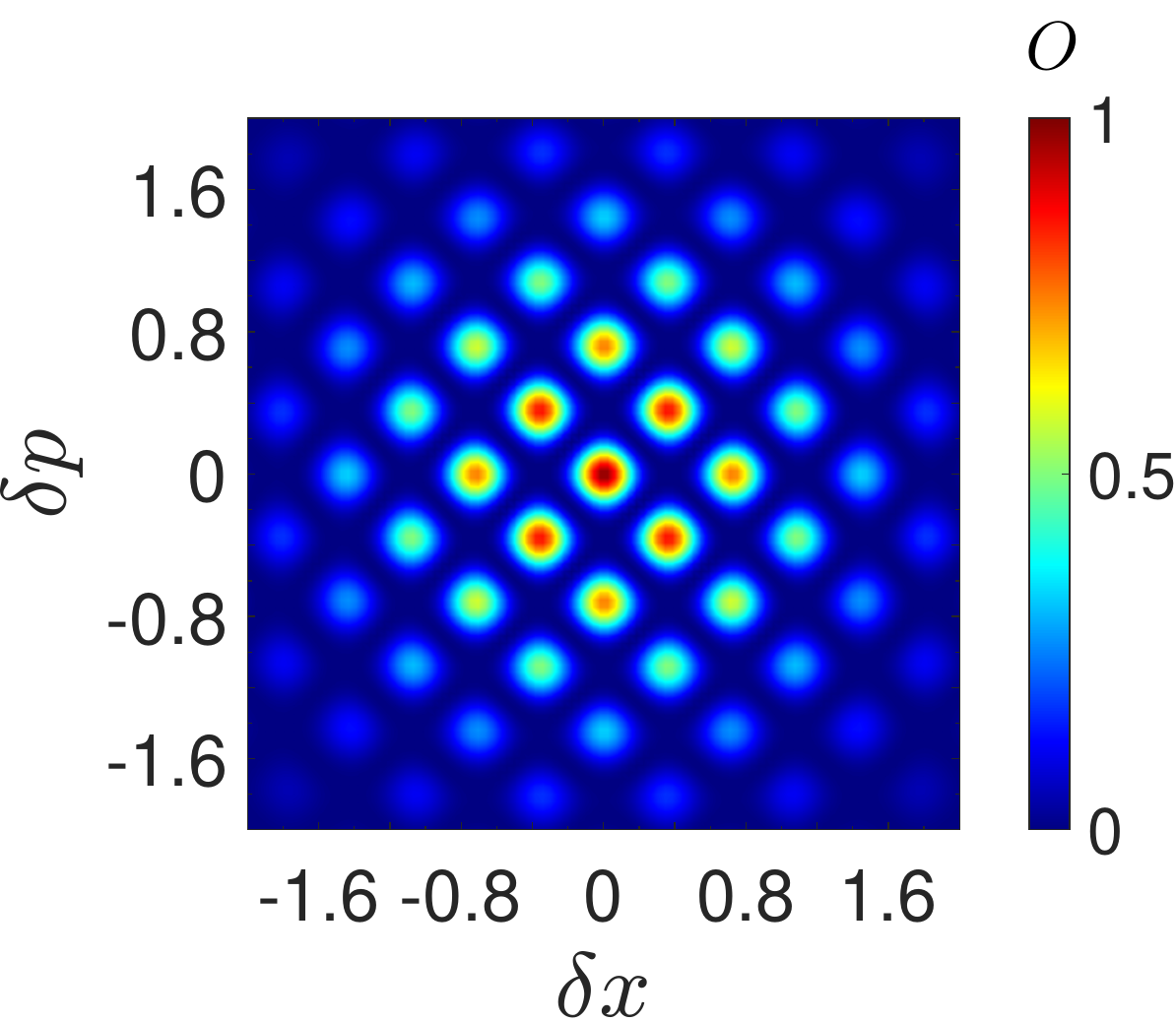}
\caption{Overlap of the pure SPASVS state with its $\delta \alpha$-displaced part with $\delta \alpha = (\delta x+\mathrm{i}\delta p)/\sqrt{2}$: (a)~$n=10$, (b)~$n=15$, and (c)~$n=20$. In all cases $r=0.5$ and $c_1=\nicefrac{1}{\sqrt{2}}$}.
\label{fig:SPASVS_pure_overlap}
\end{figure*}

For our purposes we concentrate on the second term in Eq.~\eqref{eq:wig_spa}, which contributes to the chessboardlike pattern that is visible at the phase-space origin for $n\gg 1$. 
This central interference pattern is equal to the sum of the individual Wigner functions of PASVSs as
\begin{align}\label{eq:chess1}
W_\boxplus(\bm{\zeta}):=|c_1|^2 W_{\ket{\psi^+_{\text{PA}}}}(\bm{\zeta})+|c_2|^2W_{\ket{\psi^-_{\text{PA}}}}(\bm{\zeta}).
\end{align}
The extension of an individual tile in the chessboardlike pattern is constrained concurrently along the $x$ and $p$ direction by the proper choice of weights in Eq.~(\ref{eq:chess1}). This is illustrated in Fig.~\ref{fig:SPASVS_area_versus_n}, where we show log-log plot of the extension of the central phase-space tile along the $x$ and $p$ directions versus $n$ for few $c_1$ and $c_2$ selections. The interference pattern (\ref{eq:chess1}) is much similar to the interference (\ref{eq:expr_PASV1}) of PASVS and its $\nicefrac{\pi}{2}$ rotated form in the cases depicted in Figs.~\ref{fig:SPSSVS_area_versus_n}(a) and \ref{fig:SPSSVS_area_versus_n}(c), respectively.
It is abundantly obvious from
these two examples
that changing $n$ results in a decrease in the extension of the central phase-space structure along the one-specific direction. On the other hand, the scenario shown in Fig.~\ref{fig:SPSSVS_area_versus_n}(b) represents the case where increasing $n$ causes the central phase-space tile to be constrained in all phase-space dimensions, which indicates that
the sub-Planck structures discovered for the compass state~\cite{Zurek2001} are also present in SPASVS.

\begin{figure*}
\centering
(a)~\includegraphics[width=0.25\textwidth]{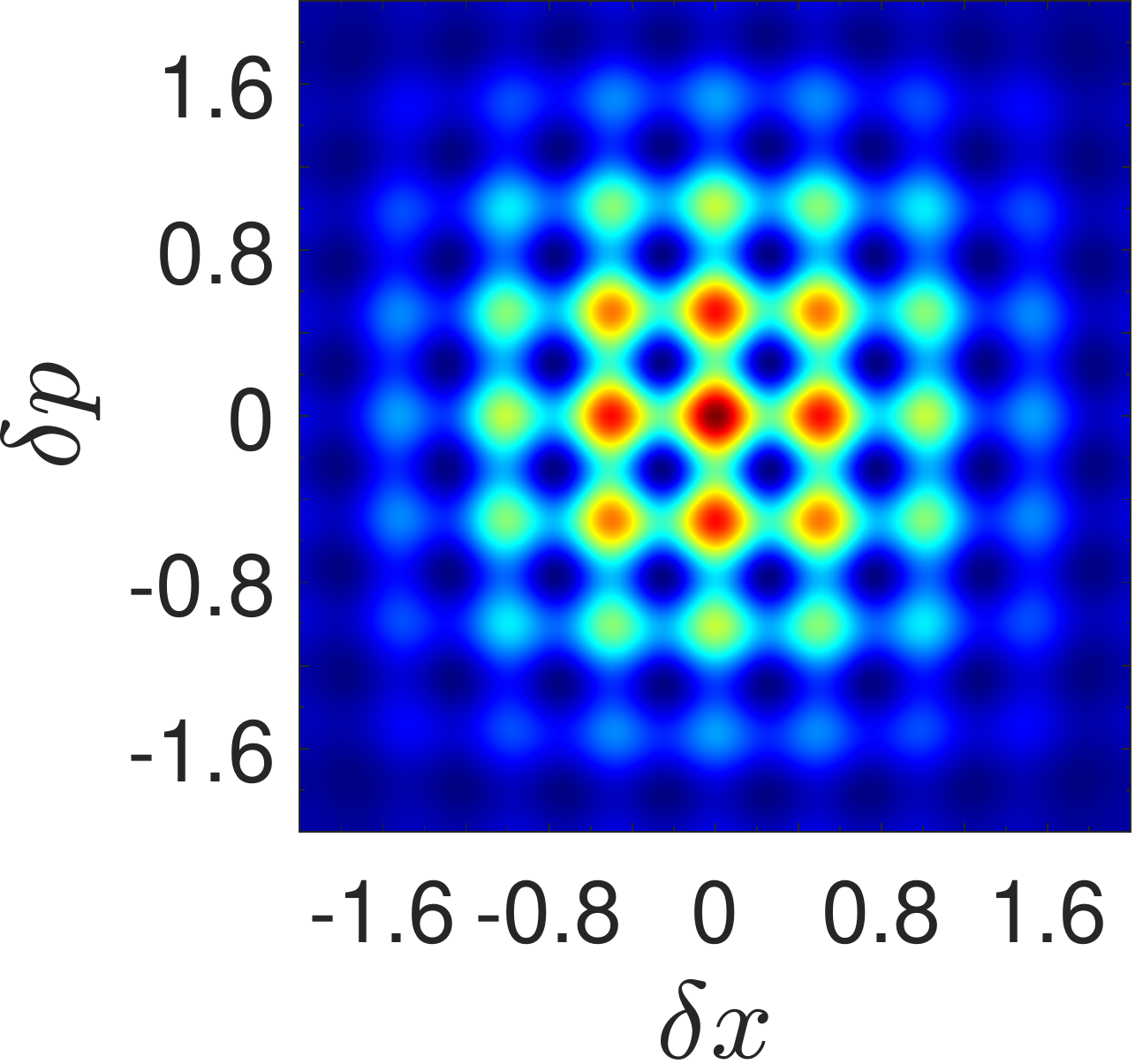}
(b)~\includegraphics[width=0.25\textwidth]{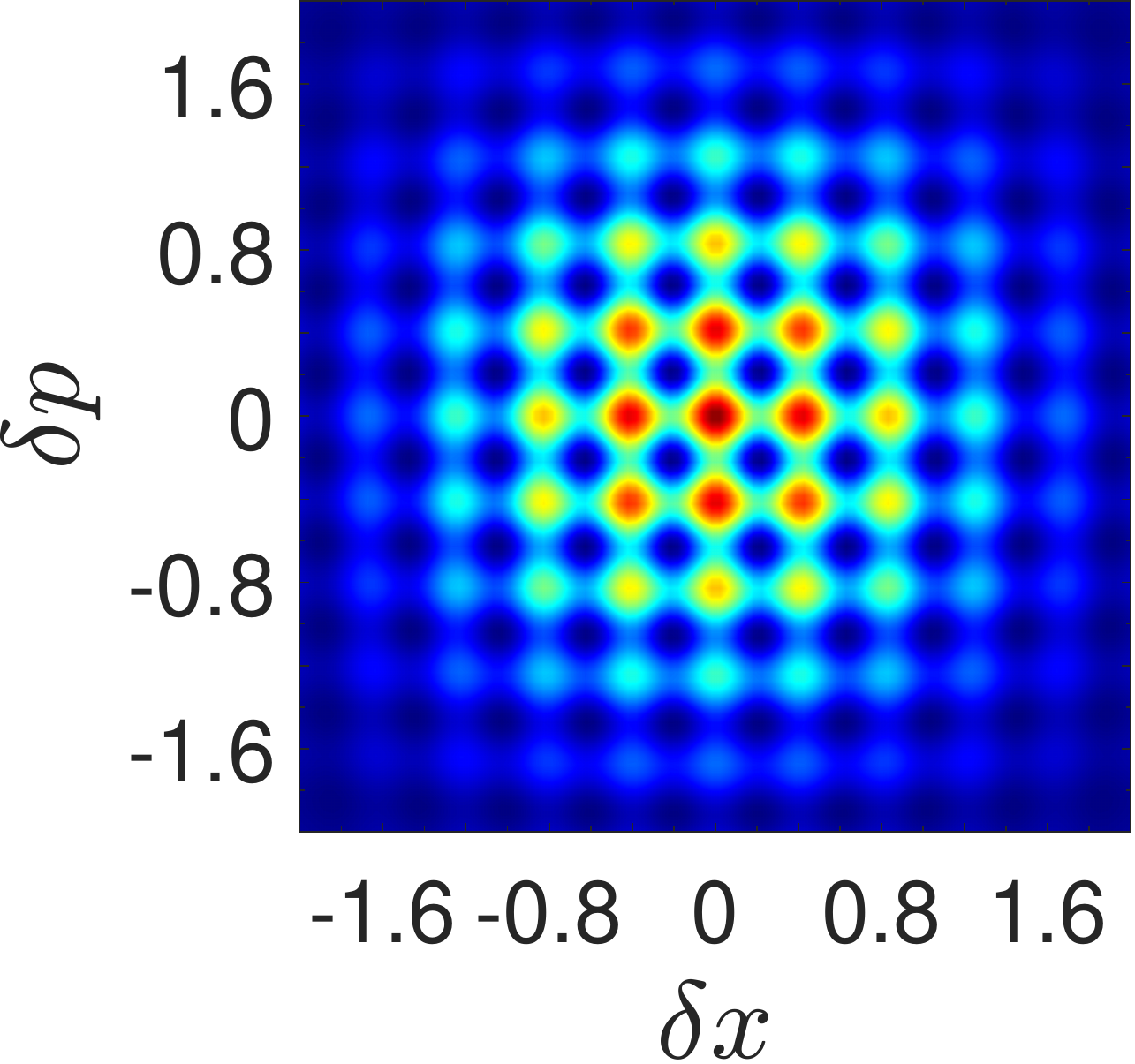}
(c)~\includegraphics[width=0.305\textwidth]{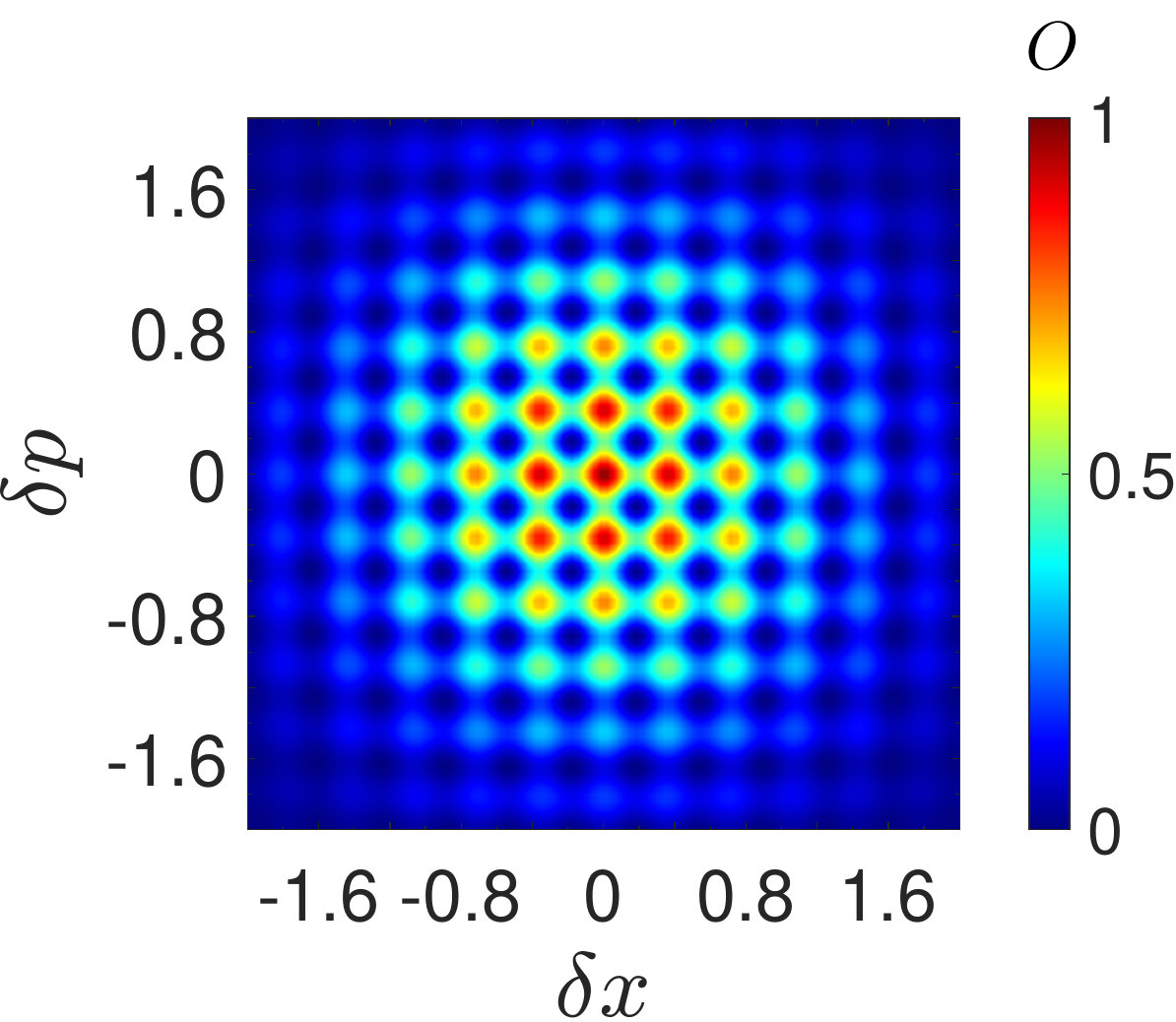}
\caption{Overlap of the mixed-state SPASVS with its $\delta \alpha$-displaced part with $\delta \alpha = (\delta x+\mathrm{i}\delta p)/\sqrt{2}$: (a)~$n=10$, (b)~$n=15$, and (c)~$n=20$. In all cases $r=0.5$ and $c_1=\nicefrac{1}{\sqrt{2}}$}.
    \label{fig:SPASVS_mixed_overlap}
\end{figure*}

\begin{figure*}
\centering
(a)~\includegraphics[width=0.25\textwidth]{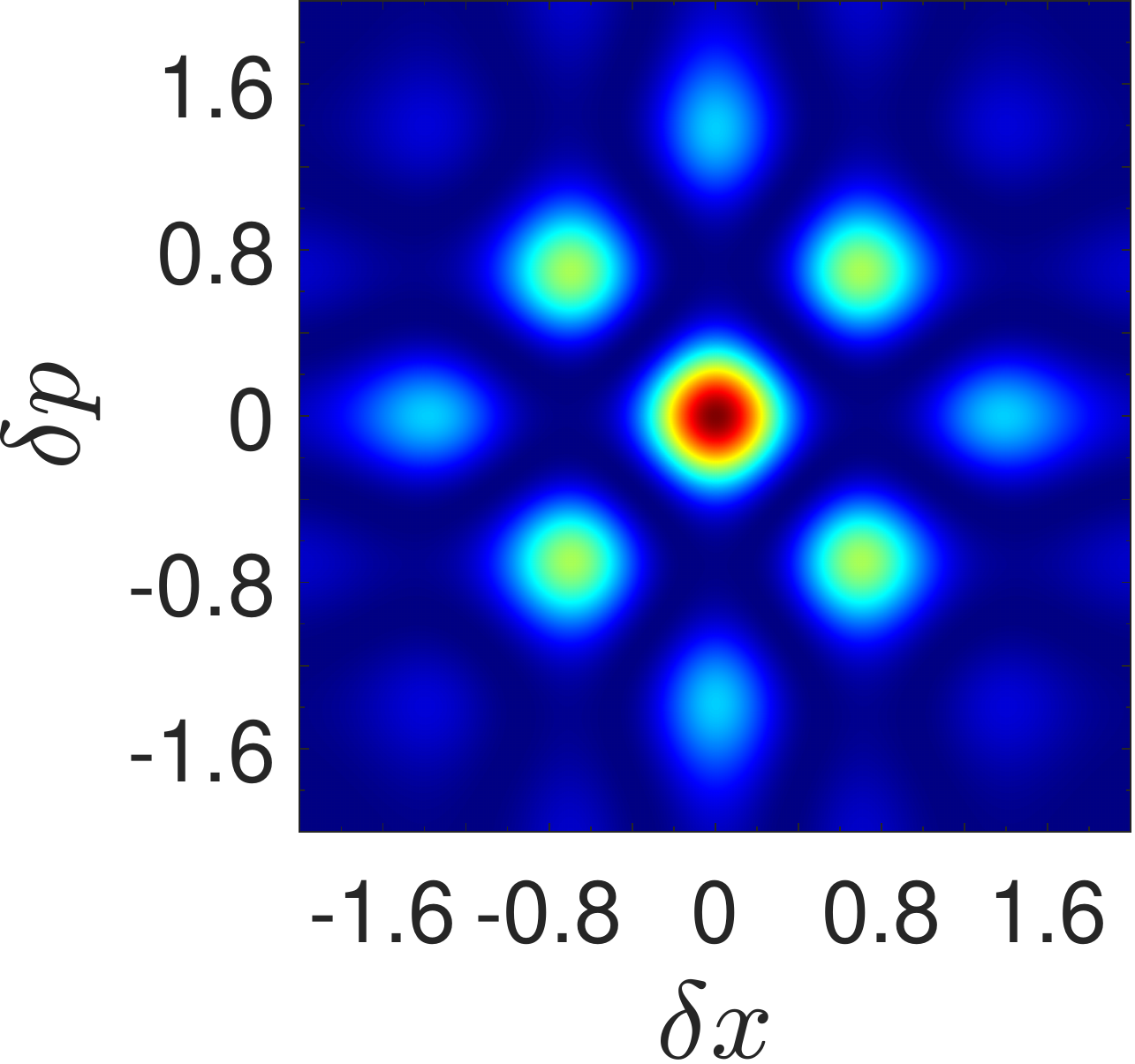}
(b)~\includegraphics[width=0.25\textwidth]{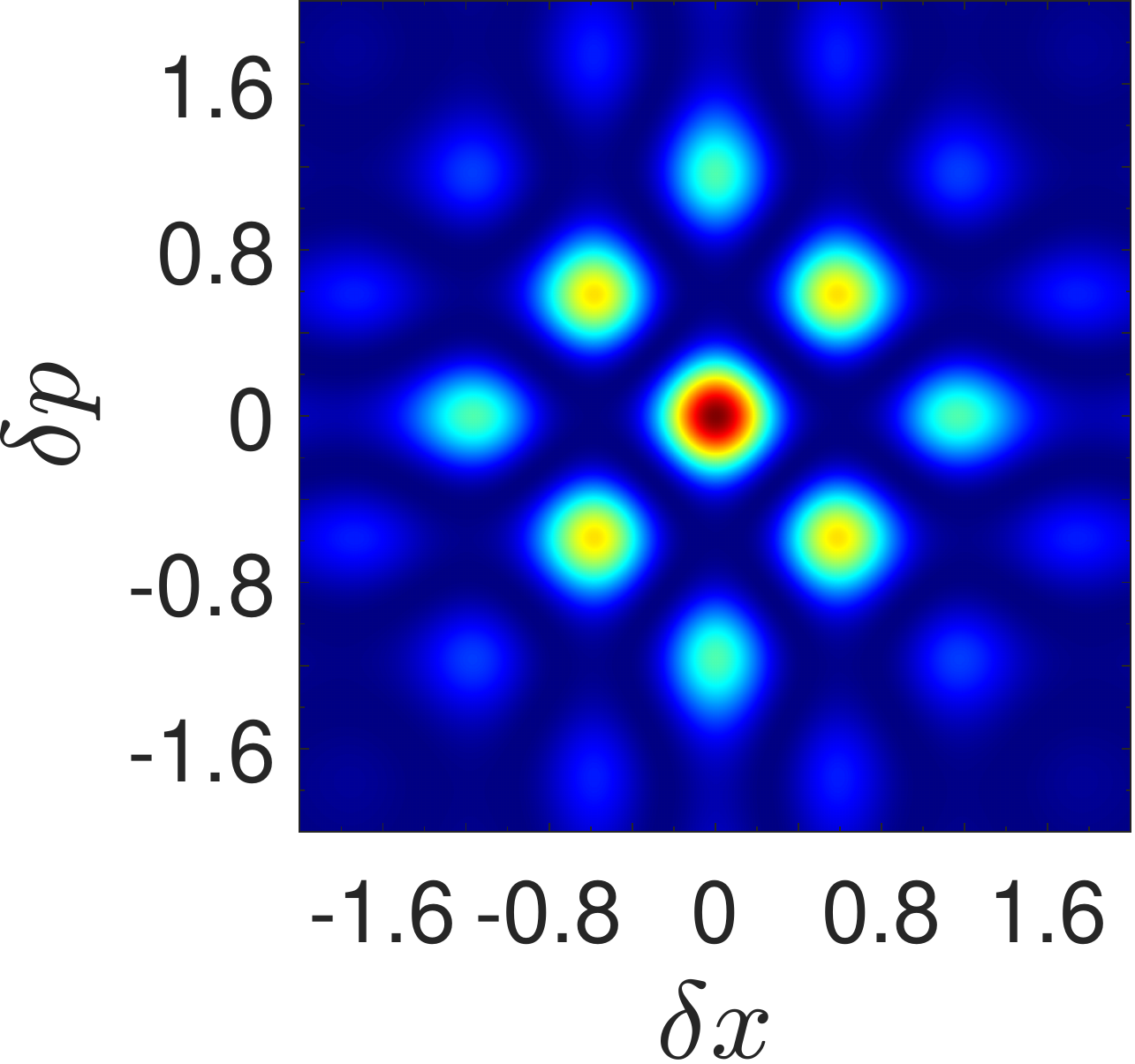}
(c)~\includegraphics[width=0.305\textwidth]{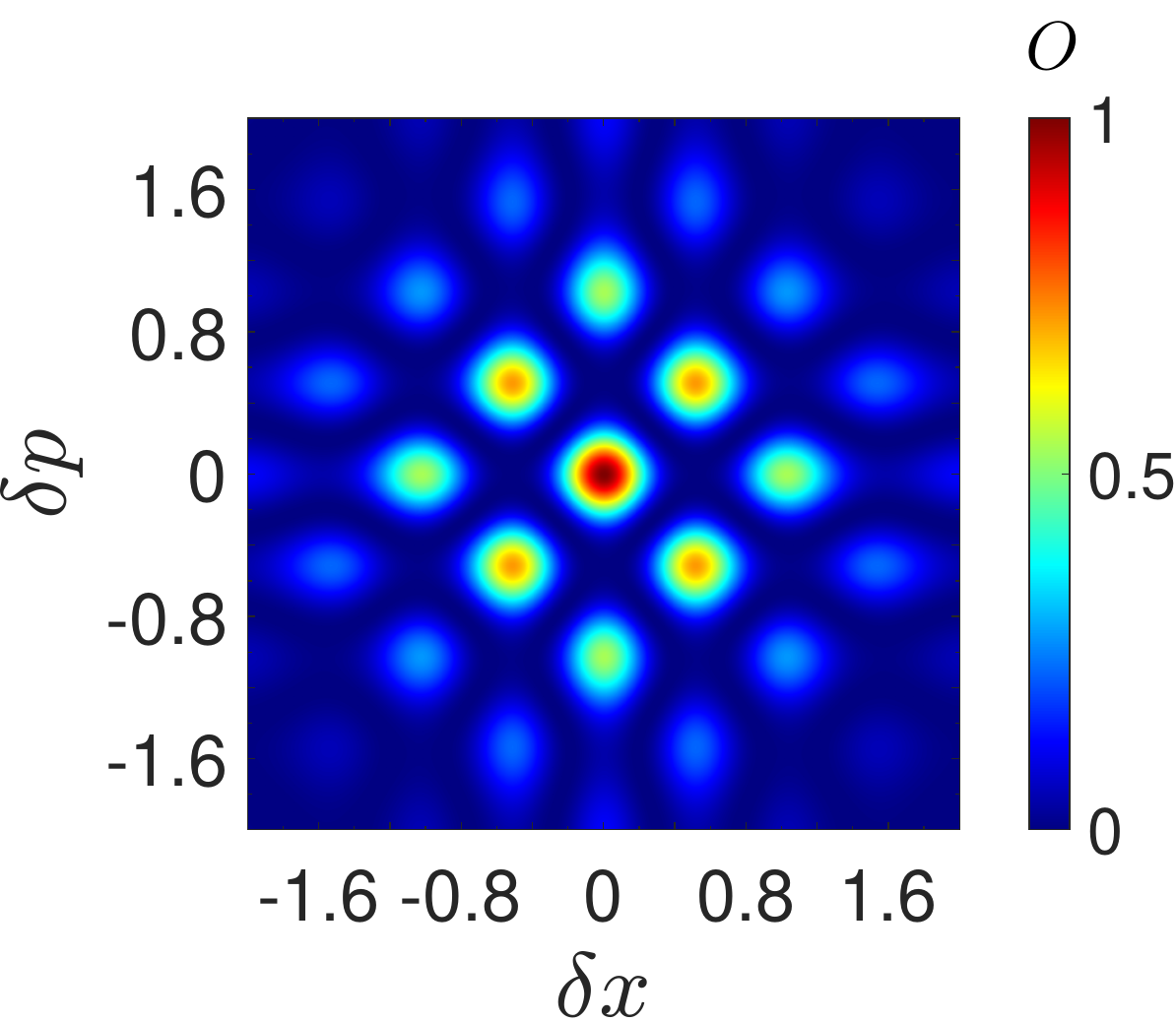}
\caption{Overlap of the pure SPSSVS state with its $\delta \alpha$-displaced part with $\delta \alpha = (\delta x+\mathrm{i}\delta p)/\sqrt{2}$: (a)~$n=10$, (b)~$n=15$, and (c)~$n=20$. In all cases $r=0.5$ and $c_1=\nicefrac{1}{\sqrt{2}}$.}
\label{fig:SPSSVS_pure_overlap}
\end{figure*}

The same sub-Planck structures are also contained by the following incoherent mixture of two PASVSs:
\begin{align}
\label{eq:rhopa}
\hat{\rho}_\text{PA}:=|c_1|^2\ket{\psi^+_{\text{PA}}}\!\bra{\psi^+_{\text{PA}}}+|c_2|^2\ket{\psi^-_{\text{PA}}}\!\bra{\psi^-_{\text{PA}}},
\end{align} and its Wigner function, which is the same as the one given in Eq.~\eqref{eq:chess1},
is shown in Fig.~\ref{fig:SPASVS_mixed_wigner}.

Similarly, for the SPSSVS (\ref{eq:sps}), the Wigner function is also written into two terms (see Appendix~\ref{appendix:appendixA} for detailed derivations):
\begin{align}\label{eq:wig_svs}
W_{\ket{\text{SPS}}}(\bm{\zeta})=2 \operatorname{Re}\left[I_{\Xi}(\bm{\zeta})\right]+W_\boxplus(\bm{\zeta}),
\end{align}
which is plotted in Fig.~\ref{fig:SPSSVS_pure_wigner}.
With
\begin{align}
\omega:=\frac{\sqrt{\tanh(2r)}}{\cosh(r)},
\end{align}
the term that contains
\begin{align}\label{eq:cross_2}
I_{\Xi}(\bm{\zeta}):=&\nonumber\frac{c_1^*c_2\exp\left(\xi\right)\left[\text{i} \tanh(2 r)\right]^n}{\pi 4^n\cosh(r)\sqrt{1+\tanh^2(r)}}\sum^n_{l=0}\frac{(n!)^2\left[2\text{i}\tanh(r)\right]^l}{[(n-l)!]^2}\\&H_{n-l}\left[-\omega\alpha_{-}\right]H_{n-l}\left[-\text{i}\omega\alpha^*_+\right],
\end{align}
causing the interference pattern that
manifests as oscillating peaks far from the phase-space origin. 
Again, we concentrate on the second term of Eq.~\eqref{eq:wig_svs}, which results in the chessboardlike pattern at the phase-space origin. This term is the sum
of the Wigner functions (\ref{eq:expr_PSSV1}) of PSSVSs, that is,
\begin{align}\label{eq:chess2}
W_\boxplus(\bm{\zeta}):=|c_1|^2W_{\ket{\psi^+_{\text{PS}}}}(\bm{\zeta})+|c_2|^2W_{\ket{\psi^-_{\text{PS}}}}(\bm{\zeta}).
\end{align}
This pattern manifests sub-Planck oscillations around the origin of the phase space by the proper choices of $c_1$ and $c_2$. This
is illustrated in Fig.~\ref{fig:SPSSVS_area_versus_n}, where we show log-log plot of the extension of the central phase-space tile. The case depicted in Fig.~\ref{fig:SPSSVS_area_versus_n}(b) is the situation in which the SPSSVS have sub-Planck structures in phase space, whereas the cases shown in Figs.~\ref{fig:SPSSVS_area_versus_n}(a) and \ref{fig:SPSSVS_area_versus_n}(c) represent the central phase-space pattern (\ref{eq:chess2}) resembling that of the sole PSSVSs.
It is interesting to note that for a given $r$ and $n$, the central tile in the chessboardlike pattern is larger than that of the SPASVS.

Additionally, we demonstrate that the following incoherent mixture of two PSSVSs likewise has the same sub-Planck structures:
\begin{align}
\hat{\rho}_\text{PS}:=|c_1|^2\ket{\psi^+_{\text{PS}}}\!\bra{\psi^+_{\text{PS}}}+|c_2|^2\ket{\psi^-_{\text{PS}}}\!\bra{\psi^-_{\text{PS}}}.
\end{align}
Similar to the case of the mixture of PASVSs~\eqref{eq:rhopa}, the
Wigner function of $\hat{\rho}_{\text{PS}}$ shown in Fig.~\ref{fig:SPSSVS_mixed_wigner}
is identical to
Eq.~\eqref{eq:chess2}.

In summary, the photon-addition or photon-subtraction operations on the superpositions of the Gaussian SVS produce the sub-Planck structures in the phase space. The same sub-Planck structures are also present in mixtures related to PASVSs or PSSVSs. The main difference between our states and the compass states~\cite{Zurek2001} is that our states are built by superposition of only two non-Gaussian SVSs rather than four coherent-state superpositions. Interestingly, adding or even subtracting photons from the superposition of the Gaussian SVS actually increases the average photon number of the resultant states, with the addition case having a greater average photon number than the subtraction case~\cite{WANG2019102}. The size of their sub-Planck structures is likewise impacted by this variation in the average photon quantity in the states. For instance, given the same number of photons used, the SPASVS and its related mixture have smaller sub-Planck structures than their equivalents in the photon-subtracted case.

\begin{figure*}
\centering
(a)~\includegraphics[width=0.25\textwidth]{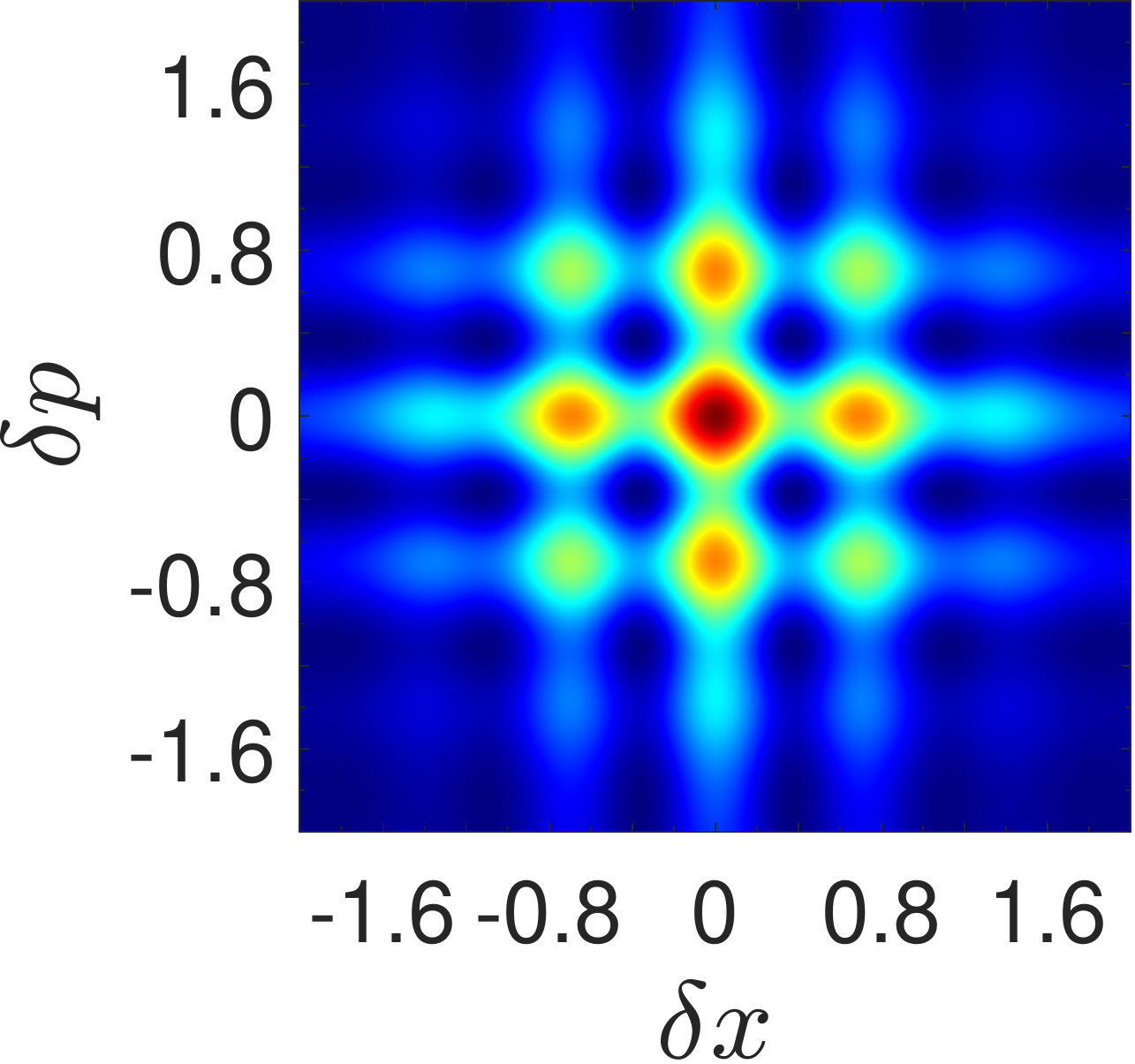}
(b)~\includegraphics[width=0.25\textwidth]{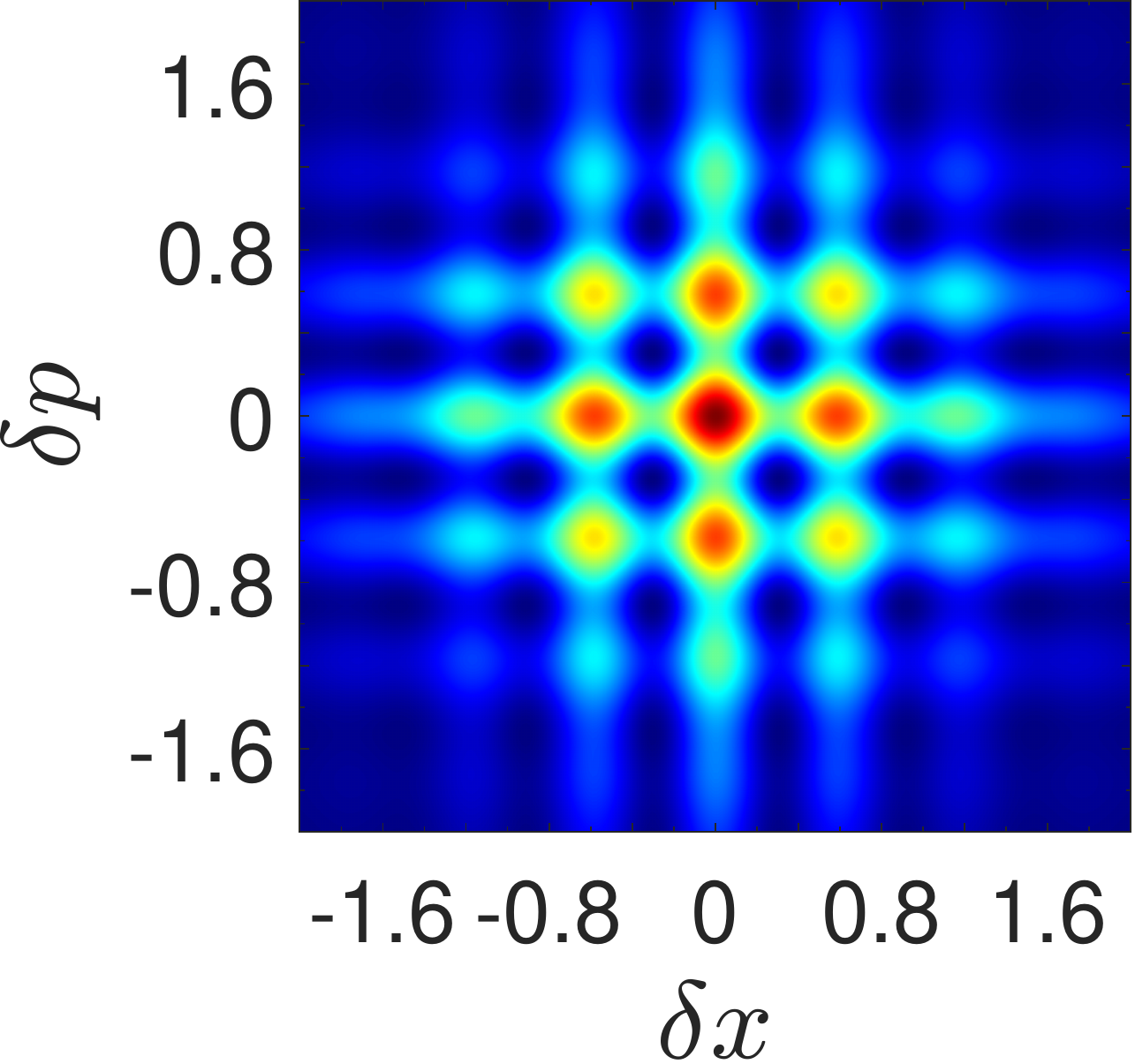}
(c)~\includegraphics[width=0.305\textwidth]{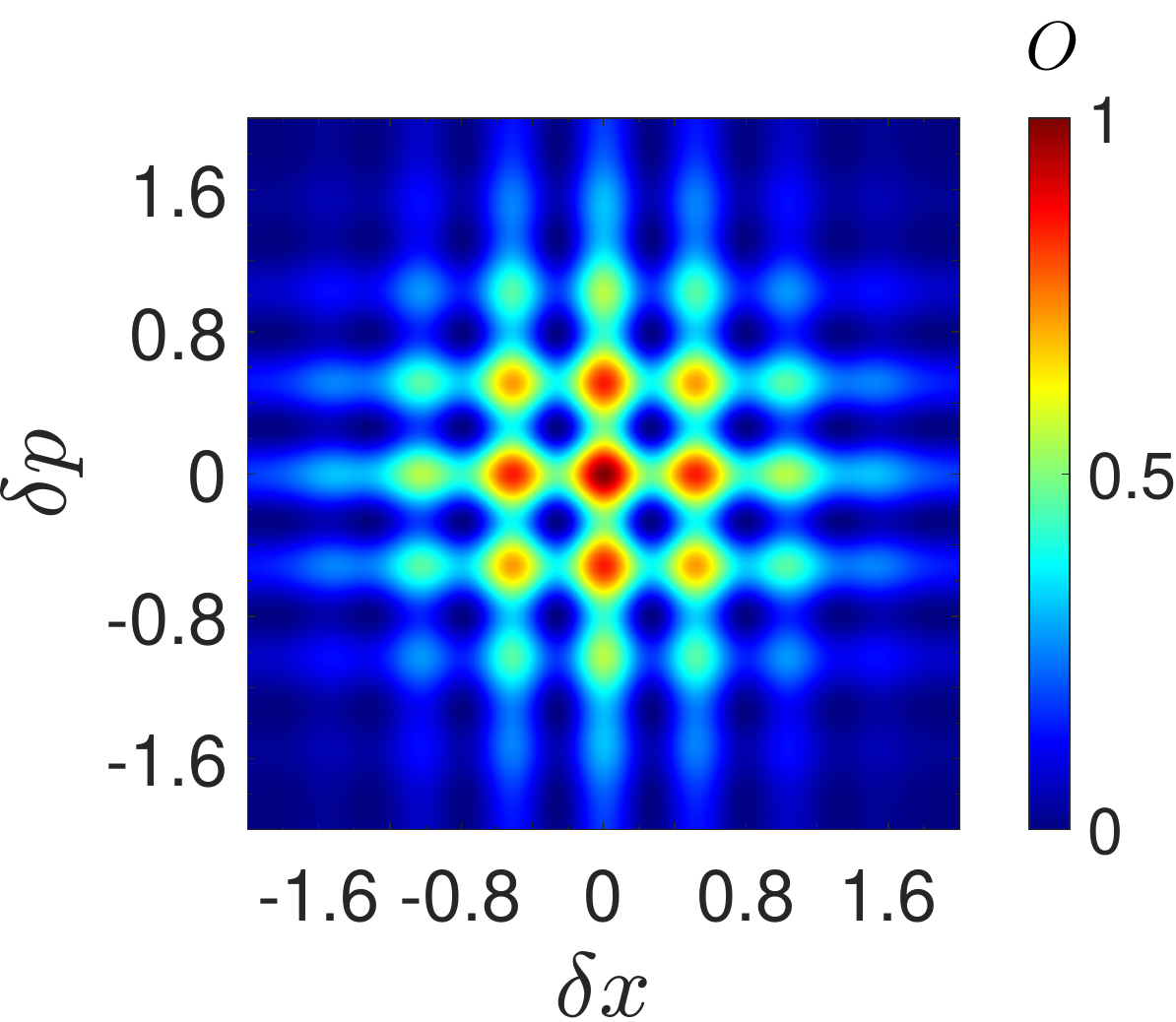}
\caption{Overlap of the mixed-state SPSSVS with its $\delta \alpha$-displaced part with $\delta \alpha = (\delta x+\mathrm{i}\delta p)/\sqrt{2}$: (a)~$n=10$, (b)~$n=15$, and (c)~$n=20$. In all cases $r=0.5$ and $c_1=\nicefrac{1}{\sqrt{2}}$.}
\label{fig:SPSSVS_mixed_overlap}
\end{figure*}

\subsection{Sub-shot noise sensitivity of our states}\label{subsec:sens}
In this section we discuss the susceptibility of our proposed states to phase-space displacement. Let us first consider SPASVS~(\ref{eq:spa}).
The overlap~(\ref{eq:overlap_HW}) for this state under the approximation
$|\delta \alpha| \ll 1$ and $n\gg1$
leads to~(see Appendix~\ref{appendix:overlaps} for the detailed derivations)
\begin{align}\label{eq:sens_spasvs}
&\nonumber O_{\text{SPA}}(\delta \alpha)\\&=\left[|c_1|^2\bra{\psi^{+}_{\text{PA}}}\hat{D}(\delta \alpha)\ket{\psi^{+}_{\text{PA}}}+|c_2|^2\bra{\psi^{-}_{\text{PA}}}\hat{D}(\delta \alpha)\ket{\psi^{-}_{\text{PA}}}\right]^2.
\end{align}
Each term of this overlap is calculated as
\begin{align}\label{eq:senspasv_1}
\bra{\psi^{\pm}_{\text{PA}}}\hat{D}(\delta \alpha)\ket{\psi^{\pm}_{\text{PA}}}=&\nonumber\frac{\left[\mp\sinh(2r)\right]^n\mathrm{e}^{-\nicefrac{|\eta_{\pm}|^2}{2}}}{4^n}\sum^n_{l=0}\frac{(n!)^2}{l![(n-l)!]^2}\\&[\mp 2\text{\text{coth}}(r)]^lH_{n-l}\left[\Theta_{\pm}\right]H_{n-l}\left[\Theta_{\pm}^*\right],
\end{align}
where
\begin{align}
\Theta_{\pm}=\text{i}\sqrt{\pm\frac{ \text{coth}(r)}{2}}\eta_{\pm},
\end{align}
and
\begin{equation}
 \eta_{\pm}=\delta\alpha\text{cosh}(r)\mp\delta\alpha^*\text{sinh}(r),
\end{equation}
with
\begin{equation}
    \delta\alpha=\delta x+\text{i}\delta p.
\end{equation}
In Fig.~\ref{fig:SPASVS_pure_overlap} we plot this overlap for $n=10$, 15, and 20.
For the large $n$, a small
displacement $|\delta \alpha|\ll1$ can turn
the SPASVS into a state
orthogonal to its original state,
and this orthogonality occurs in all phase-space directions.
We have normalized overlaps to their maximum amplitudes, $O_{{\hat{\rho}}}(0)$.

Let us now consider the
mixture of PASVSs (\ref{eq:pasvs}), for which the overlap (\ref{eq:overlap_HW}) is calculated as
\begin{align}
&\nonumber O_{{\hat{\rho}_\text{PA}}}(\delta \alpha)\\&=\left|c_1^2\bra{\psi^{+}_{\text{PA}}}\hat{D}(\delta \alpha)\ket{\psi^{+}_{\text{PA}}}\right|^2+\left|c_2^2\bra{\psi^{-}_{\text{PA}}}\hat{D}(\delta \alpha)\ket{\psi^{-}_{\text{PA}}}\right|^2.
\end{align}
We plot this overlap with $n=10$, 15, and 20 in Fig.~\ref{fig:SPASVS_mixed_overlap}. Again, we see that the overlap $O_{\hat{\rho}_{\text{PA}}}(\delta \alpha)$ disappears for the displacement $|\delta \alpha|\ll1$, but unlike the SPASVS, this orthogonality now takes place when $\delta x=\pm\delta p$ in the phase space.

Similarly, overlap (\ref{eq:overlap_HW}) for SPSSVS is obtained as
\begin{align}
&\nonumber O_{\text{SPS}}(\delta \alpha)\\&=\left[|c_1|^2\bra{\psi^{+}_{\text{PS}}}\hat{D}(\delta \alpha)\ket{\psi^{+}_{\text{PS}}}+|c_2|^2\bra{\psi^{-}_{\text{PS}}}\hat{D}(\delta \alpha)\ket{\psi^{-}_{\text{PS}}}\right]^2,
\end{align}
where
\begin{align}\label{eq:spssv_sen1}
\bra{\psi^{\pm}_{\text{PS}}}\hat{D}(\delta \alpha)\ket{\psi^{\pm}_{\text{PS}}}=&\nonumber\frac{\left[\mp\sinh(2r)\right]^n\mathrm{e}^{-\nicefrac{|\eta_{\pm}|^2}{2}}}{4^n}\sum^n_{l=0}\frac{(n!)^2}{l![(n-l)!]^2}\\&[\mp 2\text{\text{tanh}}(r)]^lH_{n-l}\left[\theta_{\pm}\right]H_{n-l}\left[\theta_{\pm}^*\right],
\end{align}
with
\begin{align}
\theta_{\pm}=\text{i}\sqrt{\pm \frac{\text{tanh}(r)}{2}}\eta_{\pm}.
\end{align}
In Fig.~\ref{fig:SPSSVS_pure_overlap} we plot this overlap with $n$ equal to 10, 15, and 20. The SPSSVS overlap plot exhibits the same behavior as the SPASVS: the overlap disappears in any direction in phase space when $|\delta\alpha|\ll1$. The only distinction is that, for the same $n$ and $r$, the central pattern of the overlap of the SPSSVS is larger than that of the SPASVS, showing that the SPSSVS is less sensitive than the SPASVS.

Finally, we consider mixtures of PSSVSs (\ref{eq:spssv}). The overlap (\ref{eq:overlap_HW}) for this state leads to
\begin{align}
&\nonumber O_{{\hat{\rho}_\text{PS}}}(\delta \alpha)\\&=\left|c_1^2\bra{\psi^{+}_{\text{PS}}}\hat{D}(\delta \alpha)\ket{\psi^{+}_{\text{PS}}}\right|^2+\left|c_2^2\bra{\psi^{-}_{\text{PS}}}\hat{D}(\delta \alpha)\ket{\psi^{-}_{\text{PS}}}\right|^2.
\end{align}
In Fig.~\ref{fig:SPSSVS_mixed_overlap}, we plot the overlap~$O_{\hat{\rho}_{\text{PS}}}(\delta \alpha)$ using the same parameter values as that of
the SPSSVS. We observe that the overlap of the mixture of the PSSVSs looks similar to the mixture of the PASVSs, with the distinction being that the mixture of PSSVSs appears to be less sensitive for a given number of $n$ and $r$, which is manifested as the larger chessboardlike pattern in the center of the phase space than that of the mixture of PASVSs.

In summary, we have demonstrated that the sensitivity associated to our proposed states depends on the quantity of photons added~(or subtracted), and it is considerably lower than the sensitivity of the coherent state when there are excessive amounts of photons added~(or subtracted). For both SPASVS and SPSSVS,  the enhanced sensitivity is unaffected by the directions of phase-space displacements.\;However, for the mixtures related to PASVSs or PSSVSs, this enhancement only takes place in particular phase-space directions. This implies that compared to mixed states, our superposition states have more potential for quantum sensing applications. Moreover, it has been found that the quantum states associated with photon-addition cases are more sensitive than their counterparts of the photon-subtraction cases.

In comparison to the original compass state~\cite{Zurek2001}, our proposed states perform similarly in terms of sensitivity to phase-space displacements, but the sensitivities of our states are impacted either by adding or subtracting photons from these states. For example, in the compass state, the sensitivity to displacements is proportional to the parameter $x_0$ (separation between coherent states), which is in fact connected to the average photon number in the states; that is, increasing $x_0$ leads to a higher average photon number in the states. In our case, enhanced sensitivity is now connected with the number of added or subtracted photons $n$ in the states, as adding or removing photons leads in a larger average photon number in the resultant states.

Our states present an alternative to compass states that have the same properties but may be prepared more feasibly. It was demonstrated that the coherent-state superpositions could be constructed deterministically using third-order Kerr nonlinearity~\cite{YS86}. However, this technique necessitates Kerr non-linearities of such magnitude that it is not practicable with currently available Kerr media. Furthermore, states of this type are particularly susceptible to loss, and because absorption cannot be ignored in currently available Kerr-media, the capacity to extract coherent-state superpositions before they decohere is severely constrained~\cite{Boyd1999}. By adding or subtracting photons from Gaussian SVSs, a comparably practical strategy for constructing coherent-state superpositions is adopted~\cite{Alexei2006,Neergaard2006,Podoshvedov2023,Takase2021,Dakna1997,Thekkadath2020,endo2023}.

\section{Summary and outlook}
\label{sec:conc}

We have shown that the Wigner function of the
SPASVS (or SPSSVS) contains 
the chessboardlike pattern around the origin of the phase space. A similar chessboardlike pattern also emerges
by the mixtures related to PASVSs and PSSVSs. The support area of the phase-space structures contained by this chessboardlike pattern varies inversely with
the photon number added (or subtracted). When a sizable number of photons are added (or subtracted), the support area of
these structures is noticeably smaller than that of the coherent state. 

The average photon numbers of our states, which are increased either by photon-addition or photon-subtraction actions on the Gaussian SVS, have an impact on the size of the sub-Planck structures in the phase space. The sub-Planck structures associated with the SPASVS are smaller than those of the SPSSVS for the same number of photons added or subtracted. This is because the photon-addition operation always leads to the higher average photon number in the resultant states. The association of the average photon number with the sub-Planck structures in our states is much similar to that of the compass states, i.e., a higher average photon number in the compass state corresponds to the smaller sub-Planck structures in the phase space.

We have demonstrated that the sensitivity of our proposed states is noticeably higher than that of the coherent state when a significant number of photons are added (or subtracted). Both the
SPASVS and SPSSVS exhibit
the
enhanced sensitivity, which is independent of the phase-space directions, indicating that they hold more promise for quantum metrology. In addition,
the difference in the sensitivities
between
the photon-addition and -subtraction cases
arose from the
different
average photon numbers in the states; photon-addition cases are demonstrated to have greater sensitivity than the subtracted cases.

It is incredibly exciting that sub-Planck structures can possibly build from photons being added to or subtracted from states. As a result, it will be able to apply a variety of ways to engineer compasslike states in association with contemporary experiments. Numerous theoretical and experimental research utilizing photon-addition or -subtraction operations from states have been developed to produce catlike states~\cite{Alexei2006,Neergaard2006,Podoshvedov2023,Takase2021,Dakna1997,Thekkadath2020,endo2023}. Theoretical research to construct the catlike states advocated in~\cite{Dakna1997} is subsequently applied in actual experiments~\cite{Neergaard2006,Alexei2006}. Another subsequent theoretical approach introduces the idea of photon subtraction from the Gaussian SVS to produce the compasslike states~\cite{Prop5}. These illustrations unequivocally demonstrate that some of these techniques can be modified to generate SPASVS and SPSSVS, which are entirely new research avenues that can be adapted in the future.
\begin{acknowledgements}
N.A.\ acknowledges the support of the postdoctoral fund of the Zhejiang Normal University under Grant No.~ZC304021937. G.X.\ acknowledges support by the National Natural Science Foundation of China under Grants No.~11835011 and No.~12174346. W.M.L.\ acknowledges support from the National Key R{\&}D Program of China under Grants No.~2021YFA1400900, No.~2021YFA0718300, and No.~2021YFA1402100, the NSFC under Grants No.~61835013, No.~12174461, and No.~12234012, and the Space Application System of China Manned Space Program.
\end{acknowledgements}

\onecolumngrid

\appendix
\section{Wigner functions of SPASVS and SPSSVS}\label{appendix:appendixA}
This section provides the main steps to drive the Wigner functions of SPASVS and SPSSVS.
\subsection{Derivations of the Wigner function of SPASVS}
Let us first consider the Wigner function of SPASVS given by Eq.~\eqref{eq:wig_spa}. The first term of this Wigner function is given by
\begin{align}\label{eq:appendix_wig1}
I_{\Xi}(\bm{\zeta})=\Braket{\psi^{+}_{\text{PA}}|\hat{\Delta}(\alpha)|\psi^{-}_{\text{PA}}},
\end{align}
where the alternative form of the displaced parity operator $\hat{\Delta}(\alpha)$ is~\cite{Ma2012}
\begin{align}
\hat{\Delta}(\alpha):=&\frac{1}{\pi^2}\mathrm{e}^{2\left|\alpha\right|^2}
\int^{\infty}_{-\infty} \,\mathrm{d}^2\beta \mathrm{e}^{
-2\alpha^{*}\beta+
2\alpha\beta^{*}}\Ket{\beta}\!\Bra{-\beta}.
\end{align}
Equation~\eqref{eq:appendix_wig1} can be rewritten as
\begin{align}
I_{\Xi}(\bm{\zeta})=\frac{(-1)^{n}\mathrm{e}^{2|\alpha|^2}}{\pi^2 \cosh (r)}
    \int^{\infty}_{-\infty}\mathrm{d}^2\beta |\beta|^{2n}
    \exp\bigg[-|\beta|^2-\frac{\tanh (r)}{2}\big(\beta^2+\beta^{*2}\big)-2\beta\alpha^{*}+2\beta^{*}\alpha\bigg].
\end{align}
We incorporate the factor $|\beta|^{2n}$ into a differential equation as
\begin{align}\label{eq:integ_use1}
I_{\Xi}(\bm{\zeta})=\frac{(-1)^{n}\mathrm{e}^{2|\alpha|^2}}{\pi^2 \cosh (r)}
    \bigg.\frac{\partial^{2n}}{\partial s^{n}\partial t^{n}}\int^{\infty}_{-\infty}\mathrm{d}^2\beta \exp\bigg[-|\beta|^2-\frac{\tanh (r)}{2}\big(\beta^{2}-\beta^{* 2}\big)-2\beta\alpha^{*}+2\beta^{*}\alpha+s\beta+t\beta^{*}\bigg]\bigg|_{s=t=0}.
\end{align}
Consider the integral formula~\cite{puri2001}
\begin{align}\label{eq:int_1}
        \int^{\infty}_{-\infty}\mathrm{d}^2 \beta \exp{\bigg[a |\beta|^2+b \beta+c \beta^*+d\beta^2+k\beta^{* 2}}\bigg]=\frac{\pi}{\sqrt{a^2-4 d k}}\exp{\bigg[\frac{-a b c+b^2 k+c^2 d}{a^2-4 d k}}\bigg],
    \end{align}
whose convergent conditions are $\operatorname{Re}\left[a\pm d\pm k\right]<0$ and $\operatorname{Re}\left[\nicefrac{(a^2-4 d k)}{a\pm d \pm k}\right]<0$. By using this integral Eq.~\eqref{eq:integ_use1} leads to

\begin{align}\label{eq:eq_eliminate1}
 I_{\Xi}(\bm{\zeta})=&\nonumber\frac{(-1)^{n}\mathrm{e}^{\xi}}{\pi \cosh (r)\sqrt{1+\tanh^2(r)}}\Bigg.\frac{\partial^{2n}}{\partial s^{n}\partial t^{n}}\exp\Bigg[\frac{\tanh(2r)}{4}s^2-\frac{\tanh(2r)}{4}t^2+\big(1+\tanh^2(r)\big)^{-1} st  -2\cosh(r)\text{sech}(2r)\alpha_-s\\&-2\cosh(r)\text{sech}(2r)\alpha^*_+t\Bigg]\Bigg|_{s=t=0},
\end{align}
with
\begin{align}
\xi:=-(\alpha^{2}-\alpha^{* 2})\tanh (2r)-2 |\alpha|^2\text{sech}(2r).
\end{align}
It is challenging to solve Eq.~\eqref{eq:eq_eliminate1} because it has $\mathrm{e}^{\gamma s t}$ terms. We employ the following sum series~\cite{Yun2010} to get rid of it.
\begin{align}\label{eq:st_rid}
\exp(Cs+Dt+Est)
=\sum_{l=0}^{\infty}\frac{E^{l}}{l!}\frac{\partial^{2l}}{\partial C^{l}\partial D^{l}}\left[\exp\left(Cs+Dt\right)\right].
\end{align}
Using this formula Eq.~\eqref{eq:eq_eliminate1} modifies as
\begin{align}
 I_{\Xi}(\bm{\zeta})=&\nonumber\frac{(-1)^{n}\mathrm{e}^\xi}{\pi \cosh (r)\sqrt{1+\tanh^2(r)}}\sum_{l=0}^{\infty}\frac{1}{l!~2^{2l}}\frac{\big[1+\tanh^2(r)\big]^{-l}}{\cosh^{2l}(r)\text{sech}^{2l}(2r)}\frac{\partial^{2l}}{\partial\alpha_+^{* l}\alpha_-^{l}}\frac{\partial^{2n}}{\partial s^{n}\partial t^{n}}\exp\bigg[\frac{\tanh(2r)}{4}s^2-\frac{\tanh(2r)}{4}t^2\\&-2\cosh(r)\text{sech}(2r)(\alpha_-s+\alpha^*_+t)\bigg]\bigg|_{s=t=0}.
\end{align}
Notice the generating function of the Hermite polynomial
\begin{equation}
    H_{n}(x)=\big. \frac{\partial^{n}}{\partial s^{n}}\exp\left(2xs-s^2\right)\big|_{s=0}
    \label{eq:eq_hermite}
\end{equation}
and its recursive relation
\begin{equation}
    \frac{\mathrm{d}^{l}}{\mathrm{d}x^{l}}H_{n}(x)=\frac{2^{l}n!}{(n-l)!}H_{n-l}(x).
    \label{eq:fracdldxlhn}
\end{equation}
The preceding equation can then be simplified in the form of Eq.~\eqref{eq:cross_1} by applying the relationships~\eqref{eq:eq_hermite} and~\eqref{eq:fracdldxlhn}.

Let us now calculate second term of the Wigner function (\ref{eq:wig_spa}). This term can be written as
\begin{align}
W_\boxplus(\bm{\zeta})=\Braket{\psi^{+}_{\text{PA}}|\hat{\Delta}(\alpha)|\psi^{+}_{\text{PA}}}+\Braket{\psi^{-}_{\text{PA}}|\hat{\Delta}(\alpha)|\psi^{-}_{\text{PA}}},
\end{align}
where
\begin{align}
\nonumber\Braket{\psi^{\pm}_{\text{PA}}|\hat{\Delta}(\alpha)|\psi^{\pm}_{\text{PA}}}=&\frac{(- 1)^{n}}{\pi^2}\frac{\mathrm{e}^{2|\alpha|^2}}{\cosh (r)}\big.\frac{\partial^{2n}}{\partial s^{n}\partial t^{n}}\int^{\infty}_{-\infty}\mathrm{d}^2\beta \exp\big[-|\beta|^2\pm\frac{1}{2}\tanh (r)(\beta^2+\beta^{*2})-2\beta\alpha^{*}+2\beta^{*}\alpha+s\beta+t\beta^{*}\big]\big|_{s=t=0}.
\end{align}
Using the integral (\ref{eq:int_1}), we get
\begin{align}
\Braket{\psi^{\pm}_{\text{PA}}|\hat{\Delta}(\alpha)|\psi^{\pm}_{\text{PA}}}=&\nonumber\frac{(-1)^{n}}{\pi}\exp\big[\pm\sinh (2r)(\alpha^{*2}+\alpha^2)-4\cosh ^2(r)|\alpha|^2 \big]\bigg.\frac{\partial^{2n}}{\partial s^{n}\partial t^{n}}\exp\bigg[\pm\frac{1}{4}\sinh (2r)(s^2+t^2)\\&+2\cosh (r)(\bar{\alpha}_{\pm} s-\bar{\alpha}_{\pm}^*t)+\cosh^2 (r)st \bigg]\bigg|_{s=t=0}.
\end{align}
Again, use of the sum series (\ref{eq:st_rid}) eliminates the factors $\mathrm{e}^{\gamma s t}$, that is,
\begin{align}
\Braket{\psi^{\pm}_{\text{PA}}|\hat{\Delta}(\alpha)|\psi^{\pm}_{\text{PA}}}=\sum_{l=0}^{\infty}\frac{(-1)^{l}}{2^{2l}l!}\frac{\partial^{2l}}{\partial \bar{\alpha}_{\pm}^{l}\partial \bar{\alpha}_{\pm}^{*l}}
   \left.\frac{\partial^{2n}}{\partial s^{n}\partial t^{n}}\exp\left[\pm\frac{\sinh (2r)}{4}\left(s^2+t^2\right)+2\cosh r\left(\bar{\alpha}_{\pm}s-\bar{\alpha}_{\pm}^*t\right) \right]\right|_{s=t=0}.
\end{align}
Then, by using the relations (\ref{eq:eq_hermite}) and (\ref{eq:fracdldxlhn}), the expression (\ref{eq:chess2}) is obtained.

\subsection{Derivations of the Wigner function of SPSSVS}
This section presents the detailed derivation of the Eq.~\eqref{eq:wig_svs}, for which the first term has the following form as below:
\begin{align}
I_{\Xi}(\bm{\zeta})=&\Braket{\psi^{+}_{\text{PS}}|\hat{\Delta}(\alpha)|\psi^{-}_{\text{PS}}}.
\end{align}
This term is calculated as
\begin{align}\label{eq:appendix_wig2}
I_{\Xi}(\bm{\zeta})=&\nonumber\frac{1}{\pi^2}\frac{\text{e}^{2|\alpha|^2}}{\cosh(r)}\frac{\partial^{2n}}{\partial s^{n}\partial t^{n}}\exp\bigg[-\frac{\tanh(r)}{2}\big(t^2-s^2\big)\bigg]\int^{\infty}_{-\infty}\mathrm{d}^2\beta\exp\bigg[-|\beta|^2-\big(\tanh(r)t+2\alpha^*\big)\beta-\big(\tanh(r)s\\&-2\alpha\big)\beta^*-\frac{\tanh(r)}{2}(\beta^2-\beta^{* 2})\bigg]\bigg|_{s=t=0}.
\end{align}

Using the integral (\ref{eq:int_1}), we obtain
\begin{align}
I_{\Xi}(\bm{\zeta})=&\nonumber\frac{\mathrm{e}^{\xi}}{\pi\cosh(r)\sqrt{1+\tanh^2 (r)}}\frac{\partial^{2n}}{\partial s^{n}\partial t^{n}}\exp\bigg[\frac{\tanh(2r)}{4}s^2-\frac{\tanh(2r)}{4}t^2+2\text{sech}(2r)\sinh(r)(\alpha^*_+ s-\alpha_- t)\\&+\text{sech}(2r)\sinh^2(r) st\bigg]\bigg|_{s=t=0}.
\end{align}
Now, we eliminate $\mathrm{e}^{\gamma st}$ terms by using Eq.~\eqref{eq:st_rid}:
\begin{align}
I_{\Xi}(\bm{\zeta})=&\nonumber\frac{\mathrm{e}^{\xi}}{\pi\cosh(r)\sqrt{1+\tanh^2 (r)}}\sum^{\infty}_{l=0}\frac{1}{l!2^{2l}\text{sech}^l(2r)}\frac{\partial^{2l}}{\partial \alpha_+^{* l}\partial \alpha_-^{l}}\frac{\partial^{2}}{\partial s^{n}\partial t^{n}}\exp\bigg[\frac{\tanh(2r)}{4}s^2-\frac{\tanh(2r)}{4}t^2\\&+2\text{sech}(2r)\sinh(r)(\alpha^*_+ s+\alpha_- t)\bigg]\bigg|_{s=t=0}.
\end{align}
Then, by using the relations (\ref{eq:eq_hermite}) and (\ref{eq:fracdldxlhn}) we obtain expression (\ref{eq:cross_2}).

Finally, we derive the second term of the Eq.~\eqref{eq:wig_svs}. This term can be written as
\begin{align}
W_\boxplus(\bm{\zeta})=\Braket{\psi^{+}_{\text{PS}}|\hat{\Delta}(\alpha)|\psi^{+}_{\text{PS}}}+\Braket{\psi^{-}_{\text{PS}}|\hat{\Delta}(\alpha)|\psi^{-}_{\text{PS}}},
\end{align}
where

\begin{align}
\Braket{\psi^{\pm}_{\text{PS}}|\hat{\Delta}(\alpha)|\psi^{\pm}_{\text{PS}}}=&\nonumber\frac{1}{\pi}\exp\big[\pm\sinh (2r)(\alpha^2+\alpha^{*2})-2\cosh (2r)|\alpha|^2\big]
    \bigg.\frac{\partial^{2n}}{\partial s^{n}t^{n}}\exp\bigg[\pm \frac{1}{4}\sinh (2r)(s^2+t^2)\\&\pm 2\sinh (r)(\bar{\alpha}_{\pm}t+\bar{\alpha}_{\pm}^{*}s)-\sinh^2 (r)st\bigg]\bigg|_{s,t=0}.
\end{align}
Again, we use Eq.~\eqref{eq:st_rid} to get rid of all $\mathrm{e}^{\gamma st}$ factors, obtaining
\begin{align}
\Braket{\psi^{\pm}_{\text{PS}}|\hat{\Delta}(\alpha)|\psi^{\pm}_{\text{PS}}}=&\nonumber\frac{1}{\pi}\exp\big[\pm\sinh (2r)(\alpha^2+\alpha^{*2})-2\cosh (2r)|\alpha|^2\big]\sum_{l=0}^{\infty}\frac{(-1)^{l}}{2^{2l}l!}
    \frac{\partial^{2l}}{\partial\bar{\alpha}_{\pm}^{l}\bar{\alpha}_{\pm}^{*l}}
    \bigg.\frac{\partial^{2n}}{\partial s^{n}t^{n}}\exp\bigg[\pm\frac{\sinh (2r)}{4}\big(s^2+t^2\big)\\&\pm 2\sinh (r)\big(\bar{\alpha}_{\pm}t+\bar{\alpha}_{\pm}^{*}s\big)\bigg]\bigg|_{s=t=0}.
\end{align}
Finally, this equation can be simplified to expression (\ref{eq:chess2}) by utilizing the relations (\ref{eq:eq_hermite}) and (\ref{eq:fracdldxlhn}).
\section*{Overlaps of SPASV and SPSSV}\label{appendix:overlaps}
In this section we calculate the overlap (\ref{eq:overlap_HW}) of SPASVS and SPSSVS. 
Note that for $n\gg1$ and $|\delta \alpha|\ll1$, the contribution of the cross terms between the states to the overlap is negligible, that is,
\begin{align}
\bra{\psi^{+}_{\text{PA}}}\hat{D}(\delta\alpha)\ket{\psi^{-}_{\text{PA}}}=0\,\text{and}\,\bra{\psi^{+}_{\text{PS}}}\hat{D}(\delta\alpha)\ket{\psi^{-}_{\text{PS}}}=0.
\end{align}
First, we drive each term of Eq.~\eqref{eq:sens_spasvs}. PASV~\eqref{eq:pasvs} can be rewritten as~\cite{Yun2010}
\begin{align}
    \ket{\psi^{\pm}_{\text{PA}}}=\hat{S}(\pm r)\left[\hat{a}^\dagger\cosh (r)\pm \hat{a}\sinh (r)\right]^n \ket{0}.
\end{align}
Then, considering the relation given by~\cite{Yun2010},
\begin{align}\label{eq:unitr}
    (f \hat{a}+g\hat{a}^{\dagger}):=\bigg(-\text{i}\sqrt{\frac{f g}{2}}\bigg)^n H_n\bigg(\text{i}\sqrt{\frac{f}{2g}}\hat{a}+\text{i}\sqrt{\frac{g}{2f}}\hat{a}^\dagger\bigg),
\end{align}
which leads to
\begin{align}
  &[\hat{a}^\dagger\cosh (r)+\hat{a}\sinh (r)]^n=\bigg[-\text{i}\sqrt{\frac{\sinh (2r)}{4}}\bigg]^n H_n\bigg[\text{i}\sqrt{\frac{\tanh (r)}{2}}\hat{a}+\text{i}\sqrt{\frac{\coth (r)}{2}}\hat{a}^\dagger\bigg],\\&[\hat{a}\cosh (r)+\hat{a}^\dagger\sinh (r)]^n=\bigg[-\text{i}\sqrt{\frac{\sinh( 2r)}{4}}\bigg]^n H_n\bigg[\text{i}\sqrt{\frac{\tanh (r)}{2}}\hat{a}^\dagger+\text{i}\sqrt{\frac{\coth (r)}{2}}\hat{a}\bigg].
\end{align}
By using these relations, we obtain
\begin{align}\label{eq:eqn11}
    \braket{\psi^{\pm}_{\text{PA}}|\hat{D}(\delta \alpha)|\psi^{\pm}_{\text{PA}}}=&\nonumber\bigg[\mp\frac{\sinh (2r)}{4}\bigg]^n \Braket{0| H_n\bigg(\text{i}\sqrt{\pm\frac{ \coth (r)}{2}}\hat{a}\bigg)\hat{D}(\eta_{\pm})H_n\bigg(\text{i}\sqrt{\pm\frac{ \coth (r)}{2}}\hat{a}^\dagger\bigg)|0},\\=&\nonumber\bigg[\mp\frac{\sinh (2r)}{4}\bigg]^n \int^{\infty}_{-\infty}\frac{\text{d}^2\alpha}{\pi}\exp\bigg(-\frac{|\alpha|^2}{2}-\frac{\alpha}{2}\eta_{\pm}^*+\frac{\alpha^*}{2}\eta_{\pm}-\frac{|\alpha-\eta_{\pm}|^2}{2}\bigg) H_n\bigg(\text{i}\sqrt{\pm\frac{\coth (r)}{2}}\alpha\bigg)\\&H_n\bigg(\text{i}\sqrt{\pm\frac{\coth (r)}{2}}(\alpha^*-\eta_{\pm}^*)\bigg),
\end{align}
where
\begin{align}
\hat{D}(\eta_{\pm})=\hat{S}^\dagger(\pm r)\hat{D}(\delta \alpha)\hat{S}(\pm r)\;\text{with}\;\eta_{\pm}=\delta \alpha \cosh (r)\mp\delta \alpha^* \sinh (r).
\end{align}
By using (\ref{eq:eq_hermite}), we get
\begin{align}
\braket{\psi^{\pm}_{\text{PA}}|\hat{D}(\delta \alpha)|\psi^{\pm}_{\text{PA}}}=&\nonumber \bigg[\mp \frac{\sinh (2r)}{4}\bigg]^n\frac{\partial^{2n}}{\partial \tau^{n}\partial t^{n}} \exp\big(-\text{i}\sqrt{\pm 2\coth (r)}~\eta_{\pm}^*\tau\big)\exp\big(-\tau^2-t^2\big)\\&\int^{\infty}_{-\infty}\frac{\text{d}^2\alpha}{\pi}\exp\bigg(-\frac{|\alpha|^2}{2}-\frac{\alpha}{2}\eta_{\pm}^*+\frac{\alpha^*}{2}\eta_{\pm}-\frac{|\alpha-\eta_{\pm}|^2}{2}+\text{i}\sqrt{\pm 2\coth (r)}~\alpha t+\text{i}\sqrt{\pm 2\coth( r)}~\alpha^* \tau\bigg)\bigg|_{\tau=t=0}.
\end{align}
Using the integral (\ref{eq:int_1}), the previous equation yields
\begin{align}
\braket{\psi^{\pm}_{\text{PA}}|\hat{D}(\delta \alpha)|\psi^{\pm}_{\text{PA}}}=&\bigg[\mp\frac{\sinh (2r)}{4}\bigg]^n\exp\bigg(-\frac{|\eta_{\pm}|^2}{2}\bigg)\frac{\partial^{2n}}{\partial \tau^{n}\partial t^{n}}\exp\bigg(-t^2+\text{i}\sqrt{\pm 2\coth (r)}~\eta_{\pm} t-\tau^2-\text{i}\sqrt{\pm 2\coth (r)}~\eta_{\pm}^*\tau\\&\mp 2\coth( r)~t\tau\bigg)\bigg|_{t=\tau=0}.
\end{align}
First, we rid out the factors $\mathrm{e}^{\gamma \tau t}$ from above equation by using (\ref{eq:st_rid}). Then, by using (\ref{eq:eq_hermite}) and (\ref{eq:fracdldxlhn}), the preceding equation is simplified to (\ref{eq:senspasv_1}).

Similarly, PSSVS can be rewritten as~\cite{Yun2010}
\begin{align}
    \ket{\psi^{\pm}_{\text{PS}}}=\hat{S}(\pm r)\left[\hat{a}\cosh (r)
\pm \hat{a}^\dagger\sinh (r)\right]^n \ket{0}.
\end{align}
The overlap
\begin{align}\label{eq:eqn11b}
    \braket{\psi^{\pm}_{\text{PS}}|\hat{D}(\delta \alpha)|\psi^{\pm}_{\text{PS}}}=&\nonumber\bigg[\mp\frac{\sinh (2r)}{4}\bigg]^n \Braket{0| H_n\bigg(\text{i}\sqrt{\pm\frac{ \tanh (r)}{2}}\hat{a}\bigg)\hat{D}(\eta_{\pm})H_n\bigg(\text{i}\sqrt{\pm\frac{ \tanh (r)}{2}}\hat{a}^\dagger\bigg)|0},\\=&\nonumber\bigg[\mp\frac{\sinh (2r)}{4}\bigg]^n \int^{\infty}_{-\infty}\frac{\text{d}^2\alpha}{\pi}\exp\bigg(-\frac{|\alpha|^2}{2}-\frac{\alpha}{2}\eta_{\pm}^*+\frac{\alpha^*}{2}\eta_{\pm}-\frac{|\alpha-\eta_{\pm}|^2}{2}\bigg) H_n\bigg(\text{i}\sqrt{\pm\frac{\tanh (r)}{2}}\alpha\bigg)\\&H_n\bigg(\text{i}\sqrt{\pm\frac{\tanh(r)}{2}}(\alpha^*-\eta_{\pm}^*)\bigg),
\end{align}
can be easily simplified to (\ref{eq:spssv_sen1}).
\twocolumngrid
\bibliography{References}

\end{document}